\documentclass[11pt,preprint]{aastex}

\usepackage{color}
\usepackage{natbib}
\usepackage{amsmath}
\tightenlines
\slugcomment{}

\newcommand \balpha     {{\boldsymbol{\alpha}}}
\newcommand \bomega     {{\boldsymbol{\omega}}}

\newcommand \bmu        {{\boldsymbol{\mu}}}

\newcommand \abs        {{\rm abs}}
\newcommand \alphaG     {\alpha_{\rm G}}

\newcommand \bahat      {\hat{\bf a}}
\newcommand \bB         {{\bf B}}
\newcommand \BB         {{\rm BB}}

\newcommand \bD         {{\bf D}}
\newcommand \bE         {{\bf E}}
\newcommand \behat      {\hat{\bf e}}
\newcommand \bh         {{\bf h}}
\newcommand \bhhat      {\hat{\bf h}}
\newcommand \bH         {{\bf H}}
\newcommand \bJ         {{\bf J}}
\newcommand \bJhat      {\hat{\bf J}}

\newcommand \bm         {{\bf m}}
\newcommand \bM         {{\bf M}}
\newcommand \bp         {{\bf p}}
\newcommand \bP         {{\bf P}}
\newcommand \bxhat      {\hat{\bf x}}
\newcommand \byhat      {\hat{\bf y}}
\newcommand \bzhat      {\hat{\bf z}}
\newcommand \bxmhat      {\hat{\bf x}_{\rm m}}
\newcommand \bymhat      {\hat{\bf y}_{\rm m}}
\newcommand \bzmhat      {\hat{\bf z}_{\rm m}}

\newcommand \bz         {{\bf z}}
\newcommand \beq        {\begin{equation}}
\newcommand \beqa	{\begin{eqnarray}}

\newcommand \calN       {{\cal N}}
\newcommand \cm         {\,{\rm cm}}

\newcommand \bdipe       {{\bf p}_{\rm e}}       
\newcommand \bdipm       {{\bf p}_{\rm m}}       

\newcommand \eeq	{\end{equation}}
\newcommand \eeqa	{\end{eqnarray}}
\newcommand \eff        {{\rm eff}}
\newcommand \erg	{\,{\rm ergs}}

\newcommand \eV 	{\,{\rm eV}}

\newcommand \ffill      {f_{\rm fill}}
\newcommand \G          {\,{\rm G}}  
\newcommand \gm         {\,{\rm g}}
\newcommand \GHz        {\,{\rm GHz}}
\newcommand \gtsim	{\gtrsim}		 
\newcommand \Ha 	{{\rm H}}

\newcommand \HH	        {{\rm H}_2}

\newcommand \inc        {{\rm inc}}

\newcommand \kHz        {\,{\rm kHz}}
\newcommand \K  	{\,{\rm K}}

\newcommand \ltsim	{\lesssim}		 
\newcommand \mat        {{\rm mat}}

\newcommand \MJy        {\,{\rm MJy}}

\newcommand \muG        {\mu{\rm G}}

\newcommand \nm         {\,{\rm nm}}     

\newcommand \Oe         {\,{\rm Oe}}     
\newcommand \ohm        {\,{\rm ohm}}

\newcommand \s	        {\,{\rm s}}
\newcommand \sr  	{\,{\rm sr}}
\newcommand \sube       {_{\rm e}}
\newcommand \subm       {_{\rm m}}
\newcommand \supe       {^{\rm (e)}}
\newcommand \supeddy    {^{\rm (eddy)}}
\newcommand \supmag     {^{\rm (mag)}}

\newcommand \yr  	{\,{\rm yr}}

\newcommand \mm         {\,{\rm mm}}

\newcommand{\oldtext}[1]{}

\countdef\decade=200
\advance\decade by \year
\countdef\hours=201
\advance\hours by \time
\divide\hours by 60
\countdef\mins=202
\advance\mins by \hours
\multiply\mins by 60
\multiply\hours by 100
\countdef\miltime=203
\advance\miltime by \hours
\advance\miltime by \time
\advance\miltime by -\mins

\begin{document}

\title{%
 	Magnetic Nanoparticles in the Interstellar
        Medium:\\
        Emission Spectrum and Polarization
	}

\author{B.T. Draine and Brandon Hensley}
\affil{Princeton University Observatory, Peyton Hall, Princeton,
       NJ 08544; {\tt draine@astro.princeton.edu}}

\begin{abstract}
The presence of ferromagnetic or ferrimagnetic nanoparticles in the
interstellar medium would give rise to magnetic dipole radiation at
microwave and submm frequencies.  Such grains may account for the
strong mm-wavelength emission observed from a number of low-metallicity
galaxies, including the Small Magellanic Cloud.
We show how to calculate the absorption and scattering cross sections
for such grains, with particular attention to metallic Fe,
magnetite Fe$_3$O$_4$, and maghemite $\gamma$-Fe$_2$O$_3$,
all potentially present in the interstellar medium.
The rate of Davis-Greenstein alignment by magnetic dissipation is also
estimated.
We determine the temperature of free-flying magnetic grains heated by
starlight and we calculate the polarization of the magnetic dipole emission from
both free-fliers and inclusions.
For inclusions, 
the magnetic dipole emission is expected to be polarized orthogonally
relative to the normal electric dipole radiation.
Finally, we present self-consistent dielectric functions for
metallic Fe, magnetite Fe$_3$O$_4$, and maghemite $\gamma$-Fe$_2$O$_3$,
enabling calculation of absorption and scattering cross sections from
microwave to X-ray wavelengths.
\end{abstract}


\section{Introduction
         \label{sec:intro}}

Observations of low-metallicity dwarf galaxies have
found surprisingly strong emission at submillimeter wavelengths
\citep[e.g.][]{Galliano+Madden+Jones+etal_2003,
       Galliano+Madden+Jones+etal_2005,
       Galametz+Madden+Galliano+etal_2009,
       Grossi+Hunt+Madden+etal_2010,
       OHalloran+Galametz+Madden+etal_2010,
       Galametz+Madden+Galliano+etal_2011},
in excess of what had been expected
based on the observed emission from dust at shorter wavelengths.
This ``submm excess'' could in principle be due to
a large mass of unexpectedly cold dust,
but in some cases the implied dust mass
dust is too large to be consistent with the gas content of the galaxy.
This is in contrast to normal metallicity spiral galaxies, where dust models
with opacities consistent with the local interstellar medium (ISM) can
reproduce the observed submm emission without invoking very cold dust
\citep[e.g.][]{Draine+Dale+Bendo+etal_2007,Aniano+Draine+Calzetti+etal_2012a}.

The Small Magellanic Cloud (SMC) is an example of this phenomenon.
Using data from COBE and WMAP, \citet{Israel+Wall+Raban+etal_2010} and
\citet{Bot+Ysard+Paradis+etal_2010} found a strong submm and mm-wave
excess in the global emission from the SMC,
and the Planck satellite
has confirmed this
\citep{Planck_LMC_SMC_2011}.  \citet{Bot+Ysard+Paradis+etal_2010} and
\citet{Planck_LMC_SMC_2011} both concluded that conventional dust
models cannot account for the observed $500\micron$ -- 3\,mm (600\,GHz
-- 100\,GHz) emission without invoking unphysically large amounts of
very cold dust.  These observations pose a strong challenge
to our understanding of interstellar dust.
If the submm excess in low-metallicity dwarfs is due to thermal
emission from dust, the dust opacity at submm frequencies must
substantially exceed that of the dust in normal-metallicity galaxies,
such as the Milky Way.

Iron is the
is the fifth most abundant element by mass (after H, He, C, and O),
assuming heavy element abundances approximately proportional
to those in the Sun
\citep{Asplund+Grevesse+Sauval+Scott_2009}.
In the ISM, typically
90\% or more of the Fe is missing from the gas phase
\citep{Jenkins_2009}, locked up in solid
grains, although as yet we know little about the nature of the Fe-containing
material.  Interstellar dust models based on amorphous silicate and
carbonaceous material 
\citep[e.g.,][]{Mathis+Rumpl+Nordsieck_1977,
Draine+Lee_1984,
Desert+Boulanger+Puget_1990,
Weingartner+Draine_2001a,
Zubko+Dwek+Arendt_2004,
Draine+Li_2007,
Draine+Fraisse_2009}
typically assume that most of the Fe missing from the gas
is incorporated in amorphous
silicate material, but it is entirely possible for much or most of the
solid-phase Fe to be in the form of metallic Fe or Fe oxides.  
A number of authors have previously proposed that the dust in the Galaxy
may include a significant population of Fe or Fe oxide particles
\citep[e.g.,][]{Schalen_1965,
       Wickramasinghe+Nandy_1971,
       Huffman_1977,
       Chlewicki+Laureijs_1988,
       Cox_1990,
       Jones_1990}.

The objective of the present study is to
consider magnetic materials as possible constituents of interstellar
dust, and to evaluate the absorption and extinction cross sections for
interstellar grains composed of such materials.  We will find that
these materials have large absorption cross sections at mm-wave
frequencies and below.

Models for the thermal emission of radiation from interstellar grains
require calculation of the absorption cross section
$C_\abs(\omega)$ for each grain type and size,
as a function of frequency $\omega$.
Calculations of $C_\abs(\omega)$ for grains
often assume the grain material to be nonmagnetic
(magnetic permeability $\mu=1$): the
material is assumed to respond only to the local
electric field.
At optical and infrared frequencies this is an excellent approximation.
However, at submm and microwave frequencies, the
magnetic response of grain material to an oscillating magnetic field
may not be negligible.

If the magnetic response enables a grain to absorb energy from
an oscillating magnetic field, then there must also be a magnetic
contribution to thermal emission.  
Magnetic dipole emission from magnetic grain materials
was discussed by \citet[][hereafter DL99]{Draine+Lazarian_1999a}, 
who concluded that 
magnetic dipole radiation might be important
at microwave frequencies if interstellar grains were composed in
part of magnetic materials.  At that time the frequencies of particular
interest were $\sim$20--60\,GHz, where
so-called ``anomalous microwave emission'' (AME) had
been detected from the ISM
\citep{Kogut+Banday+Bennett+etal_1996,
deOliveira-Costa+Kogut+Devlin+etal_1997,
Leitch+Readhead+Pearson+Myers_1997}.  
Theoretical studies 
\citep{Draine+Lazarian_1998a,Draine+Lazarian_1998b}
had shown that the observed AME could be
electric dipole radiation from rapidly-spinning
ultrasmall grains, but 
DL99 also showed that
magnetic dipole emission from magnetic grain materials could
not be ruled out as a significant contributor
to emission at these frequencies.

DL99 adopted a simple
model for the frequency-dependent absorption in magnetic materials.
The present study uses the Gilbert equation
to model the magnetic response at microwave and submm frequencies.

In Section \ref{sec:magnetic_materials} we introduce three candidate
magnetic materials: metallic iron (bcc Fe), magnetite (Fe$_3$O$_4$),
and maghemite ($\gamma$-Fe$_2$O$_3$).  
Section \ref{sec:magnetic_permeability} presents a model for the 
magnetization dynamics of these materials.
From the model, we derive the
complex polarizability tensor $\balpha^{\rm (mag)}(\omega)$ 
relating the oscillating
magnetization $\bm$ in response to the applied field $\bh$.

Because the objective is to calculate absorption cross sections $C_\abs$,
in Sections \ref{sec:polarizabilities of small particles} 
and \ref{sec:Cabs} we discuss
calculation of $C_\abs$ for small magnetic particles.
The importance of eddy currents is discussed.

The purely magnetic contribution to
the absorption cross section is calculated in Section 
\ref{sec:magnetic_absorption}.  
In Section \ref{sec:IR-Microwave_Opacities} we calculate absorption
cross sections for randomly-oriented spheres of Fe,
magnetite, and maghemite from optical to microwave
frequencies.  These calculations make use of dielectric functions
for iron, magnetite, and maghemite that are derived in Appendices
\ref{app:epsilon_Fe}-\ref{app:epsilon_maghemite}.
In Section \ref{sec:temperatures}
we estimate the temperatures for spherical grains of pure
Fe, magnetite, or maghemite illuminated by interstellar starlight.
In Section \ref{sec:IR spectrum} we calculate the emission from
small particles of metallic Fe, magnetite, and maghemite, and
the polarization of this emission is discussed 
in Section \ref{sec:polarization}.
In Section \ref{sec:discussion} we 
find that a large fraction of interstellar Fe could be in magnetic
material without violating constraints
provided by observations of microwave and submm emission, or by the
observed wavelength-dependent extinction.
The principal conclusions are
summarized in Section \ref{sec:summary}.

In a separate paper
\citep{Draine+Hensley_2012b} we show that Fe nanoparticles can
acccount for the extremely strong submm-microwave emission observed
from the SMC.

\section{\label{sec:magnetic_materials}
         Magnetic Materials}

All materials exhibit some magnetic response, but our attention here
is on materials that have unpaired electron spins that are spontaneously
ordered
even in the absence of an applied magnetic field.
These materials fall into three broad classes: ferromagnetic,
ferrimagnetic, and antiferromagnetic
[for an introduction to magnetic materials, see, e.g., 
\citet{Morrish_2001} or \citet{Coey_2010}].

The magnetization of ferromagnetic and ferrimagnetic materials occurs
because the Coulomb energy of the system is minimized for wavefunctions
where the electrons have a nonzero net spin;
this is often referred to as being due to the ``exchange interaction''.
We consider three candidate magnetic materials: metallic Fe, 
magnetite Fe$_3$O$_4$, and maghemite $\gamma$-Fe$_2$O$_3$.
At low temperatures, 
metallic Fe is {\it ferromagnetic} (all unpaired Fe spins parallel), 
while magnetite and maghemite are
{\it ferrimagnetic} (partial cancellation of Fe spins).
As we will see below, the spontaneous magnetization of these
materials leads to enhanced absorption at microwave and submm frequencies.

Two other common iron oxides, ferrous oxide (w\"ustite) FeO and hematite
$\alpha$-Fe$_2$O$_3$ are {\it antiferromagnetic}, with zero net
magnetization because the Fe spins alternate in direction from lattice
site to lattice site, resulting in perfect cancellation.  These also
have interesting properties at microwave and submm frequencies, but we
do not discuss them here.


For spontaneously magnetized single-domain samples at low temperatures,
the {\it static magnetization} $\bM_0=\bM_s$, where $\bM_s$ is the 
{\it saturation magnetization}.
For ferrimagnetic systems (see Section \ref{sec:ferri} below)
we also require the ratio $\beta\equiv |M_A|/|M_B|$ of
the magnetizations of the two opposed spin subsystems ($A$ and $B$),
and a parameter $N_{AB}$
characterizing the coupling between the two spin subsystems.


The Coulomb energy of the crystal is minimized when the unpaired
spins are aligned (i.e., it is magnetized).
Because of the lattice structure, the Coulomb energy depends on the
direction of the magnetization relative to the lattice, 
with the Coulomb energy being
minimized when the magnetization is along certain preferred directions.
The variations in Coulomb energy with direction of magnetization are
referred to as the {\it crystalline anisotropy energy}.
The
so-called {\it crystalline anisotropy field} $\bH_K$ is a fictional
magnetic field that conveniently characterizes the
``energy cost'' of small deviations of $\bM$ away from a direction which
minimizes the energy of the material; the variations in Coulomb energy
with small deviations $\theta$ of the direction of magnetization away from
the preferred direction are {\it as though}
there were an actual field $\bH_K=H_K\hat{\bf e}_{\rm easy}$ 
applied parallel to the direction
$\hat{\bf e}_{\rm easy}$
where the crystal energy is minimized,
corresponding to an effective potential
$E_K={\rm const} -\bM\cdot\bH_K$.  Thus $\bH_K$ is related to the
second derivative $\partial^2E_K/\partial\theta^2$ of the anisotropy
energy $E_K(\theta)$, evaluated at $\partial E_K/\partial\theta=0$.
For a ferromagnetic material,
$H_K=(1/M_s)\partial^2E_K/\partial\theta^2$.


\begin{table}[t]
\newcommand \texta {a}
\newcommand \textb {b}
\newcommand \TebbleCraik {c}
\newcommand \Abe {f}
\newcommand \NYD {x}
\newcommand \OD {e}
\newcommand \BKP {h}
\newcommand \Dionne {d}
\newcommand \DMS {g}
\begin{center}
\caption{\label{tab:magnetic_materials}
         Candidate Magnetic Materials}
{\footnotesize
\begin{tabular}{l c c c c c c c}
\hline
material              & $\rho$ 
                      & $V_{\rm Fe}$$^\texta$
                      & $4\pi M_s$ 
                      & $\beta$ 
                      & $\partial^2E_K/\partial\theta^2$
                      & $H_K$\,$^\textb$
                      & $N_{AB}$
\\
                      & (g$\cm^{-3}$)
                      & ($10^{-23}\cm^3$)
                      & (Oe)
                      &
                      & (ergs$\cm^{-3}$)
                      & (Oe)
                      &
\\
\hline
metallic Fe (bcc)     & 7.87                          
                      & 1.18                          
                      & 22020~$^\TebbleCraik$         
                      & 0                             
                      & $9.6\times10^5$~$^\TebbleCraik$ 
                      & ~~548                           
                      & --                            
\\ 
magnetite Fe$_3$O$_4$ & 5.18                          
                      & 2.48                          
                      & 6400$^\Dionne$                
                      & 5/9$^\OD$                     
                      & $29.2\times10^5$~$^\Abe$      
                      & ~1640                         
                      & 10800$^\Dionne$               
\\
maghemite $\gamma$-Fe$_2$O$_3$ 
                      & 4.86                          
                      & 2.73                          
                      & 4890$^\DMS$                   
                      & 3/5                           
                      & $2.4\times10^5$\,$^\BKP$      
                      & ~~617                           
                      & ~9280$^\Dionne$                
\\
\hline
\multicolumn{4}{l}{$\texta$ Volume per Fe atom}
&
\multicolumn{4}{l}{$\OD$ \citet{Ozdemir+Dunlop_1999}}
\\
\multicolumn{4}{l}{$\textb$ 
                   $H_K=[(1-\beta)/(1+\beta)M_s]\partial^2E_K/\partial\theta^2$}&
\multicolumn{4}{l}{$\Abe$ \citet{Abe+Miyamoto+Chikazumi_1976} (see text)}
\\
\multicolumn{4}{l}{$\TebbleCraik$ \citet{Tebble+Craik_1969}}
&
\multicolumn{4}{l}{$\DMS$ \citet{Dutta+Manivannan+Seehra+etal_2004}}
\\
\multicolumn{4}{l}{$\Dionne$ \citet{Dionne_2009}, Tables 4.3,4.4}
&
\multicolumn{4}{l}{$\BKP$ \citet{Babkin+Koval+Pynko_1984} (see text)}
\\
\hline
\end{tabular}
}
\end{center}
\end{table}

%
\subsection{Metallic Iron}

At $T<1185\K$ the thermodynamically favored structure for Fe is bcc.
The low-temperature saturation magnetization of bcc Fe 
is $4\pi M_s=2.202\times10^4\G$ \citep{Tebble+Craik_1969},
corresponding to $2.2\mu_B$ per Fe.
Even very small clusters of Fe atoms are ferromagnetic.
For clusters of $N=20-50$ Fe atoms, the magnetic moment per atom
is $\sim$$3\mu_B$, declining to the bulk value of $2.2\mu_B$ for
$N\gtsim 400$
\citep{Billas+Becker+Chatelain+deHeer_1993,Tiago+Zhou+Alemany+etal_2006}.
Fe spheres with diameters $\ltsim$$1.7\times10^{-6}\cm$ are
expected to be single-domain; for prolate spheroids single-domain
behavior persists to larger sizes
\citep{Butler+Banerjee_1975}.
Larger Fe particles are expected to contain more than one
magnetic domain, as this lowers the free energy.

For a crystal with cubic symmetry
(e.g., bcc Fe), the crystalline anisotropy energy
can be written \citep{Morrish_2001}
\beq
E_K = K_1(\alpha_1^2\alpha_2^2+\alpha_1^2\alpha_3^3+\alpha_2^2\alpha_3^3)
+ K_2\alpha_1^2\alpha_2^2\alpha_3^2
~~~,
\eeq
where the direction cosines $\alpha_j$ specify the direction of
$\bM$ relative to the cubic axes of the
crystal.
For bcc Fe, 
$K_1=4.8\times10^5\erg\cm^{-3}$ \citep{Tebble+Craik_1969}.
For $K_1>0$, $E_K$ is minimized for
spontaneous magnetization along one of the cubic axes
(e.g., $\alpha_1=1$, $\alpha_2=\alpha_3=0$),
with $\partial^2 E_K/\partial\theta^2=2K_1$.
The crystalline
anisotropy field is
\beq
H_K = \frac{1}{M_s}\frac{\partial^2 E_K}{\partial\theta^2}
= \frac{2K_1}{M_s} = 548\Oe~~~.
\eeq

\subsection{Magnetite {\rm Fe$_3$O$_4$}}

Magnetite Fe$_3$O$_4$
is spontaneously magnetized even at room temperature.
Of the three Fe ions in the Fe$_3$O$_4$ unit cell, one (with magnetic
moment $5\mu_B$) is in
the ``$A$'' sublattice, and two (with total magnetic moment $9\mu_B$)
are in the ``$B$'' sublattice.
The spins in the two sublattices are anti-aligned,
giving a net magnetic moment $4\mu_B$ per Fe$_3$O$_4$.
The ratio of the magnetic moments in the two sublattices
$\beta\equiv |M_A|/|M_B| = 5/9$.

Magnetite undergoes a phase transition at the Verwey transition temperature
$T_{\rm V}=119\K$, with the crystal structure changing 
from cubic (at $T>T_{\rm V}$) to monoclinic (at $T<T_{\rm V}$).
Interstellar grain temperatures will be below $T_V$.

The low-temperature saturation magnetization $4\pi M_s=6400\Oe$
\citep{Tebble+Craik_1969}.
The crystalline anisotropy energy for $T<T_{\rm V}$ is of the form
\citep{Ozdemir+Dunlop_1999}
\beq \label{eq:E_K_magnetite}
E_K = K_a\alpha_a^2 + K_b\alpha_b^2 + K_{aa}\alpha_a^4 + K_{bb}\alpha_b^4
+K_{ab}\alpha_a^2\alpha_b^2 - K_u\alpha_{111}^2
~~~,
\eeq
where $\alpha_a$, $\alpha_b$, and $\alpha_c$ are direction cosines
relative to the $[1\bar{1}0]$, $[110]$, and $[001]$ directions of the 
$T>T_{\rm V}$ cubic
lattice,
and $\alpha_{111}$ is the direction cosine relative to the $[111]$
direction.
The anisotropy constants $K_a$, $K_b$, $K_{aa}$, $K_{bb}$, $K_{ab}$, and
$K_u$ have been measured at low temperatures
\citep{Abe+Miyamoto+Chikazumi_1976}.

If the $K_u$ term is neglected,\footnote{%
   At $T=4.2\K$, $K_u/K_a=0.082$ \citep{Abe+Miyamoto+Chikazumi_1976},
   hence the $K_u$ term is a small correction in (\ref{eq:E_K_magnetite}).}
then,
in the absence of applied fields, spontaneous magnetization at $T<T_{\rm V}$
will be along the $[001]$ direction = $\hat{\bf e}_{\rm easy}$.
Let $\theta$ be the angle between $\bM$ and $\hat{\bf e}_{\rm easy}$.
If we write $\alpha_a=\sin\theta\cos\phi$, $\alpha_b=\sin\theta\sin\phi$,
and average over $\phi$, we find
\beq
\frac{\partial^2 E_K}{\partial \theta^2}
= 2\left(K_a\cos^2\phi+K_b\sin^2\phi\right)
\rightarrow (K_a+K_b) = 29.2\times10^5\erg\cm^{-3}
~~~,
\eeq
and the crystalline anisotropy field
\beq
H_K=\frac{(1+\beta)}{(1-\beta)M_s}\frac{\partial^2 E_K}{\partial\theta^2}
=1640\,{\rm Oe}
~~~.
\eeq
The interlattice coupling coefficient $N_{AB}$ will
be discussed in Section \ref{sec:ferri}.

\subsection{Maghemite $\gamma$-{\rm Fe$_2$O$_3$}}

Fe$_2$O$_3$ exists in several different crystal forms, of which two are
common in nature: hematite ($\alpha-$Fe$_2$O$_3$) and
maghemite ($\gamma-$Fe$_2$O$_3$).
Hematite is thermodynamically favored, 
but conversion of maghemite to hematite in nanoparticles
requires very high temperatures, $T \gtsim600\K$
\citep{Ozdemir+Banerjee_1984,Kido+Higashino+Kamitsuji+etal_2004}.
Maghemite is the low temperature oxidation product of
magnetite.

Maghemite is widely used in magnetic recording media. It is 
ferrimagnetic below the Curie temperature $T_{\rm C}\approx 870\K$, 
with a low temperature
magnetization $4\pi M_s=4890\Oe$ \citep{Dutta+Manivannan+Seehra+etal_2004}.
The $A$ lattice has a magnetic moment $5\mu_B$ per Fe$_2$O$_3$, and the
$B$ lattice has $25\mu_B/3$ per Fe$_2$O$_3$
\citep{Morrish_2001,Dionne_2009,Coey_2010},
giving a net
magnetic moment of $(10/3)\mu_B$ per formula unit, and
$\beta\equiv |M_A|/|M_B|=3/5$.\footnote{
  $(10/3)\mu_B$ per Fe$_2$O$_3$ implies 
  $4\pi M_s=4\pi(\rho/160{\rm \,amu})(10\mu_B/3)=7100\Oe$.
  The actual measured magnetization, $4890\Oe$, is only
  $\sim$$70\%$ of the expected value, indicating that real maghemite
  samples differ from the ideal model used here.}

The crystalline anisotropy 
for maghemite is uncertain.
For bulk maghemite at room temperature, $K_1=-4.6\times10^4\erg\cm^{-3}$
\citep{Brabers_1995}.  
\citet{Babkin+Koval+Pynko_1984} observed an increase in the anisotropy
with decreasing temperature down to $T\approx 70K$; extrapolation to
$T\approx20\K$ gives $K_1 \approx -1.8\times10^5\erg\cm^{-3}$.
For $K_1<0$, spontaneous magnetization occurs along the [111] axes,
with
\beq
\frac{\partial^2 E_K}{\partial\theta^2} = -\frac{4}{3}K_1
~~~.
\eeq
Much larger anisotropies have been
reported for maghemite nanoparticles.
\citet{Valstyn+Hanton+Morrish_1962}
found $K_1=-2.5\times10^5\erg\cm^{-3}$ at room temperature.

The crystalline anisotropy can also be studied by measurements of
the ``blocking temperature'' for superparamagnetic
behavior \citet{Morrish_2001} to determine the energy
barrier $K_{\rm eff} V$ for
magnetic reorientation in a particle of volume $V$.
For maghemite, these studies
lead to effective anisotropies that are often much larger, e.g.,
$K_{\rm eff}=2.3\times10^6\erg\cm^{-3}$ for $\sim$$7\nm$ (diameter) particles
\citep{Dutta+Manivannan+Seehra+etal_2004} at $T\approx 85\K$,
$K_{\rm eff}\approx1.2\times10^5\erg\cm^{-3}$ at $T\approx25\K$
\citep{Shendruk+Desautels+Southern+van_Lierop_2007},
$K_{\rm eff}=7\times10^6\erg\cm^{-3}$ for $\sim8\nm$ hollow particles 
at $T\approx30\K$
\citep{Cabot+Alivisator+Puntes+etal_2009},
and $K_{\rm eff}=6.0\times10^6\erg\cm^{-3}$ for $\sim$$6\nm$ particles
at $T\approx60\K$
\citep{Tsuzuki+Schaffel+Muroi+McCormick_2011}.
It is not clear why these various studies reach such different
conclusions.  The blocking temperature studies probe the entire
$E_K$ energy surface, particularly paths from the minima to saddle points.
The present study, however, is concerned with small perturbations,
and therefore we require only the local curvature of $E_K$ at the
minima.
For purposes of discussion, we will disregard the blocking temperature
results and
will take $K_1=-1.8\times10^5\erg\cm^{-3}$
based on the thin-film measurements by \citet{Babkin+Koval+Pynko_1984}, 
giving
\beq
\frac{\partial^2 E_K}{\partial\theta^2} = 2.4\times10^5\erg\cm^{-3}
\eeq
and $H_K=617\Oe$.
In view of the much larger values of $K_{\rm eff}$ obtained from
blocking temperature studies,
the adopted value of $H_K$ should be regarded as very uncertain.
The interlattice coupling coefficient $N_{AB}$ will be discussed in
Section \ref{sec:ferri}.

\section{\label{sec:magnetic_permeability}
         Magnetic Response at Microwave and Submm Frequencies}
\subsection{Ferromagnetic Resonance}

Consider a small, single-domain 
ferromagnetic sample subject to an applied field
\beqa
\bH(t) &=& H_0\bzhat + \bh(t)
\\ \label{eq:h(t)}
\bh(t) &=& \bh_0e^{-i\omega t}
~~~.
\eeqa
We will later set $H_0=0$, but retain it here for generality.
The oscillating field $\bh$ is assumed to be small.
The magnetization will include static and oscillating components:
\beqa
\bM(t) &=& \bM_0 + \bm(t)
\\
\bm(t) &=& \bm_0 e^{-i\omega t}
~~~.
\eeqa
The dynamic magnetic response of ferromagnetic materials
is a complex problem \citep[see, e.g.][]{Morrish_2001,Soohoo_1985}.
Phenomenological treatments of the damping
include the Landau-Lifshitz equation
\citep{Landau+Lifshitz_1935},
the Bloch-Bloembergen equation
\citep{Bloch_1946,Bloembergen_1950}
and the
Gilbert equation
\citep{Gilbert_1955,Gilbert_2004}.
The Bloch-Bloembergen equation has two phenomenological damping times
($\tau_1$, $\tau_2$), while the Landau-Lifshitz equation and the Gilbert
equation each have only one adjustable damping parameter.
The Bloch-Bloembergen equation has nonphysical behavior --
see Appendix \ref{app:BB_equation} --
and therefore it is not used in the present study.

In the limit of weak damping, the Landau-Lifshitz and Gilbert equations
are equivalent.
Because the Gilbert equation is more mathematically convenient,
it is often used, and we will employ it here.
\citet{Iida_1963} argued that the Gilbert equation is more physically
reasonable.

The Gilbert equation is 
\beq \label{eq:Gilbert}
\frac{d\bM}{dt} = 
\gamma \bM\times\bH_T + \alphaG\frac{\bM}{|\bM|}\times\frac{d\bM}{dt}
~~~,
\eeq
where
\beq
\gamma=-\frac{g}{2}\frac{e}{m_e c} 
= -\frac{g}{2}1.759\times10^7 \frac{\s^{-1}}{\rm gauss}
\eeq
is the ratio of magnetic moment to angular momentum
($g\approx 2$ is the usual gyromagnetic factor),
and the total effective field $\bH_T$ is discussed below.
The term $\gamma\bM\times\bH_T$ in 
(\ref{eq:Gilbert}) corresponds to precession of $\bM$
around the effective field $\bH_T$, while the second
term describes relaxation toward a solution with $\bM \parallel \bH_T$.
The dimensionless parameter
$\alphaG$ is Gilbert's phenomenological damping coefficient.

\citet{Morrish_2001} states that ``typical'' ferromagnetic absorption
has $\alphaG\approx 0.07$.  \citet{Wu+Ding+Jiang+etal_2006} studied the response of micron- and
submicron-sized Fe particles in a nonconducting, 
nonmagnetic matrix at frequencies
between 0.5 and 16.5 GHz, and reported that
their measurements were consistent with $\alphaG\approx 0.4$.
\citet{Neo+Yang+Ding_2010} used $\alphaG=0.1$ to model the behavior
of single-domain bcc Fe particles for frequencies up to $100\GHz$
(although experimental data were shown only for $\leq10\GHz$).
We will consider a range of possible
values for $\alphaG$.

The ``total'' field $\bH_T$ in (\ref{eq:Gilbert}) is an effective field.
Consider an ellipsoidal grain. 
For a uniform applied field $\bH_0+\bh$, the magnetization $\bM$
in the grain will be uniform.
The effective field can be written
\beq \label{eq:H_T}
\bH_T = \bH_0 + \bh - \bD\cdot\bM + \bH_K
~~~.
\eeq
Here $\bD$ is the ``demagnetization tensor'', 
giving the internal contribution to $\bH_T$ arising
from the ``magnetic poles'' at the surface of the sample.\footnote{
   This is completely analogous to the depolarization electric field
   arising from the surface charge associated with the discontinuity
   in the polarization field $\bP$ at the surface of a dielectric.}
For an ellipsoid, $\bD$ is diagonal, with diagonal elements 
\beq \label{eq:D_jj}
D_{jj}=4\pi L_j
~~~,
\eeq 
where $L_j$ is the same geometrical factor, or ``shape factor'', 
that arises in relating
the internal electric field to the applied electric field in a dielectric
\citep[see, e.g.,][p.\ 146]{Bohren+Huffman_1983}.
For a sphere, $L_j=1/3$.  For ellipsoids, $L_1+L_2+L_3=1$.
In the discussion below we will consider prolate spheroids with
$L_x=L_y=\frac{1}{2}(1-L_z)$.
Values of $L_x$ and $L_z$ are given in Table
\ref{tab:ferromagnetic_resonance_parameters} for various axial ratios.

At zero temperature, and with applied $\bH_0+\bh=0$, the free energy $E_K$
will depend on the
direction of the magnetization $\bM$ relative to the crystal axes.
If $\theta$ is the angle between $\bM$ and the energy-minimizing direction
$\hat{\bf e}_{\rm easy}$,
one may define a fictitious ``crystalline anisotropy field''
\beq
\bH_K \equiv \frac{1}{M_s}\frac{\partial^2 E_K}{\partial\theta^2}
\hat{\bf e}_{\rm easy}
\eeq
evaluated at the equilibrium position 
(where $\partial E_K/\partial\theta=0$).

Consider an ellipsoidal grain.
We will take $\hat{\bf e}_{\rm easy}$ and
the static (spontaneous) magnetization $\bM_0=M_0\bzhat$ ($M_0>0$)
to be in the
$\bzhat$ direction, which we take to be one of the principal axes of the
ellipsoid.
Then, to leading order in $\bm$, (\ref{eq:Gilbert}) becomes
\beqa
-i\omega m_{0x} &=& ~\gamma m_{0y}\left[H_0+H_{Kz}-D_{zz}M_0\right]
                -\gamma M_0\left[h_{0y}-D_{yy}m_{0y}\right] 
               + i\alphaG\omega m_{0y}
\\
-i\omega m_{0y} &=& \!\!-\gamma m_{0x}\left[H_0+H_{Kz}-D_{zz}M_0\right]
                +\gamma M_0\left[h_{0x}-D_{xx}m_{0x}\right] 
               - i\alphaG\omega m_{0x}
~~~.
\eeqa
To simplify, assume the sample to be a spheroid, with
$D_{yy}=D_{xx}$.  
Define
\beqa \label{eq:def_omega0}
\omega_0&\equiv& -\gamma\left[H_0+H_{Kz}-(D_{zz}-D_{xx})M_0\right]
\\
\omega_0^\prime &\equiv& \omega_0 - i\alphaG \omega
\\
\label{eq:def_omegaM}
\omega_M &\equiv& -\gamma M_0
~~~.
\eeqa
We then obtain two coupled equations
\beqa
-i\omega m_{0x} &=& -\omega_0^\prime m_{0y} + \omega_M h_{0y}
\\
-i\omega m_{0y} &=& \omega_0^\prime m_{0x} - \omega_M h_{0x}
~~~,
\eeqa
with solutions
\beqa
m_{0x} &=& \frac{\omega_0^\prime \omega_M}{(\omega_0^\prime)^2-\omega^2}h_{0x}
       - \frac{i\omega\omega_M}{(\omega_0^\prime)^2-\omega^2}h_{0y}
\\
m_{0y} &=& \frac{i\omega\omega_M}{(\omega_0^\prime)^2-\omega^2}h_{0x}
       + \frac{\omega_0^\prime \omega_M}{(\omega_0^\prime)^2-\omega^2}h_{0y}
~~~.
\eeqa
If we define
\beqa \label{eq:chi_pm,ferromagnetic}
\chi_\pm&\equiv& 
\frac{(\omega_0^\prime\pm\omega)\omega_M}
     {(\omega_0^\prime)^2-\omega^2}
= \frac{\omega_M}{\omega_0^\prime \mp \omega}
\eeqa
then the oscillating magnetization $\bm$ satisfies
\beq \label{eq:tensor_equation}
\left(
\begin{array}{c}
m_{0x} \\ m_{0y} \\ m_{0z}
\end{array}
\right)
=
\left(
\begin{array}{c c c}
\frac{1}{2}(\chi_+ +\chi_-) & -\frac{1}{2}i(\chi_+ -\chi_-) & 0 \\
\frac{1}{2}i(\chi_+-\chi_-)  & \frac{1}{2}(\chi_+ +\chi_-) & 0 \\
0 & 0 & 0
\end{array}
\right)
\left(
\begin{array}{c}
h_{0x} \\ h_{0y} \\ h_{0z}
\end{array}
\right)
~~~.
\eeq
If we define
\beqa \label{eq:circular_polarizations}
\bhhat_{\pm}
&\equiv&
\frac{1}{\sqrt{2}}\left(\bxhat \pm i\byhat\right)
~~~,
\eeqa
then an arbitrary applied field $\bh_0e^{-i\omega t}$ will produce
a magnetization
\beq \label{eq:magnetization_response}
\bm_0e^{-i\omega t} = \left[
      \chi_+ (\bh_0\!\cdot\!\bhhat_+^*) \bhhat_+
    + \chi_- (\bh_0\!\cdot\!\bhhat_-^*) \bhhat_-
      \right] e^{-i\omega t}
~~~.
\eeq
The eigenvectors $\bhhat_\pm$ correspond to circular polarization modes,
with ${\rm Re}(\bhhat_\pm e^{-i\omega t})$ 
rotating either anticlockwise ($\bhhat_+$)
or clockwise ($\bhhat_-$) around $\bM_0$.
In this linearized treatment,
the oscillating magnetization $\bm$ has no response to the component of
the applied field $\bh$ parallel to $\bM_0$ -- this is because the
magnetization along $\bM_0$ is already saturated.
The perpendicular component $h_\perp=\sqrt{h_{x}^2+h_{y}^2}$ 
can deflect the magnetization
away from $\bzhat$, leading to $m_{0x}\propto h_\perp$ and 
$m_{0y}\propto h_\perp$,
but $m_{0z}$ is second order in $h_\perp$.

\begin{table}[t]
\newcommand \TebbleCraik {a}
\newcommand \NYD {x}
\newcommand \OD {b}
\newcommand \omegacom {c}
\newcommand \DMS {d}
\begin{center}
\caption{\label{tab:ferromagnetic_resonance_parameters}
         Ferromagnetic Resonance Parameters for Metallic Fe}
{\footnotesize
\begin{tabular}{l c c c c}
\hline
shape
                      & $L_x=L_y$
                      & $L_z$
                      & $\omega_0/2\pi$ 
                      & $\omega_M/2\pi$
\\
                      &
                      &
                      & (GHz)
                      & (GHz)
\\
\hline
sphere 
                      & 0.33333
                      & 0.33333
                      & ~~1.53                          
                      & ~4.91                          
\\
1.2:1 prolate spheroid
                      & 0.35694
                      & 0.28613
                      & ~~5.90 
                      & ~4.91
\\
1.5:1 prolate spheroid
                      & 0.38351
                      & 0.23298
                      & 10.8~                          
                      & ~4.91
\\ 
2:1 prolate spheroid
                      & 0.41322
                      & 0.17356
                      & 16.3~                          
                      & ~4.91
\\ 
3:1 prolate spheroid
                      & 0.44565
                      & 0.10871
                      & 22.3~                          
                      & ~4.91
\\
4:1 prolate spheroid
                      & 0.46230
                      & 0.07541
                      & 25.4~
                      & ~4.91
\\
5:1 prolate spheroid
                      & 0.47209
                      & 0.05582
                      & 27.2~
                      & ~4.91
\\
10:1 prolate spheroid
                      & 0.48986
                      & 0.02029
                      & 30.5~
                      & ~4.91
\\
$\infty$:1 prolate spheroid
                      & 0.50000
                      & 0.00000
                      & 32.4~
                      & ~4.91
\\
\hline
\end{tabular}
}
\end{center}
\end{table}

\begin{figure}[h]
\begin{center}
\includegraphics[width=9.0cm,angle=0]{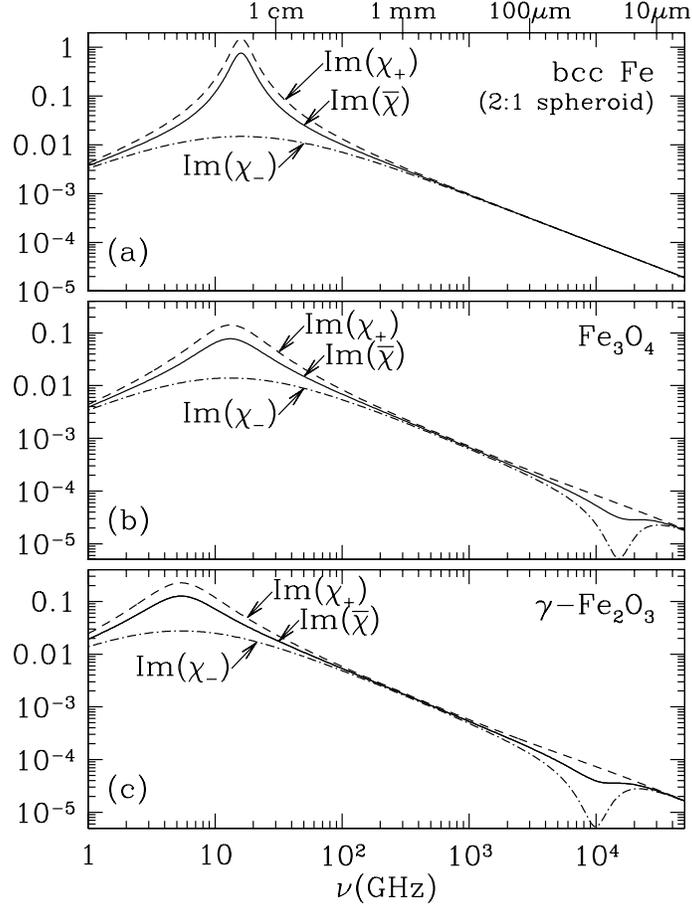}
\caption{\label{fig:chipm}
         \footnotesize
         Imaginary part of $\chi_+$, $\chi_-$, and
         $\bar{\chi}\equiv(\chi_++\chi_-)/2$
         versus frequency
         for (a) metallic Fe, 
         (b) magnetite Fe$_3$O$_4$, and 
         (c) maghemite $\gamma$-Fe$_2$O$_3$.
         Gilbert damping parameter $\alphaG=0.2$ has been assumed.
        }
\end{center}
\end{figure}

The real and imaginary parts of $\chi_\pm$ are
\beqa
{\rm Re}(\chi_\pm) &=& 
\frac{\omega_M(\omega_0\mp\omega)}{(\omega_0\mp\omega)^2+\alphaG^2\omega^2}
\\ \label{eq:Im_chi_pm}
{\rm Im}(\chi_\pm) &=&
\frac{\alphaG\omega_M\omega}{(\omega_0\mp\omega)^2+\alphaG^2\omega^2}
~~~.
\eeqa
The dissipative part of the response is measured by ${\rm Im}(\chi_\pm)$.
It is evident from (\ref{eq:Im_chi_pm}) that
$\chi_{+}$ has a resonance at 
$\omega=\omega_0$,   
corresponding
to the applied field being in resonance with free precession of $\bm$ around
the static magnetization $\bM_0$.
The resonance frequency $\omega_0$ depends on the shape, because the effective
field $\bH_T=\bH_K-\bD\cdot\bM$ includes the demagnetization field
$-\bD\cdot\bM$, which is shape-dependent.
Table \ref{tab:ferromagnetic_resonance_parameters} gives
$\omega_0$ for spheres and selected prolate spheroids.
For highly elongated spheroids, $\omega_0\rightarrow32\GHz$. 
Figure \ref{fig:chipm}a
shows
${\rm Im}(\chi_+)$ and ${\rm Im}(\chi_-)$ for 2:1 Fe spheroids.

The $\chi_\pm$ in (\ref{eq:chi_pm,ferromagnetic}) give 
the ratio\footnote{
   $\chi_\pm$ is not the same as the intrinsic magnetic susceptibility
$\chi_{\rm int}$, because the latter is the ratio of the magnetization response
to the {\it local} or {\it internal} field, 
which includes an internal depolarization field.}
of the magnetization response $\bm$ to the {\it external} applied
field $\bh$.  
If $V$ is the volume, the magnetic dipole moment of the grain is 
\beqa
\bp_{\rm m} &=& 
V\bm = V\left[
\chi_+\bhhat_+(\bhhat_+^*\cdot\bh_0)
+
\chi_-\bhhat_-(\bhhat_-^*\cdot\bh_0)
\right]
e^{-i\omega t}
\\
&=&
\balpha_{\rm m}\cdot\bh
\eeqa
where the magnetic polarizability tensor $\balpha_{\rm m}$ is
\beq \label{eq:alpha_mag}
\balpha_{\rm m} = 
V\left(\chi_+\bhhat_+\bhhat_+^* + \chi_-\bhhat_-\bhhat_-^*\right)
~~~.
\eeq
At low frequencies, we have $\chi_\perp(0) = \chi_\pm(\omega=0) =
\omega_M/\omega_0$.  For single-domain Fe spheres, we estimate
$\chi_\perp(0)\approx 3.2$.

\subsection{\label{sec:ferri}
            Ferrimagnetic and Antiferromagnetic Resonance}

Ferrimagnetic and antiferromagnetic materials have 
two oppositely-aligned spin lattices, $A$ and $B$,
with magnetizations $\bM_A$, $\bM_B$.
Let $|\bM_B| \geq |\bM_A|$.
At low temperatures, the two spin lattices are each perfectly
aligned.
Recall that $\beta \equiv |\bM_A|/|\bM_B|$,
$\bM_A=-\beta\bM_B$ and
$\bM=\bM_s=(1-\beta)\bM_B$.
Ferrimagnetic materials have $0<\beta<1$, and antiferromagnetic
materials have $\beta=1$.  

Just as for ferromagnetism, the spin alignment results
from minimization of the electronic energy.
There are now two alignments to consider: 
(1) antialignment of $\bM_A$ with $\bM_B$
and (b) alignment of the net
magnetization $\bM=\bM_A+\bM_B$ with ``easy'' directions relative
to the crystal axes.

Minimization of the energy when
the two spin systems are antialigned can be
described through fictitious fields $-N_{AB}\bM_B$ 
acting on $A$
and $-N_{AB}\bM_A$ acting on $B$.
The dimensionless coupling coefficient $N_{AB}$
has been estimated by
\citet{Dionne_2006,Dionne_2009},\footnote{
    Conventions vary.  The $\calN_{AB}$ given by 
    \citet[][Table 4.4]{Dionne_2009} is related to our $N_{AB}$ by
    $\calN_{AB}=(\rho/N_A m) N_{AB}$, where $\rho$ is the
    mass density, $N_A$ is Avogadro's number,
    and $m$ is the mass per molecule.
    Fe$_3$O$_4$ has $\calN_{AB}=240\,{\rm mole}\cm^{-3}$ and
    $(\rho/N_A m)= 0.0223\,{\rm mole}\cm^{-3}$.
    $\gamma$-Fe$_2$O$_3$ has $\calN_{AB}=282\,{\rm mole}\cm^{-3}$
    and $(\rho/N_A m)=0.0304\,{\rm mole}\cm^{-3}$.
    }
with $N_{AB}=1.08\times10^4$ for Fe$_3$O$_4$, and
$N_{AB}=9.28\times10^3$ for $\gamma$-Fe$_2$O$_3$.
As we will see, the very strong coupling between the two sublattices
results in a resonance in the infrared.

With the spin systems antialigned, the overall energy of the system
is minimized when the magnetization is along an ``easy'' direction,
which we take to be $\bzhat$: $\bM_B=M_{Bz}\bzhat$, 
$\bM_A=-\beta M_{Bz}\bzhat$.

The effective fields (or ``crystalline fields'')
$\bH_{KA}=H_{KAz}\bzhat$ 
and $\bH_{KB}=H_{KBz}\bzhat$ characterize the
energy cost of departures of the magnetization $\bM$ away
from the easy direction $\bzhat$.
We take 
$H_{KBz}= +H_K$,
$H_{KAz}= -H_K$, where
\beq
H_K =
\frac{1}{(1+\beta)} 
\frac{1}{|\bM_B|}\frac{\partial^2 E_K}{\partial\theta^2}
=
\frac{(1-\beta)}{(1+\beta)} 
\frac{1}{|\bM_s|}\frac{\partial^2 E_K}{\partial\theta^2}
~~~.
\eeq

The Gilbert equations for the two spin systems are
\beqa
\frac{d}{dt}\bM_A 
&=&
\gamma_A\bM_A\times\left(\bH_0+\bh + \bH_{KA} - N_{AB}\bM_B \right)
+ \alphaG\frac{\bM_A}{|\bM_A|}\times\frac{d\bM_A}{dt}
~~~,
\\ 
\frac{d}{dt}\bM_B 
&=&
\gamma_B\bM_B\times\left(\bH_0+\bh + \bH_{KB} - N_{AB}\bM_A \right)
+ \alphaG\frac{\bM_B}{|\bM_B|}\times\frac{d\bM_B}{dt}
~~~.
\eeqa
The contribution to the dynamics 
of the depolarization field $-\bD\cdot\bM$ is neglected, 
as $\bD\cdot\bM\sim \bM_s$ is small compared to the
crystalline fields $\bH_{KA}$, $\bH_{KB}$, and the
coupling fields $N_{AB}\bM_A$, $N_{AB}\bM_B$.

As for the ferromagnetic case, we take $\bH_0=0$, 
\beqa
\bH(t)&=&\bh_0 \, e^{-i\omega t}
~~~,
\\
\bM_A(t)&=&M_{Az}\bzhat+\bm_A \, e^{-i\omega t}
~~~,
\\
\bM_B(t)&=&M_{Bz}\bzhat+\bm_B \, e^{-i\omega t}
~~~,
\eeqa
with $M_{Bz}>0$, $M_{Az}<0$,
and linearize around the equilibrium to obtain
\beqa
-i\omega m_{Ax}
&=&
~\gamma_A m_{Ay}(H_{KAz} - N_{AB}M_{Bz}) - i\alphaG\omega m_{Ay}
-\gamma_A M_{Az}(-N_{AB}m_{By} + h_y)
\\
-i\omega m_{Bx}
&=&
~\gamma_B m_{By}(H_{KBz} - N_{AB}M_{Az}) + i\alphaG\omega m_{By}
-\gamma_B M_{Bz}(-N_{AB}m_{Ay} + h_y)
\\
-i\omega m_{Ay}
&=&
\!
-\gamma_A m_{Ax}(H_{KAz} - N_{AB}M_{Bz}) + i\alphaG\omega m_{Ax}
+\gamma_A M_{Az}(-N_{AB}m_{Bx} + h_x)
\\
-i\omega m_{By}
&=&
\!
-\gamma_B m_{Bx}(H_{KBz} - N_{AB}M_{Az}) - i\alphaG\omega m_{Bx}
+\gamma_B M_{Bz}(-N_{AB}m_{Ax} + h_x)
~~~.
\eeqa
Defining
\beqa
\omega_{0A} &\equiv& -\gamma_A\left(H_{KAz}-N_{AB}M_{Bz}\right)
\\
\omega_{0B} &\equiv& -\gamma_B\left(H_{KBz}-N_{AB}M_{Az}\right)
\\
\omega_{MA} &\equiv& -\gamma_A M_{Az}
\\
\omega_{MB} &\equiv& -\gamma_B M_{Bz}
\\
\omega_{0A}^\prime &\equiv& \omega_{0A}+i\alphaG\omega
\\
\omega_{0B}^\prime &\equiv& \omega_{0B}-i\alphaG\omega
~~~,
\eeqa
the above equations become
\beqa
-i\omega m_{Ax} 
&=&
-\omega_{0A}^\prime m_{Ay} - \omega_{MA}N_{AB}m_{By} + \omega_{MA}h_y
\\
-i\omega m_{Bx}
&=&
-\omega_{0B}^\prime m_{By} - \omega_{MB}N_{AB}m_{Ay} + \omega_{MB}h_y
\\
-i\omega m_{Ay}
&=&
~~\,\omega_{0A}^\prime m_{Ax} + \omega_{MA}N_{AB}m_{Bx} - \omega_{MA}h_x
\\
-i\omega m_{By}
&=&
~~\,\omega_{0B}^\prime m_{Bx} + \omega_{MB}N_{AB}m_{Ax} - \omega_{MB}h_x
~~~.
\eeqa
Considering circularly polarized modes
\beqa
\bh &=&
h_0 
~\bhhat_{\pm} 
\,e^{-i\omega t}
\\
\bm_\pm &\equiv& \left[(\bm_A+\bm_B)\cdot\bhhat_\pm^*\right]\bhhat_\pm
~~~,
\eeqa
we can solve to find
\beqa
m_{\pm,x}
&=&
\left[\frac{
2N_{AB}\omega_{MA}\omega_{MB}
-\omega_{MA}\omega_{0B}^\prime-\omega_{MB}\omega_{0A}^\prime
\pm (\omega_{MA}+\omega_{MB})\omega
  }
{N_{AB}^2\omega_{MA}\omega_{MB}-
\omega_{0A}^\prime\omega_{0B}^\prime - \omega^2
\pm(\omega_{0A}^\prime+\omega_{0B}^\prime)\omega 
}
\right] h_x
~~~,
\eeqa
with a similar equation for $m_{\pm,y}$.
Thus,  
\beqa
\bm_{\pm} &=& \chi_{\pm}\,h_0\,\bhhat_\pm\,e^{-i\omega t}
~~~,
\eeqa
\beqa
\label{eq:chi_pm,ferrimagnetic}
\chi_{\pm} &=& 
\frac{
2N_{AB}\omega_{MA}\omega_{MB}-
\omega_{MA}\omega_{0B}^\prime-\omega_{MB}\omega_{0A}^\prime
 \pm (\omega_{MA}+\omega_{MB})\omega}
{N_{AB}^2\omega_{MA}\omega_{MB}-\omega_{0A}^\prime\omega_{0B}^\prime-
\omega^2
\pm (\omega_{0A}+\omega_{0B})\omega}
~~~.
\eeqa
The magnetization $\bm_0e^{-i\omega t}$
in response to a general applied field $\bh_0e^{-i\omega t}$
is given by (\ref{eq:magnetization_response}), but with
$\chi_\pm$ given by (\ref{eq:chi_pm,ferrimagnetic}).
Values of $\omega_{MA}$, $\omega_{MB}$, $\omega_{0A}$, and
$\omega_{0B}$ for magnetite and maghemite are given in
Table \ref{tab:ferrimagnetic_resonance_parameters}.
Figure \ref{fig:chipm} shows ${\rm Im}(\chi_+)$ and ${\rm Im}(\chi_-)$ for
single-domain particles of magnetite Fe$_3$O$_4$ and
maghemite $\gamma$-Fe$_2$O$_3$.

\begin{table}[t]
\newcommand \TebbleCraik {a}
\newcommand \NYD {x}
\newcommand \OD {b}
\newcommand \omegacom {c}
\newcommand \DMS {d}
\begin{center}
\caption{\label{tab:ferrimagnetic_resonance_parameters}
         Ferrimagnetic Resonance Parameters}
{\footnotesize
\begin{tabular}{l c c c c c c}
\hline
material
                      & $\omega_{{\rm res},+}/2\pi$
                      & $\omega_{MA}/2\pi$
                      & $\omega_{MB}/2\pi$
                      & $\omega_{0A}/2\pi$
                      & $\omega_{0B}/2\pi$
                      & $\omega_{{\rm res},-}/2\pi$
\\
                      & (GHz) & (GHz) & (GHz)
                      & (THz)
                      & (THz)
                      & (THz)
\\
\hline
magnetite Fe$_3$O$_4$
                      & 16.1                         
                      & -4.01                        
                      & 3.21                         
                      & $-34.7$                      
                      & $19.3$                       
                      & 15.4                         
\\
maghemite $\gamma$-Fe$_2$O$_3$ 
                      & 7.83                         
                      & -4.09                        
                      & 2.71                         
                      & $-25.3$                      
                      & $15.2$                       
                      & 10.1                         
\\
\hline
\end{tabular}
}
\end{center}
\end{table}

\begin{figure}[h]
\begin{center}
\includegraphics[width=8cm,angle=0]{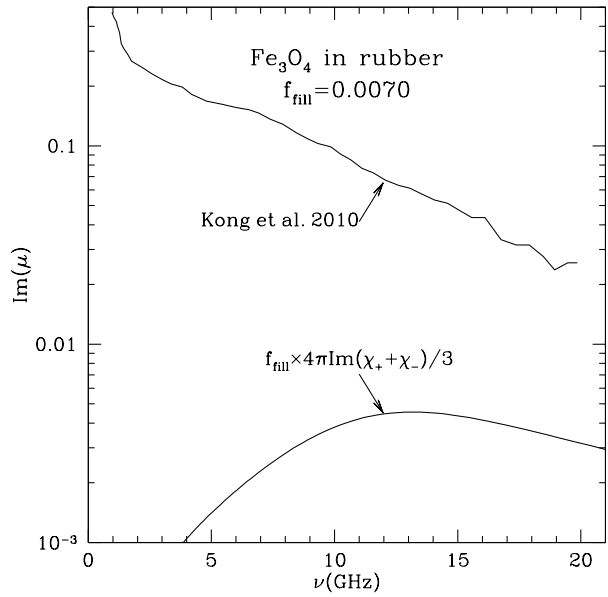}
\caption{\label{fig:fe3o4_vs_kong}
         \footnotesize
         Predicted ${\rm Im}(\mu)$ for Fe$_3$O$_4$ nanoparticles in
         nonmagnetic medium, for volume filling factor
         $\ffill=0.025$, assuming $\alphaG=0.2$.
         Also shown are
         measurements of \citet{Kong+Ahmad+Abdullah+etal_2010} for
         Fe$_3$O$_4$ particles in a rubber matrix at room temperature.
         Agreement is poor.
         The laboratory sample shows much more absorption
         at low frequencies than expected for single-domain
         nanoparticles.
         }
\end{center}
\end{figure}
\begin{figure}[h]
\begin{center}
\includegraphics[width=8cm,angle=0]{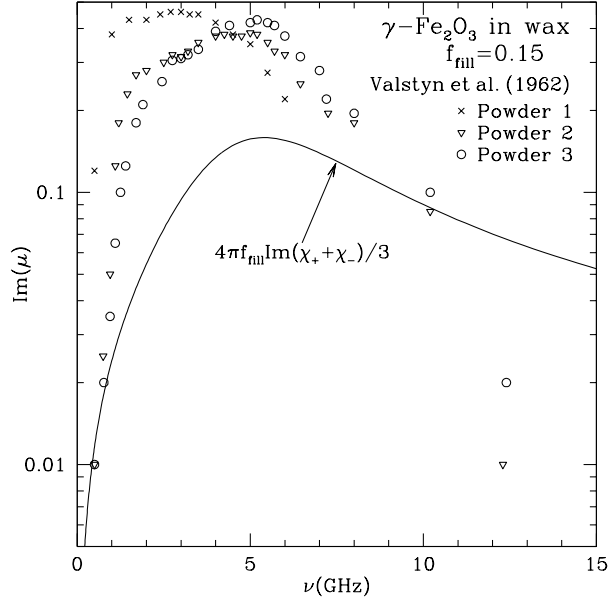}
\caption{\label{fig:fe2o3_vs_valstyn}
         \footnotesize
         Predicted ${\rm Im}(\mu)$ for $\gamma$-Fe$_2$O$_3$
         powder suspended in paraffin wax, for
         assumed $\alphaG=0.2$ and volume
         filling factor $\ffill=0.15$.
         Also shown are
         measurements of
         \citet{Valstyn+Hanton+Morrish_1962} for 3 different
         samples.
         Agreement is poor.
         }
\end{center}
\end{figure}

\subsection{Comparisons to Experimental Data}

As already noted, the Gilbert equation used here is phenomenological,
and one would like to see it tested in the laboratory.
Measurements of magnetic absorption at high frequencies are challenging.
Laboratory measurements on Fe particles in insulating matrices
are strongly affected by the conductivity of Fe unless very small 
particles are employed and
great care is taken to avoid particle coagulation.

Low conductivity materials like magnetite and maghemite are much more
suitable for laboratory work.
Figure \ref{fig:fe3o4_vs_kong} shows 1--20~GHz 
measurements on $\sim$$25\nm$ Fe$_3$O$_4$
particles with volume filling factor $\ffill=0.007$ (4~wt\%)
in a nonmagnetic, nonconducting rubber matrix
\citep{Kong+Ahmad+Abdullah+etal_2010}.
For $\ffill\ll 1$
the medium should have 
(see Appendix \ref{app:eff_med_theory}, eq.\ \ref{eq:mu_eff_sphere})
\beq
\mu\approx 1+\ffill\frac{4\pi}{3}(\chi_++\chi_-)
\eeq
if the particles are randomly oriented.
In Figure \ref{fig:fe3o4_vs_kong}
the laboratory sample exhibits much stronger absorption at
low frequencies than predicted by the present model with
$\alphaG\approx0.2$.
The reason for this is unclear.  
Electron micrography indicates particle clumping,
which complicates interpretation.
More importantly, the particles may be large enough to
be multidomain, in which case motion of domain walls will
contribute to absorption.

Figure \ref{fig:fe2o3_vs_valstyn} shows data from
\citet{Valstyn+Hanton+Morrish_1962} for three maghemite powders
dispersed in paraffin wax with volume filling factor $\ffill=0.15$.  
Typical particle sizes ranged from $120\nm$ (Powder 1)
to $\sim$$500\nm$ (Powders 2 and 3).  For 1--10~GHz, the maghemite
powders exhibit stronger absorption than predicted by the model for
single-domain absorption.  The particle sizes are large enough that
multidomain structure is likely, and the extra absorption may be due
to motion of domain walls.  Of greater concern, however, is the very
small absorption measured at the highest frequency, 12.3 GHz, well
below what was expected from the model (if $\alphaG\approx0.2$).

To test the current model, what is needed 
are measurements of absorption by 
isolated, small, single-domain particles,
at frequencies up to $\sim$$300\GHz$.
We have been unable to find such data for
Fe, Fe$_3$O$_4$, or $\gamma-$Fe$_2$O$_3$ at frequencies above
$20\GHz$, and even at lower frequencies the data are sparse.

\section{\label{sec:polarizabilities of small particles}
         Polarizabilities of Small Particles}

Magnetic effects appear to be of potential importance 
only for $\lambda\gtsim 500\micron$
($\nu \ltsim 600\GHz$).   We may therefore assume the particle
to be small compared to the wavelength, in which case the electromagnetic
response of the particle is predominantly through the electric and
magnetic dipole moments $\bdipe(t)$ and $\bdipm(t)$ 
induced by the incident electromagnetic field.
The electric and magnetic dipole moments can be written
\beqa
\bdipe &=& \balpha\sube\cdot\bE_\inc ~~~,
\\
\bdipm &=& \balpha\subm\cdot\bH_\inc ~~~,
\eeqa
where $\bE_\inc$ and $\bH_\inc$ 
are the applied electric and magnetic fields, both
assumed to be $\propto e^{-i\omega t}$,
and $\balpha\sube(\omega)$ and $\balpha\subm(\omega)$ 
are the electric and magnetic
polarizability tensors for the grain.
We will assume the grain to be a spheroid, 
with symmetry axis and static magnetization parallel to $\bzhat$.

\subsection{Electric Polarizability}

We assume the dielectric function to be isotropic (i.e., a scalar).
The electric polarizability tensor for the ellipsoidal particle has eigenvalues
\citep{van_de_Hulst_1957}
\beq
\alpha_{{\rm e},j}(\omega) 
= \frac{V}{4\pi}\frac{\epsilon-1}{(\epsilon-1)L_j + 1}
~~~~~,~~~~~j=1 - 3~~~,
\eeq
where $\epsilon(\omega)=\epsilon_1+i\epsilon_2$ 
is the dielectric function, and
$L_j$ is the same geometrical factor as
appears in (\ref{eq:D_jj}).

\subsection{Magnetic Polarizability}
\subsubsection{Magnetization}

A small particle can acquire a net 
magnetic moment $\bdipm$ from aligned electron
spins, and also from electric currents circulating within the body of the grain.
Therefore, we separate the magnetic polarizability 
tensor $\balpha\subm$ into two components:
\beq
\balpha\subm(\omega) = \balpha\subm\supmag(\omega) + 
\balpha\subm\supeddy(\omega)~~~;
\eeq  
$\balpha\subm\supmag\bH_\inc$ is the magnetic moment contributed by
magnetization of the grain material itself (i.e., alignment of electron
spins), and $\balpha\subm\supeddy\bH_\inc$ is the
magnetic moment generated by the eddy currents (zero for nonconducting
materials).

In principle, $\balpha\subm\supmag$ and $\balpha\subm\supeddy$ should
be solved for self-consistently.
Alignment of the electron spins
will be in response to both $\bH_\inc$ and the field generated
by the eddy currents; the eddy currents are induced by
$\partial\bB/\partial t$, which will include a contribution from
the oscillating magnetization $\bm$.  
Nevertheless, we will provisionally
assume that $\balpha\subm\supmag$ can be calculated neglecting eddy currents,
and that
$\balpha\subm\supeddy$ can be calculated neglecting the magnetic response
of the grain.  The validity of this approximation will be verified below.


\subsubsection{\label{sec:eddy currents}
         Eddy Currents}

The eddy current contribution is estimated from the solution
for eddy currents in a nonmagnetic sphere of radius $a$
\citep{Landau+Lifshitz+Pitaevskii_1993}:
$\balpha\subm\supeddy$ is diagonal, with diagonal elements
\beq \label{eq:alpha^eddy}
\alpha_{{\rm m}}\supeddy(\omega,a)
=
\frac{3V}{8\pi}\left[\frac{3}{y^2} - \frac{3}{y}\cot y - 1\right]
~~~,
\eeq
where $y^2 \equiv \epsilon (\omega a/c)^2$.
The dielectric function
$\epsilon = 1 + \delta\epsilon^{\rm (bound)}+4\pi i\sigma/\omega$, where
$\delta\epsilon^{\rm (bound)}$ is the contribution
of the bound electrons, 
and $\sigma$ is the electrical conductivity of the grain
material.
While (\ref{eq:alpha^eddy}) is for a spherical shape,
we will use it as an estimate for spheroids,
setting $a=(3V/4\pi)^{1/3}$.

The limiting case of a perfect conductor has ${\rm Im}(y)\rightarrow\infty$,
and $\alpha_{{\rm m},j}\supeddy\rightarrow -3V/8\pi$.
In this limit, the magnetic field generated by eddy currents on the surface 
completely cancel the applied magnetic
field $\bh$ within the sphere.
For finite conductivity, the eddy currents are not confined to the
surface, and the magnetic field in the interior is not uniform.
We take the typical field in the interior to be
\beqa \label{eq:heff}
\bh_{\rm eff} &\approx& 
\bh_0 \left( 1 - \phi_{\rm eddy}\right) e^{-i\omega t}
\\ \label{eq:phi_eddy}
\phi_{\rm eddy}(\omega,a) &\equiv& -
\frac{8\pi\alpha_{\rm m}\supeddy(\omega,a)}{3V} ~~~.
\eeqa
Our estimate (\ref{eq:heff}) for effects of eddy currents
is exact in the limits $\alpha_{\rm m}\supeddy=0$ 
(nonconducting: $\phi_{\rm eddy}=0$) and
$\alpha_{\rm m}\supeddy\rightarrow-3V/8\pi$ 
(perfect conductor: $\phi_{\rm eddy}=1$), 
and should be a reasonable approximation of the shielding
for intermediate values of $\alpha_{\rm m}\supeddy$.

Shielding by eddy currents will act to lower the magnetization of the
grain material.
We will take the contribution of magnetization
to the magnetic polarizability tensor to be
\beq
\balpha_{\rm m}\supmag \approx 
\left[1-{\rm Re}(\phi_{\rm eddy})\right]V
\left(\chi_+\bhhat_+\bhhat_+^* + \chi_-\bhhat_-\bhhat_-^*\right) ~~~.
\eeq
If ${\rm Re}(\phi_{\rm eddy}) \ltsim 0.2$, 
the eddy currents do not reduce the magnetic field within the sphere
by more than $\sim$$20\%$.  

The electrical conductivity for interstellar metallic Fe particles
with Ni impurities is estimated in Appendix \ref{app:epsilon_Fe}.
Figure \ref{fig:Fe polarizability/volume} shows 
$4\pi|\alpha_{{\rm m},j}\supeddy|/V$ calculated
for Fe spheres with radii $a=0.1,\, 0.3,\, 1.0\micron$.
From the figure, we see that for $a\leq0.3\micron$,
the magnetization response of an Fe grain is not substantially
affected by eddy currents for frequencies $\nu\ltsim 500\GHz$.

Eddy currents do not appreciably affect the magnetization of submicron
grains of magnetite and maghemite, because the electrical conductivity
of these materials is much smaller than the conductivity of metallic Fe.

\begin{figure}[ht]
\begin{center}
\includegraphics[width=10.0cm,angle=0]{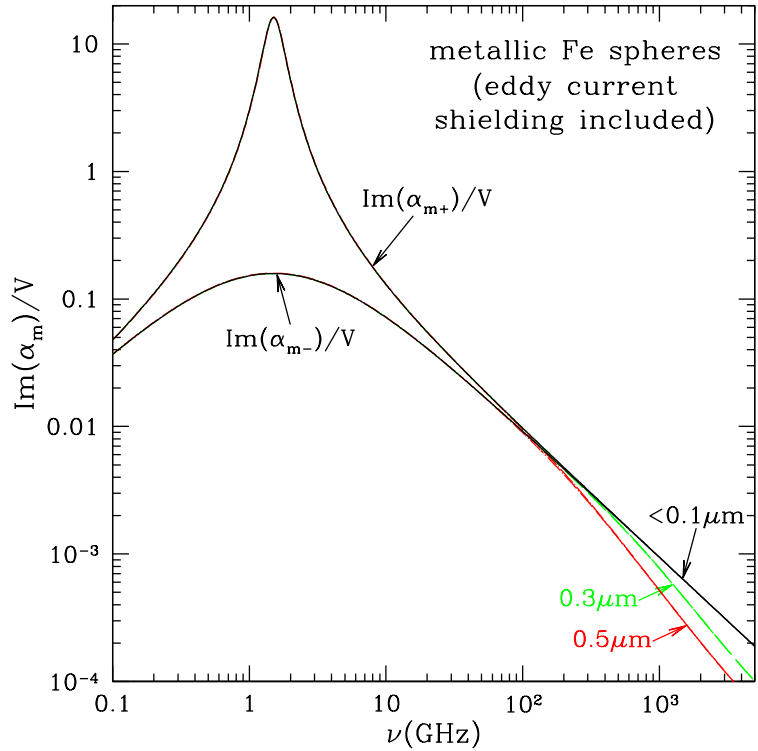}
\caption{\footnotesize
  \label{fig:Fe polarizability/volume}
  Dissipative part
  ${\rm Im}(\alpha_{\rm m})$ 
  of magnetic polarizability arising from magnetization (solid curves)
  in Fe grains,
  including the shielding effects of
  eddy currents.
  The Gilbert equation with $\alphaG=0.2$ is assumed,
  and $\alpha_{\rm m}$ is shown for both circular polarizations.
  For $a\ltsim0.3\micron$ eddy current shielding is unimportant
  for $\nu< 500\GHz$.
  }
\end{center}
\end{figure}

\section{\label{sec:Cabs}
         Absorption by Small Particles: The Dipole Limit}

For a uniform sphere
illuminated by a monochromatic plane wave, 
the electromagnetic scattering problem can be
solved exactly \citep{Mie_1908,Debye_1909}
provided the material is isotropic, i.e., can be characterized
by a scalar dielectric function $\epsilon(\omega)$ 
and a scalar magnetic permeability $\mu(\omega)$.
This mathematical solution is commonly referred to as ``Mie theory''.
Many numerical implementations of Mie theory assume the material to be
nonmagnetic ($\mu=1$), although
a code for general scalar $\epsilon$ and $\mu$ has
been implemented by \citet{Milham_1994}.
The magnetic response (\ref{eq:magnetization_response})
is, however, highly anisotropic, therefore the Mie theory solution
is not directly applicable to the present problem.
Lacking an exact solution, we seek an approximate
treatment.

If the grain is small compared to $c/\omega$, then
$\bE_\inc$ and $\bH_\inc$ can be approximated as uniform over the
body of the grain.
Let $\bdipe(t)$ and $\bdipm(t)$ be the electric and magnetic dipole moment
of the grain.
The time-averaged rate at which the incident wave
does work on the grain is 
\beqa 
\Bigl\langle\frac{dW}{dt}\Bigr\rangle &=& 
\Bigl\langle{\rm Re}(\bE_\inc)\cdot 
\frac{d\,{\rm Re}(\bdipe)}{dt}\Bigr\rangle + 
\Bigl\langle{\rm Re}(\bH_\inc)\cdot 
\frac{d\,{\rm Re}(\bdipm)}{dt}\Bigr\rangle
\\ \label{eq:<dW/dt>}
 &=& 
\bigl\langle{\rm Re}(\bE_\inc)\cdot {\rm Re}(-i\omega \bdipe)\bigr\rangle + 
\bigl\langle{\rm Re}(\bH_\inc)\cdot {\rm Re}(-i\omega \bdipm)\bigr\rangle
~~~,
\eeqa
where $\langle ...\rangle$ denotes a time-average.
If $\behat_{j}$ and $\bhhat_{j}$ are normalized
eigenvectors of $\balpha\sube$ and
$\balpha\subm$, respectively, with eigenvalues $\alpha_{{\rm e},j}$ and 
$\alpha_{{\rm m},j}$ (for $j=1,2,3$) then (\ref{eq:<dW/dt>}) becomes
\beq
\Bigl\langle \frac{dW}{dt}\Bigr\rangle =
\frac{\omega}{2}
\sum_{j=1}^3
\left[
{\rm Im}(\alpha_{{\rm e},j}) |\bE_\inc^*\cdot \behat_j|^2 +
{\rm Im}(\alpha_{{\rm m},j}) |\bH_\inc^*\cdot \bhhat_j|^2
\right]
~~~.
\eeq
For unpolarized, isotropic illumination,
we set $|\bE_\inc^*\cdot\behat_j|^2=(1/3)E_0^2$ and
$|\bH_\inc^*\cdot\bhhat_j|^2=(1/3)H_0^2=(1/3)E_0^2$.
The absorption cross section is
\beqa
\langle C_\abs\rangle &=& \frac{\langle dW/dt\rangle}{c E_0^2/8\pi} =
\langle C_\abs\supe\rangle + \langle C_\abs\supmag\rangle
\\
\langle C_\abs\supe\rangle &=&
\frac{4\pi\omega}{3c} ~
{\rm Im}\left[{\rm Tr}(\balpha\sube)\right] 
=
\frac{\omega V}{3c} 
\sum_{j=1}^3 
\frac{\epsilon_2}
{\left[(\epsilon_1-1)L_j + 1\right]^2 + 
 \left[L_j\epsilon_2\right]^2
}
\\
\label{eq:Cabs^mag}
\langle C_\abs\supmag\rangle &=&
\frac{4\pi\omega}{3c} ~
{\rm Im}\left[{\rm Tr}(\balpha\subm)\right] = 
\frac{4\pi\omega V}{3c} {\rm Im}\left(\chi_+ + \chi_-\right)
+ \frac{4\pi\omega}{c} {\rm Im}\left(\alpha\subm^{\rm (eddy)}\right)
~~~,
\eeqa
where ${\rm Tr}$ denotes the trace of a matrix, equivalent to the
sum of eigenvalues.

\section{\label{sec:magnetic_absorption}
         Magnetic Absorption Cross Section}

\begin{figure}[ht]
\vspace*{-1.0em}
\begin{center}
\includegraphics[width=7.0cm,angle=0]{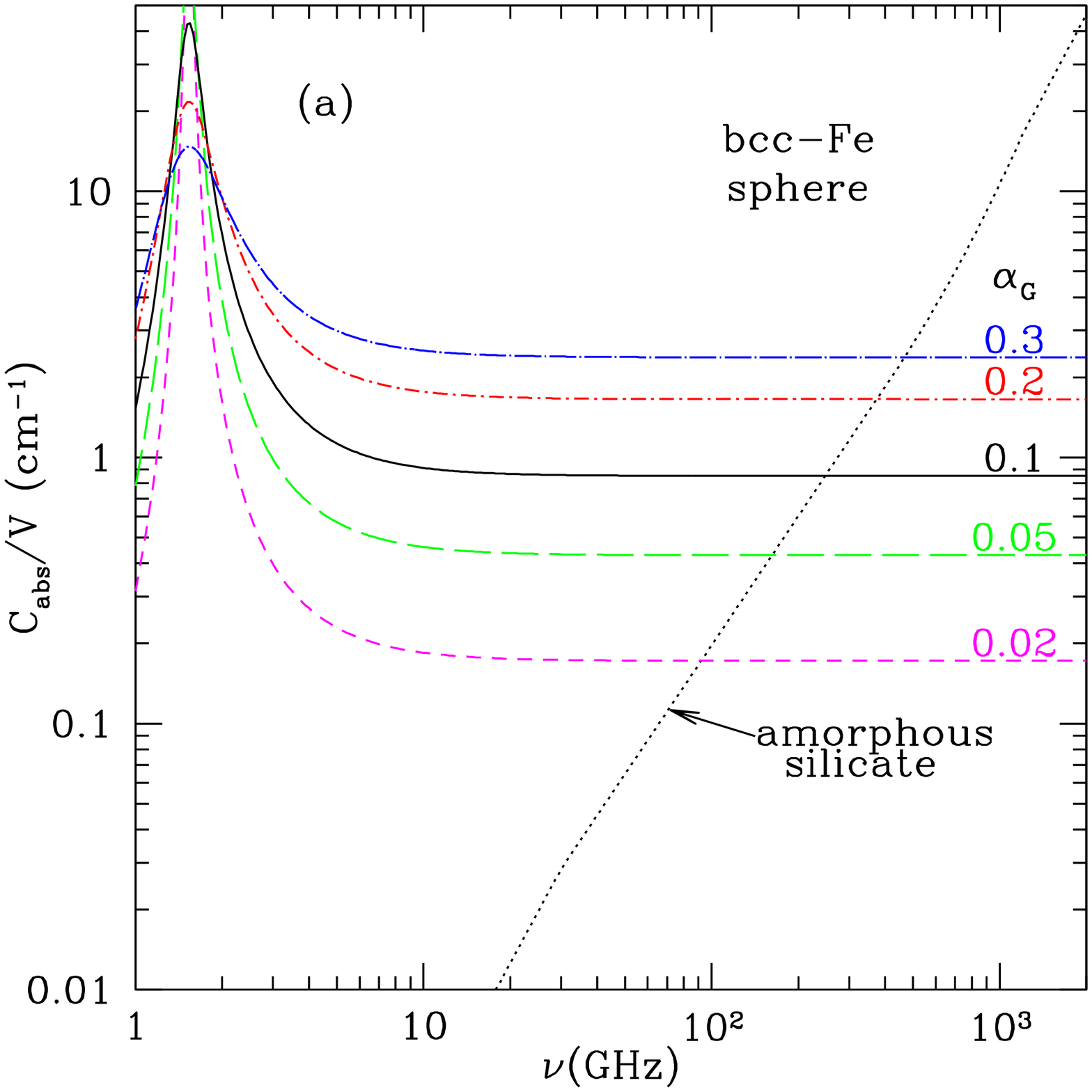}
\includegraphics[width=7.0cm,angle=0]{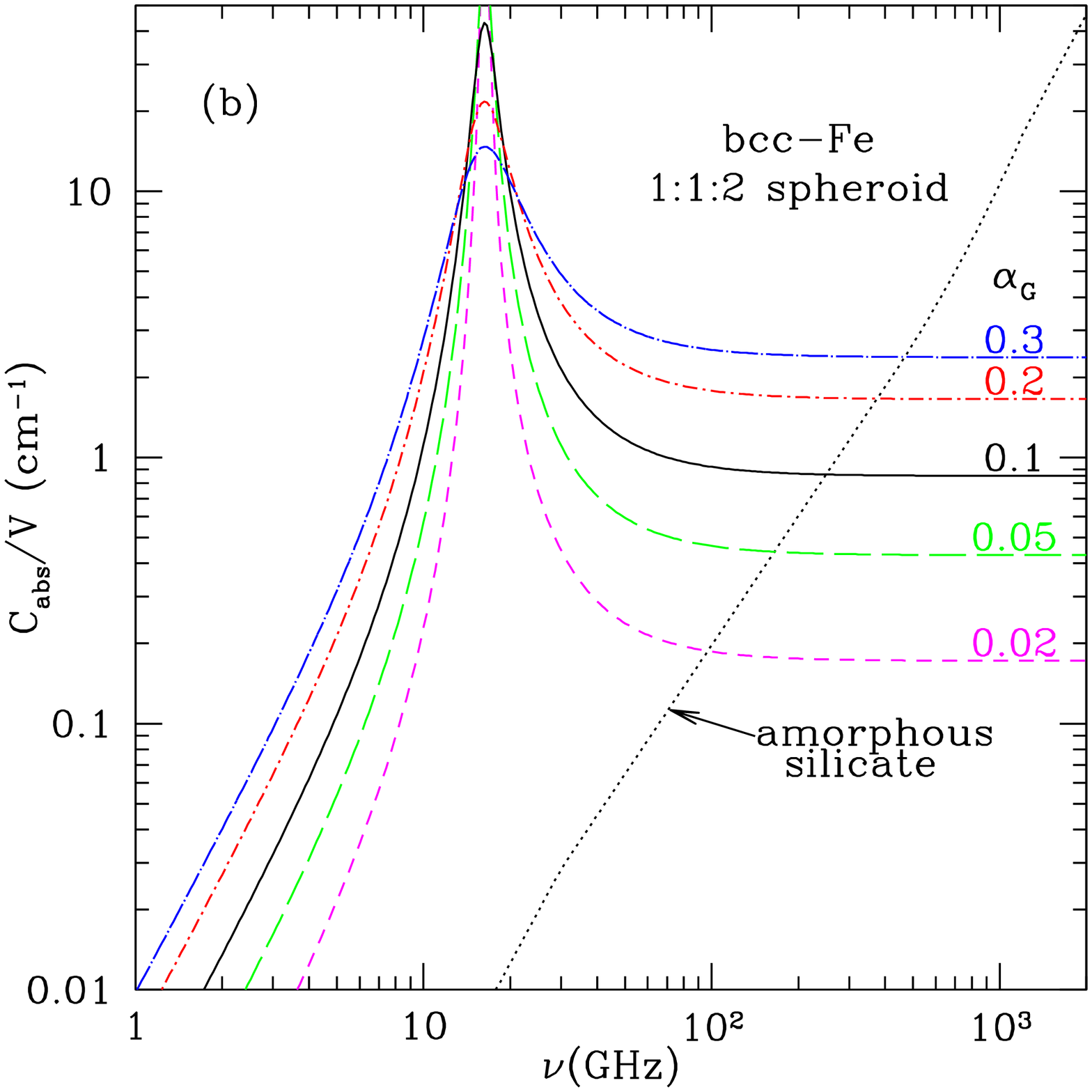}\\
\includegraphics[width=7.0cm,angle=0]{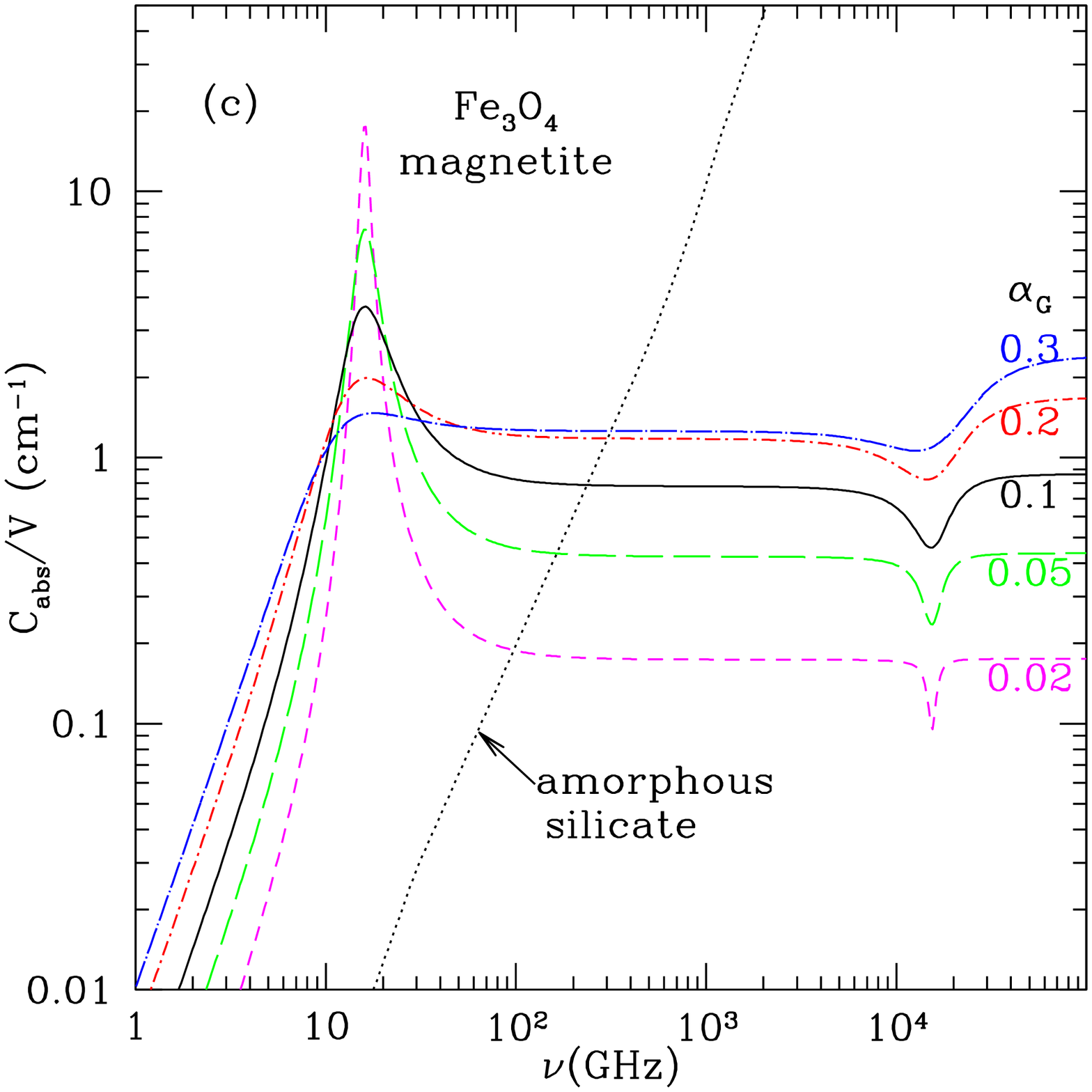}
\includegraphics[width=7.0cm,angle=0]{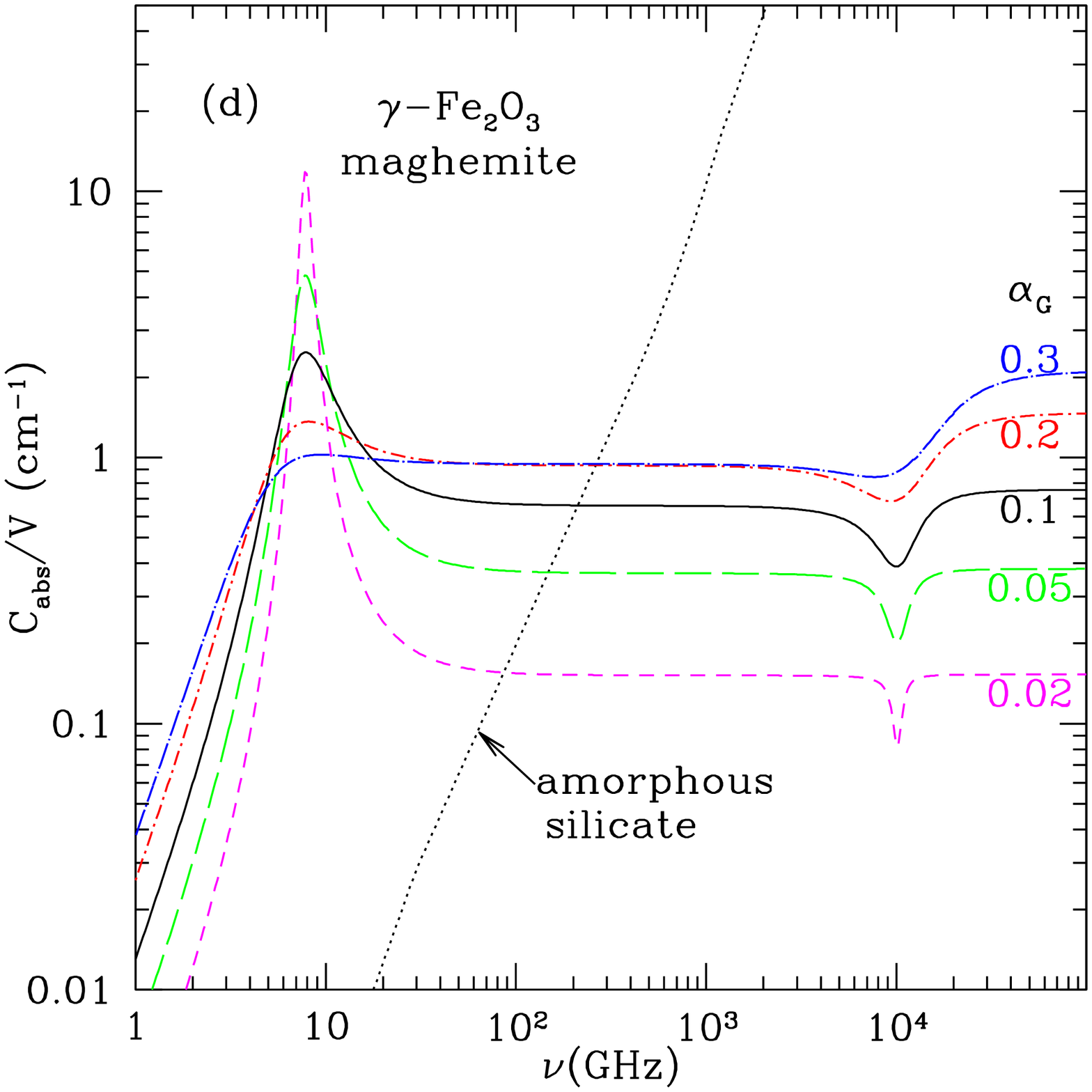}
\caption{\label{fig:magnetic_Cabs}
         \footnotesize
         Magnetic absorption cross section $C_\abs$ per volume $V$ calculated 
         from (\ref{eq:Cabs^mag}) for
         (a) Fe sphere, 
         (b) Fe 1.5:1 prolate spheroid,
         (c) Fe$_3$O$_4$ magnetite, and
         (d) $\gamma$-Fe$_2$O$_3$ maghemite (see text).
         For ferrimagnetic materials, $C_\abs/V$ is shape-independent.
         For comparison, the dotted line shows $C_\abs/V$ for
         small spheres of amorphous silicate.
         }
\vspace*{-1.0em}
\end{center}
\end{figure}

\subsection{Ferromagnetic Material}

For a ferromagnetic material with $\chi_\pm$ given by
(\ref{eq:chi_pm,ferromagnetic}), we have
\beqa \label{eq:ferromagnetic_Cabs} 
\bigl\langle C_\abs\supmag(\omega)\bigr\rangle &=&
\frac{8\pi}{3}\frac{V}{c}
\frac{\alphaG\omega^2\omega_M\left[\omega_0^2+(1+\alphaG^2)\omega^2\right]}
     {\left[\omega_0^2-(1+\alphaG^2)\omega^2\right]^2+
            4\alphaG^2\omega_0^2\omega^2}
     ~~~.  
\eeqa 
with a peak near
\beq
\omega_{{\rm res},+} \equiv 
\frac{\omega_0}{\sqrt{1+\alphaG^2}} \approx \omega_0
\eeq
The resonance at $\omega_{{\rm res},+}$ 
is associated with the $\bhhat_+$ circular polarization mode.
The height and width
of the resonant absorption are determined by the damping coefficient
$\alphaG$.  The resonance frequency $\omega_{{\rm res},+}$
depends on the eigenvalues of
the demagnetization tensor $\bD$ (see eq.\ \ref{eq:D_jj}), 
and hence on the particle shape.
Table \ref{tab:ferromagnetic_resonance_parameters} gives 
$\omega_{0}$ for Fe
spheres and spheroids with selected shapes.
Figures \ref{fig:magnetic_Cabs}a and {\ref{fig:magnetic_Cabs}b 
show $C_\abs/V$ for Fe spheres (absorption peak at $\sim$$1.5\GHz$)
and 2:1 prolate spheroids (absorption peak at $\sim$$16\GHz$).

The asymptotic behavior for ferromagnetic material is
\beq \label{eq:Cabs_ferromagnetic_only} \langle C_\abs\supmag
\rangle\rightarrow \left\{
\begin{array}{l l}
\frac{8\pi V}{3c}\frac{\alphaG\omega_M\omega^2}{\omega_0^2} 
& {\rm for~}\omega\ll\omega_0
\\
\frac{8\pi V}{3c}\frac{\alphaG\omega_M}{(1+\alpha_G^2)}
& {\rm for~}\omega\gg \omega_0
\end{array}
\right.
~~~.
\eeq

\subsection{Ferrimagnetic Material}

For ferrimagnetic materials, with $\chi_\pm$ given by 
(\ref{eq:chi_pm,ferrimagnetic}), $C_\abs(\omega)$ has more complicated
behavior.
For $\alphaG\ll 1$ there are two local extrema, near 
\beq
\omega_{{\rm res},\pm} = 
\left[\frac{(\omega_{0A}+\omega_{0B})^2}{4}
-\omega_{0A}\omega_{0B}+N_{AB}^2\omega_{MA}\omega_{MB}\right]^{1/2}
\pm\frac{(\omega_{0A}+\omega_{0B})}{2}
~~~.
\eeq
The frequency $\omega_{{\rm res},+}$ for resonance when
$\bh\propto\bhhat_+$ (i.e., applied field rotating anticlockwise around
the static magnetization $\bM$) is the frequency
of free precession of
the magnetizations $\bM_A$ and $\bM_B$ around the $\bzhat$ axis, with $\bM_A$
and $\bM_B$ remaining antiparallel -- 
it corresponds
to ordinary ferromagnetic resonance, with $\bm_{A}=-\beta \bm_{B}$.

The frequency $\omega_{{\rm res},-}$ associated with
polarization $\bhhat_-$ is much higher -- at THz frequencies --
and corresponds to precession with $\bM_B$ and $\bM_A$ no longer
antiparallel, but instead with ${\rm Re}(\bm_A)\cdot{\rm Re}(\bm_B) > 0$.
With $\bM_A$ and $\bM_B$ no longer antiparallel, the
precession is driven primarily by the
strong coupling $N_{AB}\bM_A\cdot\bM_B$.


The asymptotic behavior for ferrimagnetic material is
\beq \label{eq:Cabs_ferrimagnetic_only}
\langle C_\abs^{\rm (mag)} \rangle\rightarrow \left\{
\begin{array}{l l}
\frac{8\pi V}{3c} \alphaG A\omega^2
 & {\rm for~}\omega\ll\omega_{{\rm res},+}
\\
\frac{8\pi V}{3c} \alphaG B
& {\rm for~}\omega\gg \omega_{{\rm res},-}
\end{array}
\right.
~~~,
\eeq
where
\beq 
A \equiv
\frac{(2N_{AB}\omega_{MA}\omega_{MB}-
       \omega_{MA}\omega_{0B}-\omega_{MB}\omega_{0A})
       (\omega_{0B}\!-\!\omega_{0A})}
      {\left(N_{AB}^2\omega_{MA}\omega_{MB}-\omega_{0A}\omega_{0B}\right)^2}
      -\frac{(\omega_{MB}-\omega_{MA})}
       {(N_{AB}^2\omega_{MA}\omega_{MB}-\omega_{0A}\omega_{0B})}
\eeq
\beq
B \equiv \frac{(\omega_{MB}-\omega_{MA})}{(1+\alphaG^2)}
~~~.
\eeq
Figures \ref{fig:magnetic_Cabs}c and
\ref{fig:magnetic_Cabs}d show the magnetic contribution to the
absorption cross section
per unit volume calculated for small particles of
magnetite Fe$_3$O$_4$, and maghemite $\gamma$-Fe$_2$O$_3$.

The dip in absorption at the upper resonance frequency $\omega_{{\rm res},-}$ is
notable in Figures \ref{fig:magnetic_Cabs}c and \ref{fig:magnetic_Cabs}d,
but we will see below that at these frequencies electric dipole
absorption is dominant, and this feature in the magnetic dipole
absorption is probably not observable.

\section{\label{sec:IR-Microwave_Opacities}
         Opacities}

Above we have discussed the response of metallic Fe, magnetite, and maghemite
to oscillating electric and magnetic fields in the dipole limit,
$a\ll \lambda$.
We now calculate the absorption cross section, as a function of
wavelength, including both electric and magnetic effects, over a broad
range of frequencies.

We will treat the grains as spherical and calculate
$C_\abs^{\rm MT}$ using Mie theory with $\mu=1$.  To this we add
the magnetic dipole absorption calculated in the magnetic dipole limit,
including the eddy current correction.
Thus, for randomly-oriented grains, we take
\beq
C_\abs = C_\abs^{\rm MT} + \frac{4\pi\omega V}{3c}
{\rm Im}\left[\left(1-\phi_{\rm eddy}\right)\left(\chi_++\chi_-\right)\right]
~~~.
\eeq
The magnetization response for Fe is strongly dependent on the
shape of the Fe particles.
Interstellar Fe nanoparticles -- whether free-flying or inclusions within
larger grains --
are presumably nonspherical.
To compute the magnetization
contribution to the absorption, 
we will assume the Fe to be in 2:1 prolate spheroids (magnetized
along the symmetry axis).

\begin{figure}[t]
\newcommand \figwidth {8.7cm}
\begin{center}
\vspace*{-1.0em}
\includegraphics[angle=0,width=\figwidth]{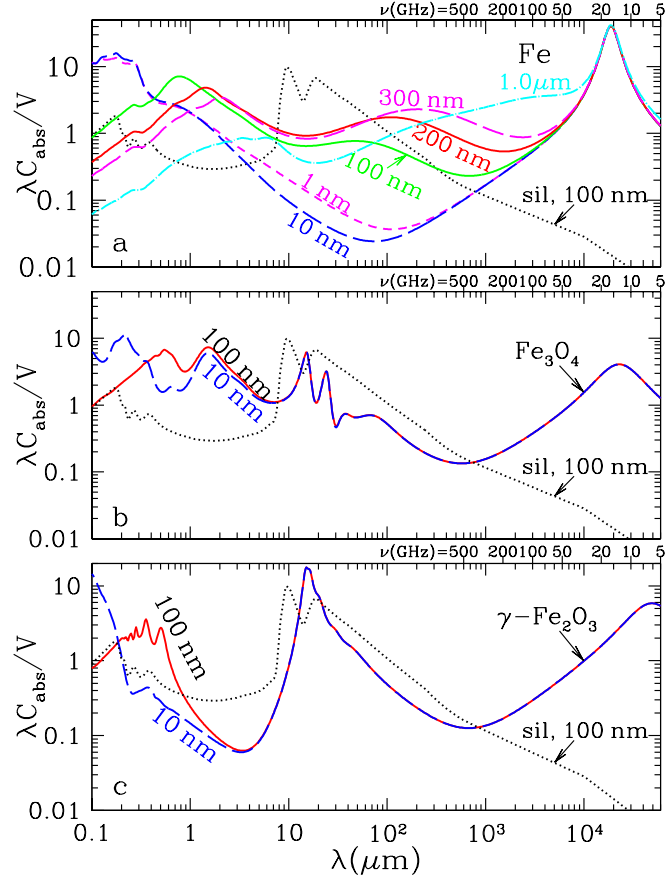}
\caption{\label{fig:lambda*Qabs/a}
         \footnotesize
         $\lambda C_\abs/V$ as a function of wavelength $\lambda$.
         (a) Fe spheres with radii $a=1$, 10, 100, 200, 300, 
         and $10^3\nm$.
         The increase in opacity near $\lambda=100\micron$ as $a$
         increases from $10\nm$ to $300\nm$ is due to the
         increased dissipation by eddy currents.
         The eddy current absorption peak shifts to longer wavelengths
         with increasing grain size.
         For $a<100\nm$, absorption for $\lambda > 1000\micron$
         ($\nu < 300\GHz$) is dominated by 
         magnetic absorption, with a ferromagnetic resonance peak at
         $\sim$$15\GHz$ (the ferromagnetic absorption has been
         calculated for 2:1 prolate spheroids -- see text).
         (b) Magnetite spheres with radii
         $a=10$ and  $100\nm$.
         The ferrimagnetic resonance peak is at $15\GHz$.
         (c) Maghemite spheres with radii
         $a=10$ and  $100\nm$.
         Because of the low electrical conductivity, eddy current
         absorption is unimportant for magnetite and maghemite.
         The ferrimagnetic resonance peak is at $8\GHz$.
         Also shown is $\lambda C_\abs/V$ for $a=100\nm$ amorphous
         silicate spheres.
         }
\vspace*{-1.0em}
\end{center}
\end{figure}

Figure \ref{fig:lambda*Qabs/a}a shows $\lambda C_\abs/V$ as a function of
$\lambda$ for
Fe spheres with selected radii ranging from $a=1\nm$ to $1.0\micron$.
For comparison, we also show $\lambda C_\abs/V$ for
$a=100\nm$ amorphous silicate spheres.
Because 
amorphous silicate is nonconducting, electric dipole
absorption dominates in the FIR, and $C_\abs\supe/V$ is independent of
$a$ for $2\pi a/\lambda \ll 1$.  
By contrast, the eddy current absorption in the Fe particles
causes $C_\abs\supmag/V$ at far-infrared and submm frequencis 
to be sensitive to $a$ 
(see also Figure \ref{fig:Fe polarizability/volume}).
For $a\gtsim 100\nm$, ``magnetic dipole'' absorption due to
eddy current dissipation is important for $\lambda\gtsim 30\micron$.
True magnetic absorption dominates only at very long wavelengths:
for $a=100\nm$, magnetic absorption dominates for
$\lambda > 1\,$mm ($\nu<300\GHz$).

From Figure \ref{fig:lambda*Qabs/a}a
we see that at $\lambda=10^3\micron=1\mm$ ($\nu=300\GHz$)
an $a=100\nm$ Fe sphere 
has $C_\abs/V$ that
is a factor $\sim$2 times larger than for astrosilicate,
whereas for $a=300\nm$, $C_\abs/V$ exceeds astrosilicate
by a factor $\sim$10, because of eddy currents in the conducting
particle.
Therefore, if a significant fraction of the interstellar Fe 
is in particles of such size, it
could noticeably affect the overall submm and mm-wave emission.

\section{\label{sec:temperatures}
         Grain Temperatures}

\begin{figure}[t]
\begin{center}
\includegraphics[angle=0,width=8.0cm]{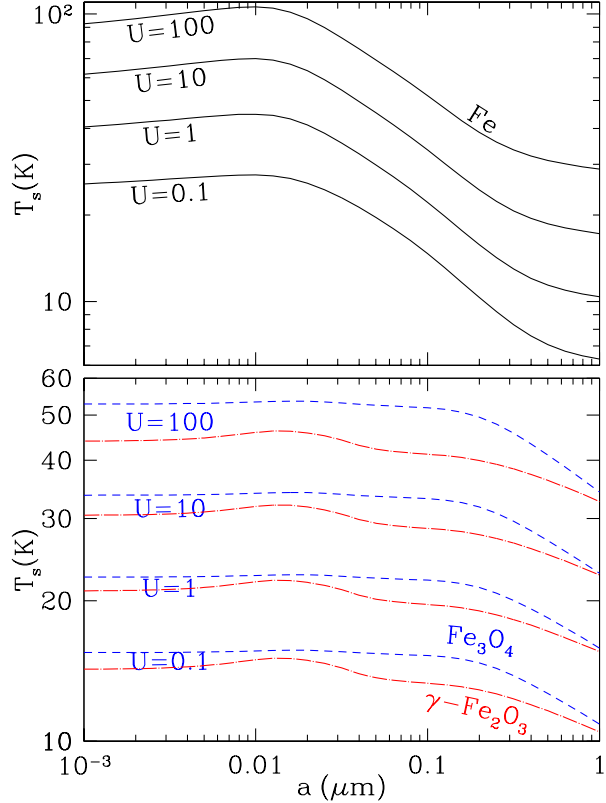}
\caption{\label{fig:Tvsa}\footnotesize
         Steady-state 
         temperature $T_s$ as a function of radius $a$ for spheres of
         Fe, Fe$_3$O$_4$,
         and Fe$_2$O$_3$ grains heated by radiation with the spectrum
         of the local ISRF \citep{Mathis+Mezger+Panagia_1983} and
         intensity $U=0.1, 1, 10, 10^2$ times that of the local ISRF.
         }
\end{center}
\end{figure}

With absorption cross sections $C_\abs(\omega)$ calculated as described
above, we can evaluate the rate 
\beq
P_{\rm heat}(a) = \int d\nu \, C_\abs(\nu) u_\nu c
\eeq
at which a grain of
radius $a$ will absorb energy from the interstellar radiation field
with specific energy density $u_\nu$.
We assume $u_\nu$ to
have the spectrum estimated by
\citet[][herafter MMP83]{Mathis+Mezger+Panagia_1983}
for the
local interstellar radiation field (ISRF) multiplied by
a factor $U$
($U=1$ corresponds to the solar-neighborhood ISRF).
We then solve for the ``steady-state'' 
temperature $T_s$ for which the time-averaged
thermal emission is equal to $P_{\rm heat}$:
\beq
\int d\nu\, C_\abs(\nu) 4\pi B_\nu(T_s) = P_{\rm heat}
~~~.
\eeq
The actual grain temperature will fluctuate around $T_s$.  For
$a\gtsim 0.01\micron$ the temperature 
fluctuations are small enough that they can
be neglected here.

Figure \ref{fig:Tvsa} shows the steady-state grain temperature $T_s$
as a function of radius $a$ for Fe, magnetite, and maghemite grains heated
by radiation with the MMP83 spectrum.

\section{\label{sec:IR spectrum}
         Infrared Emission Spectrum}

\begin{figure}[t]
\begin{center}
\includegraphics[width=8.0cm]{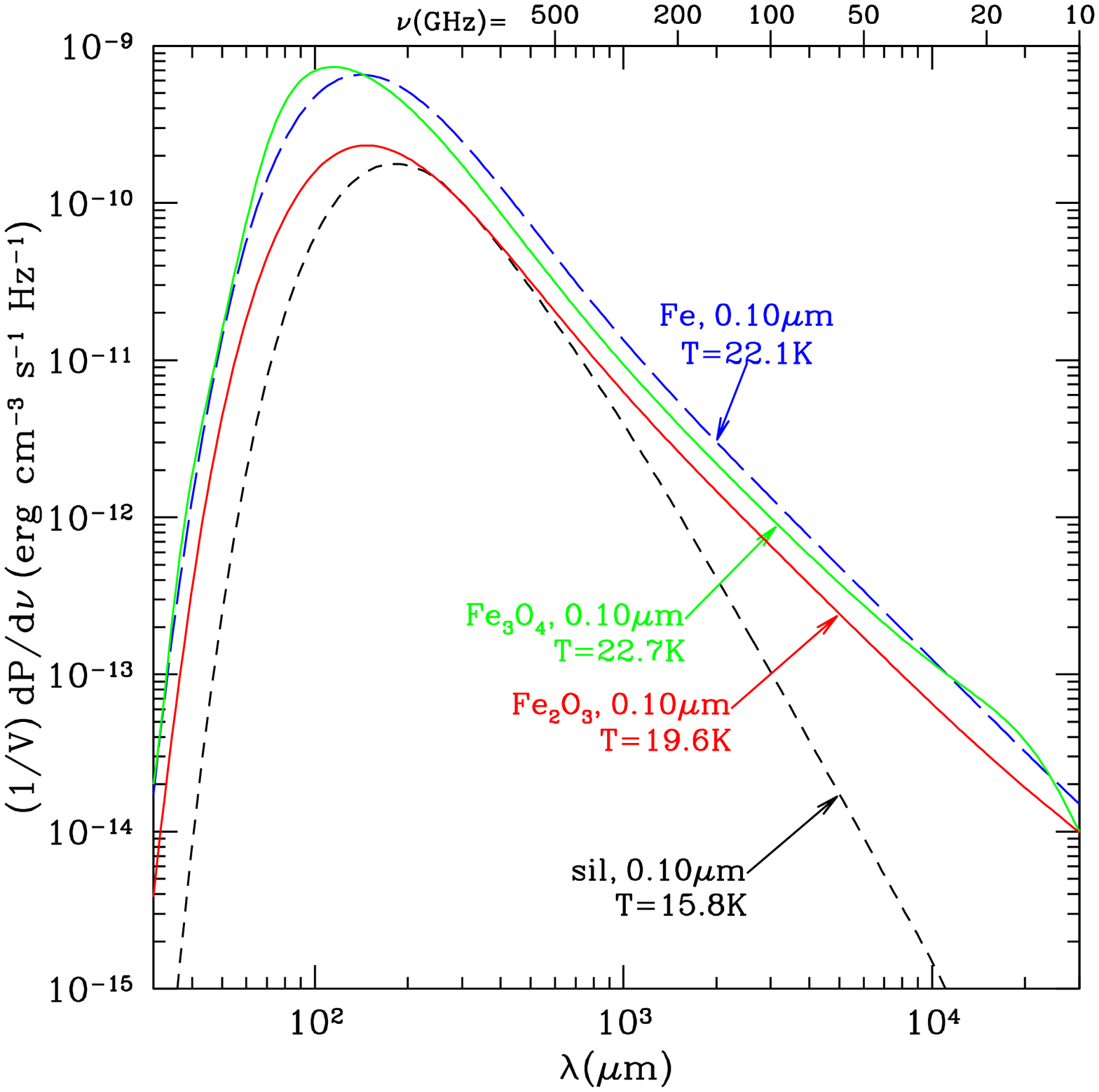}
\includegraphics[width=8.0cm]{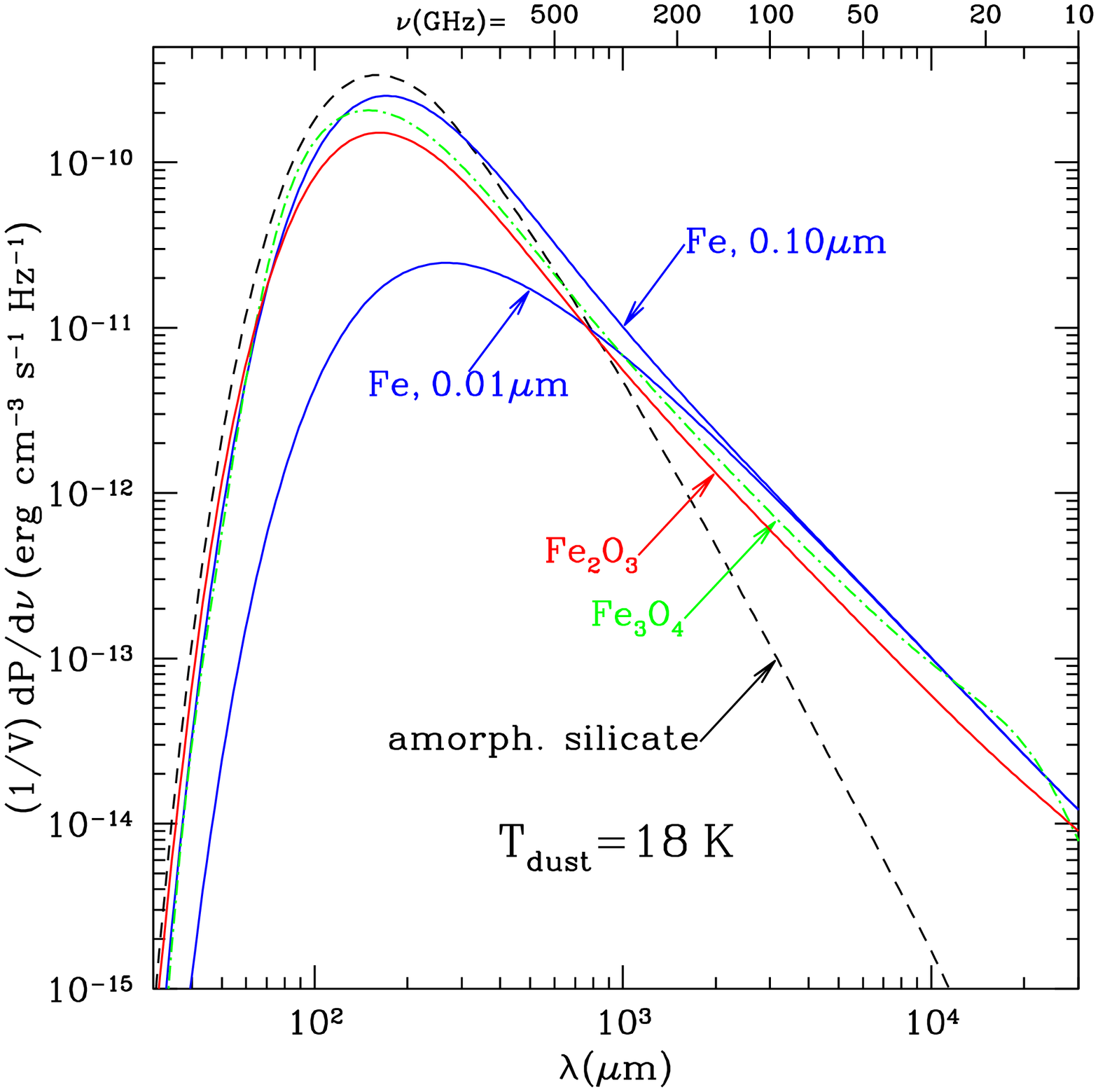}
\caption{\label{fig:IR_emission}\footnotesize
         (a) Emissivity per grain volume of $a=0.10\micron$ grains of 
         astrosilicate,
         metallic Fe, magnetite Fe$_3$O$_4$, and maghemite
         $\gamma-$Fe$_2$O$_3$,
         heated by the MMP83 radiation field.
         (b) Emissivity per grain volume of small grains of astrosilicate,
         metallic Fe, magnetite, and maghemite at $T=18\K$.
         }
\end{center}
\end{figure}

Figure \ref{fig:IR_emission}a shows the emission spectrum, per unit
grain volume, for $a=0.1\micron$ spheres of
Fe, Fe$_3$O$_4$, and $\gamma$-Fe$_2$O$_3$ 
heated by
the MMP83 starlight spectrum with the intensity ($U=1$)
estimated for the Solar neighborhood.
Because of their enhanced absorption cross sections in the
optical and near-IR, the Fe, Fe$_3$O$_4$, and $\gamma$-Fe$_2$O$_3$
grains are somewhat warmer than amorphous silicate grains heated by the same
$U=1$ radiation fields, with grain temperatures of $\sim$20K versus
the 15.8K temperature for the amorphous silicate grain, but all
have emission spectra peaking near $\sim$150$\micron$.
At longer wavelengths $\lambda \gtsim 1\mm$
($\nu < 300\GHz$), the magnetic grains radiate more strongly than
the amorphous silicate, as expected from the cross sections
shown in Figure \ref{fig:lambda*Qabs/a}.

The magnetic grains may be inclusions in larger grains 
(rather than free-fliers),
in which case it is more appropriate to compare the emissivities
of the different grain materials at a common temperature.
Figure \ref{fig:IR_emission}b shows the emission per unit grain
volume at $T=18\K$ for the same four grain materials.
At 300 GHz ($\lambda=10^3\micron$), the power radiated per grain volume by the
magnetic grains ranges from 1.2 (for $\gamma$-Fe$_2$O$_3$) to 2.2 
(for Fe$_3$O$_4$) times 
larger than the emission from the amorphous silicate.
At 100 GHz the difference between amorphous silicate and the magnetic
materials is even more pronounced:
factors of 5 (for $\gamma$-Fe$_2$O$_3$), 8 (for Fe$_3$O$_4$)
and 9 (for Fe).

\section{\label{sec:polarization}
         Polarization}

\subsection{\label{sec:DG_alignment}
            Davis-Greenstein Alignment}

Let $\bh_{\rm ism}$ be the static interstellar magnetic field; $\bh_{\rm ism}$
is very weak compared to either $4\pi\bM_0$ or the
crystalline anisotropy field $\bH_K$.
However, as pointed out by \citet{Davis+Greenstein_1951}, $\bh_{\rm ism}$
can exert systematic torques on a spinning grain.

Consider a magnetic particle spinning in space at some rotational frequency
$\bomega$.
Grain rotation can be excited by many processes.
At a minimum, elastic collisions with gas atoms
will excite ``Brownian'' rotation with
\beq
\frac{\omega}{2\pi} \approx \frac{1}{2\pi}\left(\frac{15 kT}{8\pi\rho a^5}\right)^{1/2}
\approx 20 \kHz \left(\frac{T}{10^2\K}\right)^{1/2}
\left(\frac{5\gm\cm^{-3}}{\rho}\right)^{1/2}
\left(\frac{0.1\micron}{a}\right)^{5/2}
~~~.
\eeq
However, grains with $a\gtsim0.1\micron$ appear likely to be spinning
suprathermally, as a result of systematic torques due to
formation of $\HH$ on the grain surface and emission of photoelectrons
\citep{Purcell_1979} as well as radiative torques due to starlight
\citep{Draine+Weingartner_1996}.
Suprathermal rotation of small $a\ltsim 0.05\micron$ grains is thought
to be suppressed by ``thermal flipping'' \citep{Lazarian+Draine_1999a}.

Let $\psi$ be the angle between $\bomega$ and $\bh_{\rm ism}$.
The grain's spontaneous magnetization $\bM_0$ need not be aligned with
its angular velocity $\bomega$; let
$\Theta$ be the angle between $\bM_0$ and $\bomega$.
Let $\bxhat_m$, $\byhat_m$, $\bzhat_m$ be unit vectors ``frozen into''
the magnetic material, with $\bzhat_m\parallel\bM_0$.
The magnetic field can be separated into two components, one that appears
stationary to the rotating grain, and another, denoted $\bh_{{\rm ism},\omega}$,
oscillating with frequency $\omega$.
Choose $\bxhat_m$ to be perpendicular to the $\bomega-\bM_0$ plane,
and $\byhat_m=\bzhat_m\times\bxhat_m$.  If
${\rm Re}(\bh_{{\rm ism},\omega})$ is in the $\bxhat_m$ direction at $t=0$
then the magnetic
field in ``grain coordinates'' is
\beqa
\label{eq:H0_in_grain}
\bh_{\rm ism} &=& 
h_{\rm ism}\cos\psi\cos\Theta \, \bzhat_m
+
{\rm Re}\left(\bh_{{\rm ism},\omega}\right)
\\
\bh_{{\rm ism},\omega}&=& h_{\rm ism}\sin\psi ~
\left(
-\bxhat_m + i\cos\Theta \, \byhat_m - i\sin\Theta \, \bzhat_m
\right)
e^{-i\omega t}
~~~.
\eeqa
The grain is spinning in a static field.
Following Davis \& Greenstein, 
assume that the magnetic dissipation in the grain material
is the same as in a stationary sample in a rotating magnetic field
corresponding to $\bh_{{\rm ism},\omega}$.
As before, the magnetic 
eigenmodes are $\bhhat_\pm=(\bxhat_m\pm i\byhat_m)/\sqrt{2}$,
and the time-averaged energy dissipation rate is (see eq.\ \ref{eq:<dW/dt>})
\beqa
\Bigl\langle\frac{dW}{dt}\Bigr\rangle &=&
\frac{\omega}{2} V 
\left[{\rm Im}(\chi_+)|\bh_{{\rm ism},\omega}\cdot\bhhat_+^*|^2+
      {\rm Im}(\chi_-)|\bh_{{\rm ism},\omega}\cdot\bhhat_-^*|^2
\right]
\\
&=&
\frac{1}{4}\omega V h_{\rm ism}^2 
\sin^2\psi
\left[
(1-\cos\Theta)^2{\rm Im}(\chi_+) + (1+\cos\Theta)^2{\rm Im}(\chi_-)
\right]
\\ \label{eq:dWdt_approx}
&\approx&
\frac{1}{2}\omega V h_{\rm ism}^2 \sin^2\psi \left(1+\cos^2\Theta\right)
{\rm Im}\left(\frac{\chi_++\chi_-}{2}\right)
\\
&=& K V h_{\rm ism}^2 \omega^2 \sin^2\psi 
\\
K(\Theta) &\equiv& 
\frac{{\rm Im}(\chi_++\chi_-)}{2\omega} ~
\left(\frac{1+\cos^2\Theta}{2}\right)
\\
&\approx& \frac{\alphaG\omega_M}{2\omega_0^2}
(1+\cos^2\Theta) 
~~~,
\eeqa
where the approximation (\ref{eq:dWdt_approx}) takes
${\rm Im}(\chi_+-\chi_-)\ll{\rm Im}(\chi_++\chi_-)$ 
because the grain rotational
frequency $\omega \ll \omega_0$.

The grain, spinning in a static field, is converting rotational
kinetic energy into heat.
Davis and Greenstein showed that the associated torques
would
act to reduce the transverse component of the angular momentum,
bringing the grain rotation axis into alignment with the
local magnetic field.
If $K$ is constant,
the dissipation causes $\tan\psi$ to
decay exponentially:
\beqa
\tan\psi &=& \tan\psi(t=0) ~ e^{-t/\tau_{\rm DG}}
\\
\tau_{\rm DG} &=& \frac{I}{VK h_{\rm ism}^2} 
\approx \frac{2\rho a^2}{5 K h_{\rm ism}^2}
\\
&=& 5.1\times10^5\yr \left(\frac{\rho}{5\gm\cm^{-3}}\right)
\left(\frac{a}{0.1\micron}\right)^2
\left(\frac{10^{-13}\s}{K}\right)
\left(\frac{5\muG}{h_{\rm ism}}\right)^2 ~~~,
\eeqa
where $I$ is the moment of inertia.
For metallic Fe in a 2:1 prolate spheroid,
\beq
K \approx 1.47\times10^{-12} \alphaG (1+\cos^2\Theta) \s ~~~.
\eeq
If $\alphaG\approx 0.2$, this gives a rate of magnetic dissipation
that is only slightly larger than what is estimated for normal paramagnetic
materials, $K\approx 10^{-13}(18\K/T_{\rm gr})\s$
\citep{Jones+Spitzer_1967}.
However, as discussed in Appendix \ref{app:BB_equation},
at low or intermediate frequencies, we might expect the effective
value of $\alphaG$ to be
$\alphaG\approx 1/\omega\tau_2$, where $\tau_2$ is the spin-spin
relaxation time.
For a typical
spin-spin relaxation time of $\sim10^{-10}\s$, we would then have
$\alphaG\approx 8\times10^4 [20\kHz/(\omega/2\pi)](10^{-10}\s/\tau_2)$,
suggesting that $K\sim10^{-7}\s$
for grains where 
metallic Fe is a significant fraction of the
grain volume,
giving a Davis-Greenstein
alignment time $\tau_{\rm DG}\approx 1 \yr$: extremely
rapid alignment of the grain angular momentum with the local magnetic
field.

Rapid alignment by ferromagnetic inclusions, originally proposed
by \citet{Jones+Spitzer_1967}, has been previously discussed by various authors,
including
\citet{Duley_1978},
\citet{Mathis_1986},
\citet{Goodman+Whittet_1995},
and \citet{Martin_1995}.

\subsection{Polarization of the Magnetic Dipole Emission}
\subsubsection{Single-Domain Magnetic Grains}

Single-domain magnetic grains, if their directions of static
magnetization are aligned, will produce strong polarization, as
previously noted by DL99.
Let us assume that the grains are prolate spheroids, magnetized along
the long axis.  The grains are assumed to be spinning around an axis
perpendicular to the symmetry axis.
Without loss of generality, let $\bxhat$ point toward the observer,
let $\bzhat$ be orthogonal to $\bxhat$ and $\bH_0$,
and let $\byhat=\bzhat\times\bxhat$.
Then $\bH_0$ is in the $\bxhat\byhat$ plane.

\begin{figure}[t]
\begin{center}
\includegraphics[width=9.0cm,angle=0]{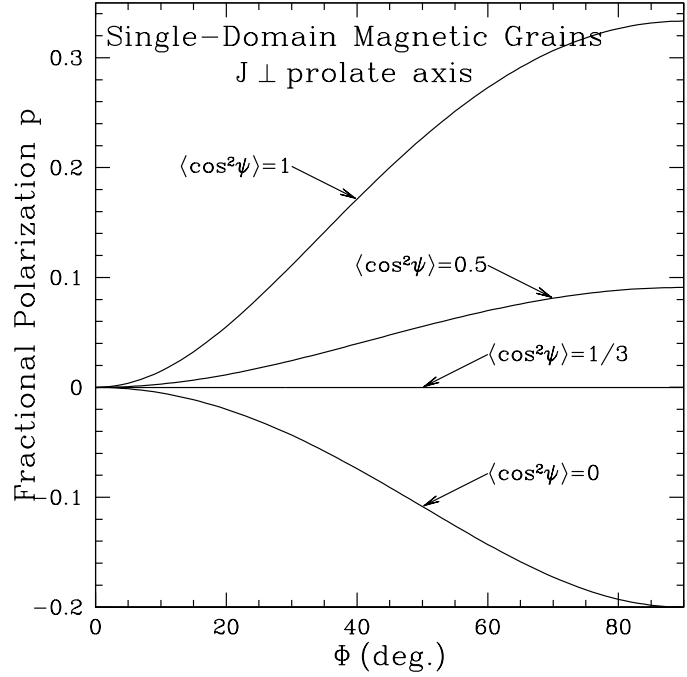}
\caption{\label{fig:fpola} \footnotesize Degree of linear polarization
  for magnetic dipole emission from free-flying, partially-aligned
  single-domain magnetic grains, as a function of the angle $\Phi$
  between the line-of-sight and the magnetic field $\bH_0$ defining
  the alignment direction.  Results are shown for perfect alignment
  $\langle\cos^2\psi\rangle=1$, and partial alignment with
  $\langle\cos^2\psi\rangle=0.5$.  The polarization is zero for random
  alignment ($\langle\cos^2\psi\rangle=1/3$), and negative
  for $\langle\cos^2\psi\rangle<1/3$.}
\end{center}
\end{figure}

If $\bzmhat$ is a unit vector along the
prolate symmetry axis (the direction of magnetization), then
the magnetic absorption cross section for linearly-polarized 
radiation with $\bH_\inc=H_\inc \bhhat e^{-i\omega t}$ 
can be written
\beq
C_\abs\supmag = \frac{2\pi}{c}\omega V {\rm Im}(\chi_++\chi_-)
\left[1- \langle(\bhhat\cdot\bzmhat)^2\rangle \right]
~~~.
\eeq
The brackets $\langle ...\rangle$ denote averaging over the grain
rotation and precession.
Let $C_{\abs,\perp}$ and $C_{\abs,\parallel}$ be $C_\abs$ 
for $\bE_\inc\perp\byhat$ (i.e., $\bH_\inc\parallel\byhat$)
and 
$\bE_\inc\parallel\byhat$ (i.e., $\bH_\inc\parallel\bzhat$), respectively:
\beqa
C_{\abs,\perp}\supmag &=& \frac{2\pi}{c}\omega V {\rm Im}(\chi_++\chi_-)
\left[1- \langle(\byhat\cdot\bzmhat)^2\rangle \right]
\\
C_{\abs,\parallel}\supmag &=& \frac{2\pi}{c}\omega V {\rm Im}(\chi_++\chi_-)
\left[1- \langle(\bzhat\cdot\bzmhat)^2\rangle \right]
~~~.
\eeqa
The fractional polarization of the emitted radiation is just
\beqa
p &=& \frac{C_{\abs,\perp}\supmag-C_{\abs,\parallel}\supmag}
         {C_{\abs,\perp}\supmag+C_{\abs,\parallel}\supmag}
\\
  &=& \frac{\langle(\bzhat\cdot\bzmhat)^2\rangle-
          \langle(\byhat\cdot\bzmhat)^2\rangle}
         {2-\langle(\bzhat\cdot\bzmhat)^2\rangle
           -\langle(\bzhat\cdot\bzmhat)^2\rangle}
~~~.
\eeqa
Let $\Phi$ be the angle between the line-of-sight $\bxhat$ and $\bH_0$,
and let $\psi$ be the ``alignment angle'' 
between the grain angular momentum (assumed to be $\perp$ to the
long axis) and $\bH_0$.
Then (see Appendix \ref{app:polarization_for_spinning_grain})
\beq \label{eq:polarization}
p = \frac{\sin^2\Phi\left(3\langle\cos^2\psi\rangle -1\right)} 
   {5+\langle\cos^2\psi\rangle\left(1-3\cos^2\Phi\right)}
~~~.
\eeq
The numerator is proportional to the familiar ``Rayleigh reduction factor''
$(3/2)(\langle\cos^2\psi\rangle-1/3)$, which varies from 0 to 1
as $\langle\cos^2\psi\rangle$ varies from $1/3$ to 1.
The polarization $p$ is plotted as a function of $\Phi$ 
in Figure \ref{fig:fpola}, for
selected values of
$\langle\cos^2\psi\rangle$, including 1 (perfect alignment),
$1/3$ (random alignment), and $\langle\cos^2\psi\rangle=0$ (perfect
antialignment).

Single-domain grains aligned with the long axis (and magnetization)
$\bz\subm\perp\bH_0$ give $p > 0$.
The assumed
perfect alignment of the grain's magnetization with the principal
axis of smallest moment of inertia results in very large polarizations,
ranging from $-0.2$ to $1/3$
depending on the values of $\langle\cos^2\psi\rangle$ and $\sin^2\Phi$.

\subsubsection{Randomly-Oriented Magnetic Inclusions}

Magnetic material might be primarily in the form of ``inclusions'' in
larger grains.  If the nonspherical inclusions are themselves
perfectly oriented relative to the principal axes of the host grain,
then (\ref{eq:polarization}) will approximate the polarization of the
magnetic dipole emission.

\begin{figure}[t]
\begin{center}
\includegraphics[width=9.0cm,angle=0]{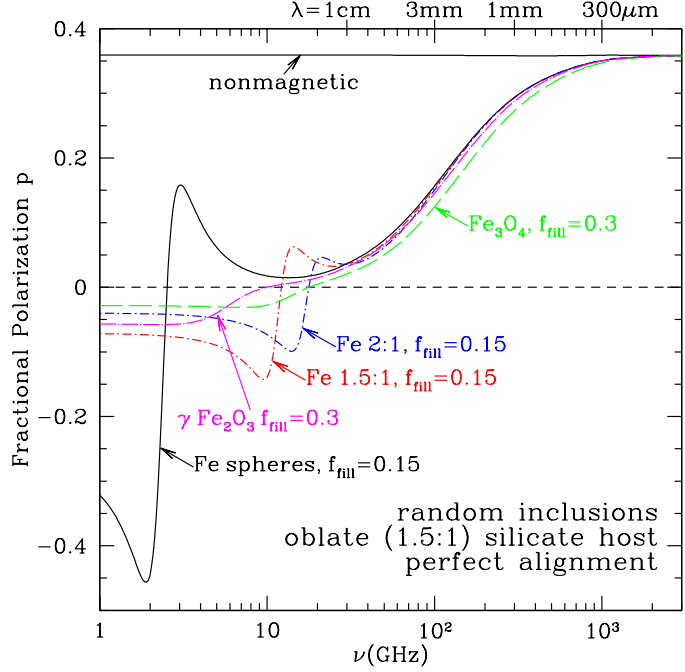}
\caption{\label{fig:fpolb} \footnotesize Degree of linear polarization
  for emission from perfectly aligned 1.5:1 oblate spheroids
  composed of amorphous silicate only (curve labelleled ``nonmagnetic'')
  or amorphous silicate grains with randomly-oriented single-domain
  magnetic inclusions.
  Magnetic inclusions considered are
  Fe spheres and spheroids (1.5:1 and 2:1) with volume filling
  factor $\ffill=0.15$, Fe$_3$O$_4$ magnetite with $\ffill=0.3$,
  and $\gamma$-Fe$_2$O$_3$ maghemite with $\ffill=0.3$.
  At high frequencies, electric
  dipole emission dominates, polarized with $\bE$ along the long axes ($p>0$).
  For grains with magnetic inclusions,
  magnetic dipole emission dominates at low frequency, polarized
  with $\bH$ along the long axes ($p<0$).
  The
  detailed frequency dependence depends on the nature of the magnetic
  inclusions, but in all cases the degree of polarization drops by a
  factor $\sim$4 as the frequency is reduced from $300\GHz$ to $50\GHz$.
  }
\end{center}
\end{figure}

If, as seems more likely, the inclusions are randomly-oriented
relative to the principal axes of the host, and the inclusions have a
sufficiently low volume filling factor $\ffill$, then, using
effective medium theory (see Appendix \ref{app:eff_med_theory})
we may
approximate the magnetic permeability of the grain as a sum over the
magnetic response of the individual inclusions (assumed to be spherical --
see eq.\ \ref{eq:mu_eff_sphere}):
\beq 
\mu \approx 1 +
\ffill\frac{4\pi}{3}\left(\chi_+ + \chi_-\right) ~~~.
\eeq 
The host material is assumed to be nonmagnetic.
For simplicity, we
let both inclusions and host have the same complex dielectric function
$\epsilon(\omega)$.  We assume the grain to be spinning around
$\bahat_1$, the principal axis of largest moment of inertia; $\bomega$
precesses around $\bH_{\rm ism}$, with an angle $\psi$ between
$\bomega$ and $\bH_{\rm ism}$.  Let $\Phi$ be the angle between the
static magnetic field and the line-of-sight.  For a grain with shape
factors $L_1,L_2,L_3$ (for principal axes $\bahat_1$, $\bahat_2$,
$\bahat_3$), the absorption cross section in the dipole limit is
\beqa
\nonumber C_{\abs,\perp} &=& V\frac{\omega}{c} \bigg[ {\rm
    Im}(\epsilon) \left( \frac{\langle(\bzhat\cdot\bahat_1)^2\rangle}
  {|1+L_1(\epsilon-1)|^2} +
  \frac{\langle(\bzhat\cdot\bahat_2)^2\rangle} {|1+L_2(\epsilon-1)|^2}
  + \frac{\langle(\bzhat\cdot\bahat_3)^2\rangle}
  {|1+L_3(\epsilon-1)|^2} \right) +
  \\ \label{eq:CabsE||z_for_random_incl} && \hspace*{2em}{\rm Im}(\mu)
  \left( \frac{\langle(\byhat\cdot\bahat_1)^2\rangle}
       {|1+L_1(\mu-1)|^2} +
       \frac{\langle(\byhat\cdot\bahat_2)^2\rangle} {|1+L_2(\mu-1)|^2}
       + \frac{\langle(\byhat\cdot\bahat_3)^2\rangle}
       {|1+L_3(\mu-1)|^2} \right) \bigg] \\ \nonumber
C_{\abs,\parallel} &=& V\frac{\omega}{c} \bigg[ {\rm Im}(\epsilon)
  \left( \frac{\langle(\byhat\cdot\bahat_1)^2\rangle}
       {|1+L_1(\epsilon-1)|^2} +
       \frac{\langle(\byhat\cdot\bahat_2)^2\rangle}
            {|1+L_2(\epsilon-1)|^2} +
            \frac{\langle(\byhat\cdot\bahat_3)^2\rangle}
                 {|1+L_3(\epsilon-1)|^2} \right) +
                 \\ \label{eq:CabsE||y_for_random_incl}
                 && \hspace*{2em}{\rm Im}(\mu) \left(
                 \frac{\langle(\bzhat\cdot\bahat_1)^2\rangle}
                      {|1+L_1(\mu-1)|^2} +
                      \frac{\langle(\bzhat\cdot\bahat_2)^2\rangle}
                           {|1+L_2(\mu-1)|^2} +
                           \frac{\langle(\bzhat\cdot\bahat_3)^2\rangle}
                                {|1+L_3(\mu-1)|^2} \right) \bigg]
~~,~~~~ 
\eeqa
where the orientational averages are
(see Appendix \ref{app:polarization_for_spinning_grain})
\beqa
\langle(\byhat\cdot\bahat_1)^2\rangle &=&
\sin^2\Phi\langle\cos^2\psi\rangle +
\frac{1}{2}\cos^2\Phi\langle\sin^2\psi\rangle
\\ \langle(\byhat\cdot\bahat_2)^2\rangle =
\langle(\byhat\cdot\bahat_3)^2\rangle &=&
\frac{1}{4}\cos^2\Phi(1+\langle\cos^2\psi\rangle) +
\frac{1}{2}\sin^2\Phi\langle\sin^2\psi\rangle
\\ \langle(\bzhat\cdot\bahat_1)^2\rangle &=&
\frac{1}{2}\langle\sin^2\psi\rangle
\\ \langle(\bzhat\cdot\bahat_2)^2\rangle =
\langle(\bzhat\cdot\bahat_3)^2\rangle &=&
\frac{1}{4}\left(1+\langle\cos^2\psi\rangle\right) ~~~.  
\eeqa
The fractional polarization of the emission is 
\beqa \label{eq:pol_frac}
p &\equiv& \frac{C_{\abs,\perp} - C_{\abs,\parallel}} {C_{\abs,\perp} +
  C_{\abs,\parallel}} = \frac{2N}{D}[1+3\cos(2 \psi)] \sin^2\Phi 
\\ \nonumber
N &=&  B_1B_2B_3\left[A_1(A_2+A_3)-2A_2A_3\right]
{\rm Im}(\epsilon)-
\\
&&A_1A_2A_3 \left[B_1 (B_2+B_3)-2 B_2B_3\right] {\rm Im}(\mu )
\\ \nonumber
D &=& 
\Big\{10 A_2A_3+11 A_1 (A_2+A_3)+
\\ \nonumber
&&
\left[A_1 (A_2+A_3)-2A_2A_3\right] 
\left(\cos(2\psi)+\cos(2\Phi)\left[1+3\cos(2\psi)\right]\right)\Big\}
B_1B_2B_3 {\rm Im}(\epsilon) +  
\\ \nonumber
&&
\Big\{10 B_2B_3+11 B_1 (B_2+B_3)+
\\
&&\left[B_1 (B_2+B_3)-2 B_2B_3\right] 
\left(\cos(2\psi)+\cos(2\Phi)\left[1+3\cos(2\psi)\right]\right)\Big\}
A_1A_2A_3 {\rm Im}(\mu)
\\
A_j&\equiv& |1+L_j(\epsilon-1)|^2
\\
B_j&\equiv& |1+L_j(\mu-1)|^2
\eeqa

What volume filling factors might be expected?  Forsterite Mg$_2$SiO$_4$
($\rho=3.25\gm\cm^{-3}$)
has a volume per Mg $V_{\rm Mg}=3.6\times10^{-23}\cm^3$.  Solar abundances
have Mg/Fe$\approx1.26$ \citep{Asplund+Grevesse+Sauval+Scott_2009}.
Thus if all of the Mg were in Mg$_2$SiO$_4$,
and all of the Fe were
in magnetic inclusions, the volume filling factor of the inclusions
would be $\ffill=V_{\rm Fe}/(1.26V_{\rm Mg}+V_{\rm Fe})=0.20$ for metallic Fe,
$0.35$ for Fe$_3$O$_4$, and $0.38$ for $\gamma$-Fe$_2$O$_3$ (using
the values for $V_{\rm Fe}$ from Table \ref{tab:magnetic_materials}).

Figure \ref{fig:fpolb} shows $p$ calculated using
eqs.\ (\ref{eq:CabsE||y_for_random_incl}-\ref{eq:pol_frac}) for 1.5:1
oblate spheroids ($L_1=0.446$, $L_2=L_3=0.277$) 
of amorphous silicate, containing randomly-oriented
magnetic inclusions 
with volume filling factor
$\ffill=0.15$ for metallic Fe, or
$\ffill=0.3$ for Fe$_3$O$_4$ or $\gamma$-Fe$_2$O$_3$.  
Electric dipole emission dominates at high
frequencies, and (for perfect alignment) $p$ is large, $p\gtsim0.2$.
With decreasing frequency, the importance of the magnetic dipole
contribution increases and $p$ decreases.
For all of the examples with magnetic inclusions,
$p$ decreases by a factor 
$\sim$4 as the frequency varies from $300\GHz$ to $50\GHz$.
Note, however, that other emission processes may be important
at $\ltsim50\GHz$, such as electric dipole radiation from spinning dust
\citep{Draine+Lazarian_1998a,Draine+Lazarian_1998b,Hoang+Lazarian+Draine_2011}, 
which is expected to be minimally
polarized \citep{Lazarian+Draine_2000}, or free-free emission, which
is unpolarized.

It may be possible to observe the predicted drop in polarization
by observing at frequencies $\gtsim 100\GHz$, where spinning dust emission,
free-free emission, and synchrotron emission will be negligible.
For example, {\it Planck} will measure the polarization of
dust emission at $143\GHz$ and $217\GHz$.  In regions where magnetic
dipole emission contributes an appreciable fraction of the total flux,
one expects the polarization at $143\GHz$ to be lower.
Figure \ref{fig:fpolb} has 
$p(143\GHz)/p(217\GHz)=0.75$ and $0.78$ for
grains with inclusions of Fe$_3$O$_4$ and $\gamma$-Fe$_2$O$_3$, respectively.
This is a small effect.

Interpretation is further complicated by the possibility that the FIR
and submm emission comes from more than one type of grain, with differing
degrees of polarization, such as in the models
discussed by \citet{Draine+Fraisse_2009}.  In models where the silicate
grains are aligned, but the carbonaceous grains are not, 
\citet{Draine+Fraisse_2009} predicted that the polarization would decrease
with increasing frequency -- opposite to what is predicted for grains with
magnetic inclusions.  This effect would possibly overwhelm the polarization
signature of magnetic dipole emission at frequencies $\nu>100\GHz$.

\section{\label{sec:discussion}
         Discussion}


\subsection{\label{sec:extinction}
            Constraints from the Observed Extinction}

\begin{figure}[t]
\begin{center}
\includegraphics[width=8.0cm,angle=0]{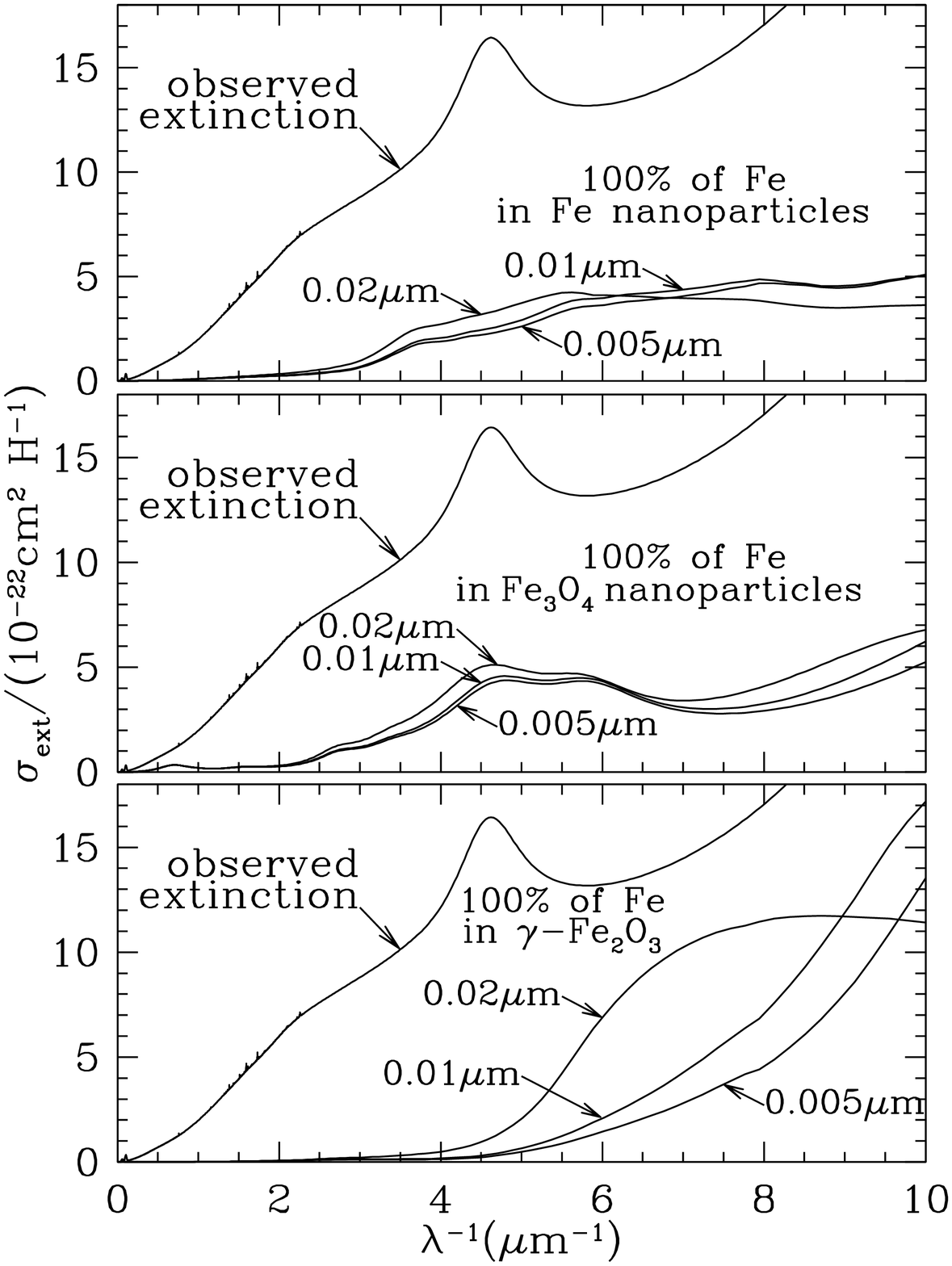}
\caption{\label{fig:extinction}
         \footnotesize
         Observed $R_V=3.1$ extinction curve
         \citep{Fitzpatrick_1999}, and the extinction contribution
         if 100\% of Fe were in metallic Fe,
         Fe$_3$O$_4$, or $\gamma$-Fe$_2$O$_3$
         nanoparticles with radii $a=0.005\micron$, $0.01\micron$,
         or $0.02\micron$.
         The observed extinction may be incompatible with 100\% of
         the Fe being in
         $a=0.02\micron$ $\gamma$-Fe$_2$O$_3$ particles, but the
         other cases shown here appear to be compatible with the
         observed extinction in the visible and ultraviolet.}
\end{center}
\end{figure}

If present as inclusions, magnetic grains could be accommodated within the
grain population without significantly affecting the optical-UV extinction.
If present as free-flying nanoparticles, their contribution to the
extinction could potentially be significant.  Figure \ref{fig:extinction}
shows the contribution to the extinction if 100\% of the Fe is in
spheres 
with radii $a=0.005$, $0.010$, and $0.020\micron$
composed of metallic Fe, Fe$_3$O$_4$, or $\gamma$-Fe$_2$O$_3$.
The extinction contributed by metallic Fe or Fe$_3$O$_4$ 
never exceeds $\sim$$40\%$ of the observed extinction at optical-UV wavelengths,
and the calculated extinction is quite smooth, lacking sharp features that
might be recognizable.
In the case of $\gamma$-Fe$_2$O$_3$ particles, 100\% of the Fe in
$0.02\micron$ particles is probably incompatible with the observed extinction,
but smaller particles appear to be permitted.

\subsection{\label{sec:comparison_with_emission}
            Constraints from the Observed IR-Microwave Emission}

\begin{figure}[t]
\begin{center}
\includegraphics[angle=0,width=9.0cm,angle=270]{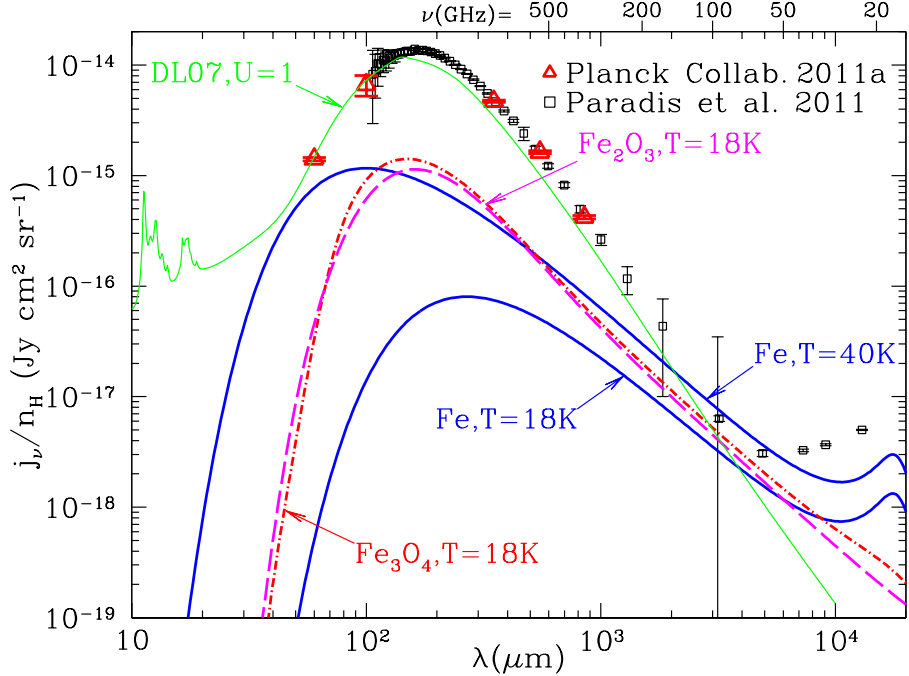}
\caption{\label{fig:fe_emit}
         \footnotesize
         Emission per H calculated if 100\% of the interstellar Fe
         is in $a=0.01\micron$ particles of
         metallic Fe at $T=18\K$, metallic Fe at $T=40\K$,
         magnetite Fe$_3$O$_4$ at $T=18\K$,
         maghemite $\gamma$-Fe$_2$O$_3$ at $T=18\K$.
         Also shown are observational determinations of
         the emission per H from
         COBE-FIRAS and WMAP \citep{Paradis+Bernard+Meny+Gromov_2011}
         (see text)
         and from IRAS and Planck measurements in the North Ecliptic Pole
         region \citep{Planck_DISM_2011} (see text).
         Also shown for comparison is
         emission calculated for the DL07 dust model
         \citep{Draine+Li_2007}
         heated by the local starlight radiation field 
         \citep{Mathis+Mezger+Panagia_1983}.
         The observed emission from the diffuse ISM does not appear
         to limit the fraction of Fe that could be in
         nanoparticles of metallic Fe, magnetite, or maghemite.
         }
\end{center}
\end{figure}

Figure \ref{fig:fe_emit} shows 
the predicted emissivity per H nucleon if 100\% of the interstellar Fe is in
particles of either metallic Fe, magnetite, or maghemite.
For the magnetite and maghemite we assume $T\approx18\K$,
characteristic of the bulk of the dust in the diffuse ISM.
For metallic Fe, we assume particle size $a\approx0.01\micron$ (so that
eddy currents are not important) and we consider two temperatures:
$T\approx18\K$, appropriate if the metallic Fe is present as
inclusions in larger grains, and $T\approx40\K$, appropriate if the
Fe is in free-flying nanoparticles heated by interstellar starlight
(see Figure \ref{fig:Tvsa}).

Figure \ref{fig:fe_emit} also shows the mean $110\micron \leq \lambda \leq
1.3\cm$ emission/H measured by COBE-FIRAS and WMAP for regions
with $|b|>6^\circ$ and $I_\nu(240\micron)>18\MJy\sr^{-1}$
\citep{Paradis+Bernard+Meny+Gromov_2011},\footnote{
    To obtain the emission per H, we
    take this region to have
    $\langle N_\Ha\rangle=3\times10^{21}\cm^{-2}$.
    \citet{Paradis+Bernard+Meny+Gromov_2011} state that
    $\langle N({\rm H\,I})\rangle=2\times10^{21}\cm^{-2}$, which we increase
    by a factor 1.5 to allow for both 21-cm self absorption and
    $\HH$.}
as well as the emissivity for low-velocity \ion{H}{1} measured
by IRAS (60 and 100$\micron$) and Planck (350, 550, and 850$\micron$)
around the North Ecliptic Pole \citep{Planck_DISM_2011}.

From Figure \ref{fig:fe_emit} we see that as much as 100\% of the Fe could
be in single-domain nanoparticles of metallic Fe, magnetite, or maghemite
without exceeding the observed
emission from the diffuse ISM of the Milky Way.

\subsection{Comparison with DL99}

Above we have concluded that the observed emission from the ISM
does not preclude the bulk of the Fe being in single-domain
Fe nanoparticles,
whereas DL99 argued that no more than 5\% of the Fe was allowed to
be in metallic form.
These differing conclusions arise from different models used for
the magnetic response of Fe at microwave frequencies.

DL99 modeled the magnetic response using a scalar susceptibility
$\chi(\omega)$ with the frequency dependence of a damped harmonic oscillator.
The resonance frequency $\omega_0$
was identified with the precession frequency of
spins in the local magnetic field produced by the other spins in the
system, and the damping time was set to $\tau_0=1/(2\omega_0)$.
This choice of $\tau_0$
is probably reasonable for
paramagnetic materials, because the local field fluctuates on a
time scale related to the precession period of nearby spins.
For ferromagnetic or ferrimagnetic materials, however, it is not clear
that this damping time is appropriate.

Using this simple damped oscillator form for the magnetic susceptibility,
DL99 concluded that nanoparticles of metallic Fe would have
a strong absorption peak near 80~GHz,
due to the magnetic analogue of
a Fr\"ohlich resonance
when ${\rm Re}(4\pi\chi)=-3$.
Based on observations of dust-correlated emission
by the COBE-DMR experiment \citep{Kogut+Banday+Bennett+etal_1996},
DL99 argued that no more than
$\ltsim5\%$ of interstellar Fe could be in metallic Fe particles.

The present study employs a more realistic dynamical model
for the magnetic response,
and reaches a different conclusion.
The assumption of a scalar magnetic susceptibility used by DL99
provides a good phenomenological
description of paramagnetism or
the magnetization of bulk ferromagnetic materials
at low frequencies 
(where the dynamic permeability $\mu(\omega)$ arises from motion of
domain walls), but it 
does not describe the response
of magnetic systems at frequencies $\nu\gtsim 10\GHz$, where
precession of the spins is important.
The present study uses a
tensor susceptibility derived from the Gilbert equation (\ref{eq:Gilbert}).
For single-domain Fe particles, the magnetization response
(\ref{eq:chi_pm,ferromagnetic}) has an absorption peak
in the 1.5--20~GHz range, depending on shape, and the thermal emission
is not as strong as estimated by DL99.

\subsection{\label{sec:validity_of_Gilbert_eq.}
            Validity of the Gilbert Equation}

Our model for the magnetic response of Fe, maghemite, and magnetite
is based on the Gilbert equation (\ref{eq:Gilbert}), with the dissipation
characterized by an adjustable parameter $\alphaG$.
We have adopted $\alphaG\approx0.2$ for purposes of discussion, 
but
the existing experimental literature employs a range of values of
$\alphaG$.  
If the Gilbert equation is used, then $\alphaG\approx0.2$ gives about the
highest values possible for the absorption at $\nu\gtsim 100\GHz$
(see Figure \ref{fig:magnetic_Cabs}).
If $\alphaG$ were to be much smaller (e.g., $\alphaG=0.05$)
the opacity would be reduced (by about a factor 3 in going from
$\alphaG=0.2$ to $0.05$), and 
the microwave-submm emissivity would be reduced by the same factor.

Quite aside from the question of what value to use for the Gilbert
damping parameter $\alphaG$, we must recognize that 
the Gilbert equation uses a prescription for the dissipation that
is simple and mathematically convenient, but not based on an underlying
physical model.  Empirical evidence
for the accuracy of the Gilbert equation at high frequencies is scant.
Laboratory measurements of electromagnetic absorption
in Fe and Fe oxide nanoparticles at frequencies up to 500 GHz are
required to validate use of the Gilbert equation at these frequencies.

The present treatment of dynamic magnetization assumed that the
material was perfectly-ordered and pure, but
interstellar grain materials are likely to be non-crystalline
and impure.  
We suspect that lattice defects and impurities will mainly lead to
broadening of the resonance appearing near $\omega_0$ (for ferromagnetic
materials) and $\omega_{\rm res,\pm}$ for ferrimagnetic materials, but
this should be studied experimentally.  
Measurements of the frequency-dependent magnetic polarizability tensor
$\balpha\subm(\omega)$ for nanoparticles composed of amorphous Fe-rich
oxides or silicates over the frequency range $1-500\GHz$ would
be of great value
to test the models that have been put forward here.

\section{\label{sec:summary}
         Summary}

A substantial fraction of the Fe in the ISM may be in magnetic
materials such as metallic iron, magnetite Fe$_3$O$_4$, or
maghemite $\gamma$-Fe$_2$O$_3$.
We discuss the implications on the thermal emission from interstellar dust,
and the polarization of this emission.
The principal conclusions are as follows:
\begin{enumerate}

\item We obtain 
(in Sections 
\ref{sec:magnetic_permeability} and
\ref{sec:polarizabilities of small particles})
the frequency-dependent magnetic polarizability
tensor $\balpha\subm(\omega)$ of
single-domain ferromagnetic and ferrimagnetic particles, and
evaluate this for three candidate materials: metallic Fe, magnetite
Fe$_3$O$_4$, and maghemite $\gamma$-Fe$_2$O$_3$.

\item Eddy currents are shown to have only a small effect on the
magnetic absorption for $\nu\ltsim500\GHz$.

\item We show (in Section \ref{sec:Cabs})
how to calculate the absorption cross section $C_\abs(\omega)$
for magnetic particles that are small compared to the wavelength.

\item In order to be able to calculate $C_\abs(\omega)$ 
we obtain self-consistent dielectric functions for
Fe (Appendix \ref{app:epsilon_Fe}), 
magnetite (Appendix \ref{app:epsilon_Fe3O4}), and 
maghemite (Appendix \ref{app:epsilon_maghemite}),
based on a combination of theory and experiment.


\item We confirm the finding of \citet{Fischera_2004} for Fe grains 
that magnetic
dipole absorption arising from eddy currents can result in very
large absorption cross sections in the FIR and submm for
radii $a\gtsim 0.05\micron$.

\item The Gilbert equation implies that small
metallic Fe, magnetite, and maghemite grains have opacities that
scale approximately as 
$\kappa \approx \nu^0$ for $10\GHz \ltsim \nu \ltsim 100\GHz$
(see Figs.\ \ref{fig:magnetic_Cabs} and \ref{fig:lambda*Qabs/a}).

\item We calculate the grain temperature $T$, as a function of radius,
for particles of Fe, magnetite,
and maghemite heated by starlight (Fig.\ \ref{fig:Tvsa}).
Small ($a\ltsim0.03\micron$) Fe grains have $T\approx 40 U^{0.2}\K$.

\item If the Gilbert equation with $\alphaG\approx0.2$ applies at the
$\sim$$20\kHz$ rotational frequencies of $\sim0.1\micron$ grains, then
Davis-Greenstein alignment should take place on $\sim$$10^5\yr$
time scales, as estimated for ordinary paramagnetic dissipation
\citep{Jones+Spitzer_1967}.
However, it is possible that the Gilbert equation with 
$\alphaG\approx0.2$ may significantly underestimate magnetic dissipation
at the rotation frequencies of grains, in which case
grains with a significant fraction of the volume contributed by
ferromagnetic or ferrimagnetic material may undergo rapid alignment
by the Davis-Greenstein mechanism.

\item We calculate the expected polarization of the magnetic
dipole emission from aligned free-flying magnetic nanoparticles.
The polarization has the ``normal'' sense ($\bE\perp\bB_{\rm ISM}$)
and can be large (see Figure \ref{fig:fpola}).

\item We calculate the expected polarization of the
emission from a nonspherical silicate host with
randomly-oriented magnetic inclusions.
The magnetic dipole emission becomes important at low frequencies,
and can result in a reversal of the polarization direction
(see Figure \ref{fig:fpolb}).

\item We show (Fig.\ \ref{fig:fe_emit}) that up to 
100\% of interstellar Fe
could be in grains of
metallic Fe, magnetite, and maghemite without exceeding the observed
emission from interstellar dust at frequencies $\nu>30\GHz$.

\end{enumerate}

\acknowledgements

We thank 
Vincent Guillet 
and 
Alex Lazarian 
for helpful comments.
This research made use of NASA's Astrophysics Data System Service,
and was supported in part
by NSF grant AST 1008570.
BH acknowledges support from
a NSF Graduate Research Fellowship under
Grant No.\ DGE-0646086.

\begin{appendix}

\section{\label{app:BB_equation}
         Bloch-Bloembergen Equation}

Let $\bM(t)$ be the magnetization.
Equations for $d\bM/dt$ all include
a term $-\gamma\bM\times\bH_T$ for the torque arising from
applied field and crystalline anisotropies.  This torque is clearly
nondissipative, and additional terms must be added 
to represent the effects of dissipation.

Many studies of paramagnetic resonance employ the
Bloch-Bloembergen equation \citep{Bloch_1946,Bloembergen_1950}:
\beq \label{eq:Bloch-Bloembergen_equation}
\left(\frac{d\bM}{dt}\right)_\BB =
\gamma \bM\times\bH_T  - \frac{\bM_\parallel-\bM_0}{\tau_1}
- \frac{\bM_\perp}{\tau_2} ~~~,
\eeq
where $\bM_\parallel\equiv(\bM\cdot\bM_0)\bM_0/|M_0|^2$,
and $\bM_\perp \equiv \bM-\bM_\parallel$ 
are the components of the magnetization
parallel and perpendicular to the stationary magnetization $\bM_0$ that
would be present if the oscillating field $\bh_0$ were set to zero
and the magnetization were allowed to reach equilibrium.
The time $\tau_1$ is interpreted as the spin-lattice
relaxation time, the time required for the spin system to exchange
heat with the lattice.
The time $\tau_2$ is interpreted as the spin-spin relaxation time, the
time for one spin to be perturbed by nearby spins.
The Bloch-Bloembergen equation appears on the surface to be reasonable,
and is frequently used.  Here we show that this equation
is unphysical.

Consider a weak periodic driving field of the form (\ref{eq:h(t)}).
Linearizing, we obtain
\beqa
-i\omega m_{0x} &=& ~\gamma m_{0y}\left[H_0+H_{Kz}-D_{zz}M_0\right]
                -\gamma M_0\left[h_{0y}-D_{yy}m_{0y}\right] 
                -m_{0x}\tau_2^{-1}
\\
-i\omega m_{0y} &=& \!\!-\gamma m_{0x}\left[H_0+H_{Kz}-D_{zz}M_0\right]
                +\gamma M_0\left[h_{0x}-D_{xx}m_{0x}\right] 
               - m_{0y}\tau_2^{-1}
~~~.
\eeqa
To simplify, assume the sample to be a spheroid, with
$D_{yy}=D_{xx}$.
With the definitions of $\omega_0$ and $\omega_M$ from
(\ref{eq:def_omega0}, \ref{eq:def_omegaM}) and
\beq
q \equiv \omega \left(1 - \frac{1}{i\omega\tau_2}\right) ~~~,
\eeq
the oscillating magnetization $\bm_0$  obeys the tensor equation
(\ref{eq:tensor_equation}),
but with
\beq \label{eq:chi_pm,BB}
\chi_\pm^\BB \equiv \frac{(\omega_0\pm q)\omega_M}{\omega_0^2-q^2} =
\frac{\omega_M}{\omega_0\mp q}
~~~.
\eeq
As before, the eigenvectors are the two circular polarizations
(\ref{eq:circular_polarizations}), with the magnetization given
by (\ref{eq:magnetization_response}).
For the two circular polarizations, the magnetization response is
given by $\chi_+$ and $\chi_-$.

Dissipation is given by ${\rm Im}(\chi)$: straighforward algebra yields
\beq \label{eq:BB_Im_chi}
{\rm Im}(\chi_\pm^\BB) = 
\frac{\pm~ \omega_M\tau_2^{-1}}{(\omega_0\mp\omega)^2 + \tau_2^{-2}}
~~~.
\eeq
Thus, for anticlockwise circular polarization $\bhhat_+$,
the Bloch-Bloembergen equation has positive dissipation (${\rm Im}(\chi_+)>0$).
However, for the clockwise circular polarization $\bhhat_-$, 
(\ref{eq:BB_Im_chi}) has
${\rm Im}(\chi_-)<0$: the
dissipation is {\it negative}, which is clearly unphysical.
The Bloch-Bloembergen equation is frequently applied to
studies of magnetic resonance.  
This unphysical aspect of the Bloch-Bloembergen equation
has been previously noted \citep{Lax+Button_1962,Berger+Bissey+Kliava_2000},
it is surprising that it does not appear to be widely recognized.

The Gilbert equation (\ref{eq:Gilbert}) was proposed by
\citep{Gilbert_1955,Gilbert_2004} as a simple way of including
dissipation in the dynamics of ferromagnetic resonance.
If we try to require these to have similar response functions $\chi_\pm$
for some frequency $\omega$,
we would have
\beq
\chi_\pm^{\rm G}(\omega) \approx \chi_\pm^\BB(\omega)
~~~.
\eeq
Comparison of (\ref{eq:chi_pm,ferromagnetic}) and
(\ref{eq:chi_pm,BB}) implies that we would need to have
\beq
\alphaG\omega \approx \pm \tau_2^{-1}
~~~.
\eeq
Clearly this cannot be satisfied simultaneously for both circular
polarization modes.  If we limit consideration
to the $\bhhat_+$ circular polarization mode, we would have
$\alphaG\omega \approx \tau_2^{-1}$: the
assumption of constant $\alphaG$ corresponds to a Bloch-Bloembergen
spin-spin relaxation time $\tau_2\propto\omega^{-1}$.
In other words, the Gilbert equation implies that for high driving
frequencies $\omega$
the relaxation time is shorter than for low driving frequencies.
This would presumably arise from a spectrum of relaxation processes with
a broad range of characteristic time scales, with the ``effective''
time scale for a given driving frequency $\omega$ scaling as $1/\omega$.

As seen above, the Bloch-Bloembergen equation 
(\ref{eq:Bloch-Bloembergen_equation}) is unphysical,
implying negative dissipation for the $\bhhat_-$ circular polarization.
Whether the phenomenological representation of dissipation in the
Gilbert equation provides a good approximation to real materials
is a question that can only be answered experimentally.
We have not been able to find experimental data that provide a
convincing answer.

\section{\label{app:epsilon_Fe}
         Dielectric Function and Conductivity of Metallic Fe}

To calculate emission and absorption by Fe grains, we require the
complex dielectric function $\epsilon(\omega)$ and
complex magnetic permeability $\bmu(\omega)$.
Because we need to calculate heating of the grain by starlight as well
as thermal emission, we require the dielectric function from the
ultraviolet to microwave.  We undertake to construct the complete
dielectric function from microwave to X-rays.

Fe is body-centered cubic, and the dielectric
function is a scalar $\epsilon = \epsilon_1 + i\epsilon_2$,
if magneto-optical effects are ignored.
The real part $\epsilon_1$ will be obtained from the imaginary
part $\epsilon_2$ using the Kramers-Kronig relation
\citep{Landau+Lifshitz+Pitaevskii_1993}
\beq \label{eq:KK_integral}
\epsilon_1(\omega) = 1 +  \frac{2}{\pi} P 
\int_0^\infty \frac{x\epsilon_2(x)}{x^2-\omega^2} dx ~~~,~~~
\eeq
where $P$ indicates that the principal value is to be taken.

\begin{figure}[ht]
\begin{center}
\includegraphics[angle=0,width=7.0cm]{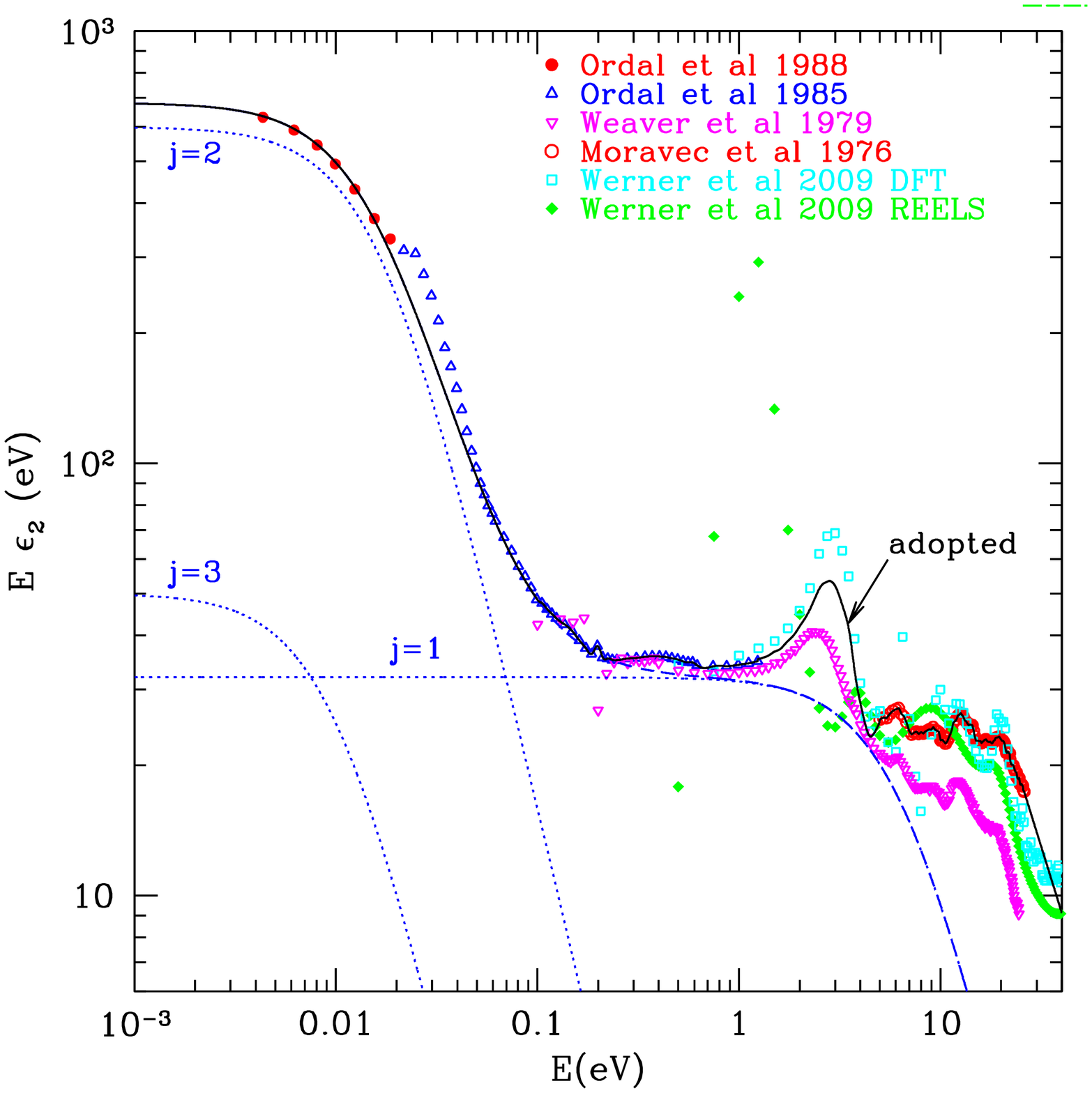}
\includegraphics[angle=0,width=7.0cm]{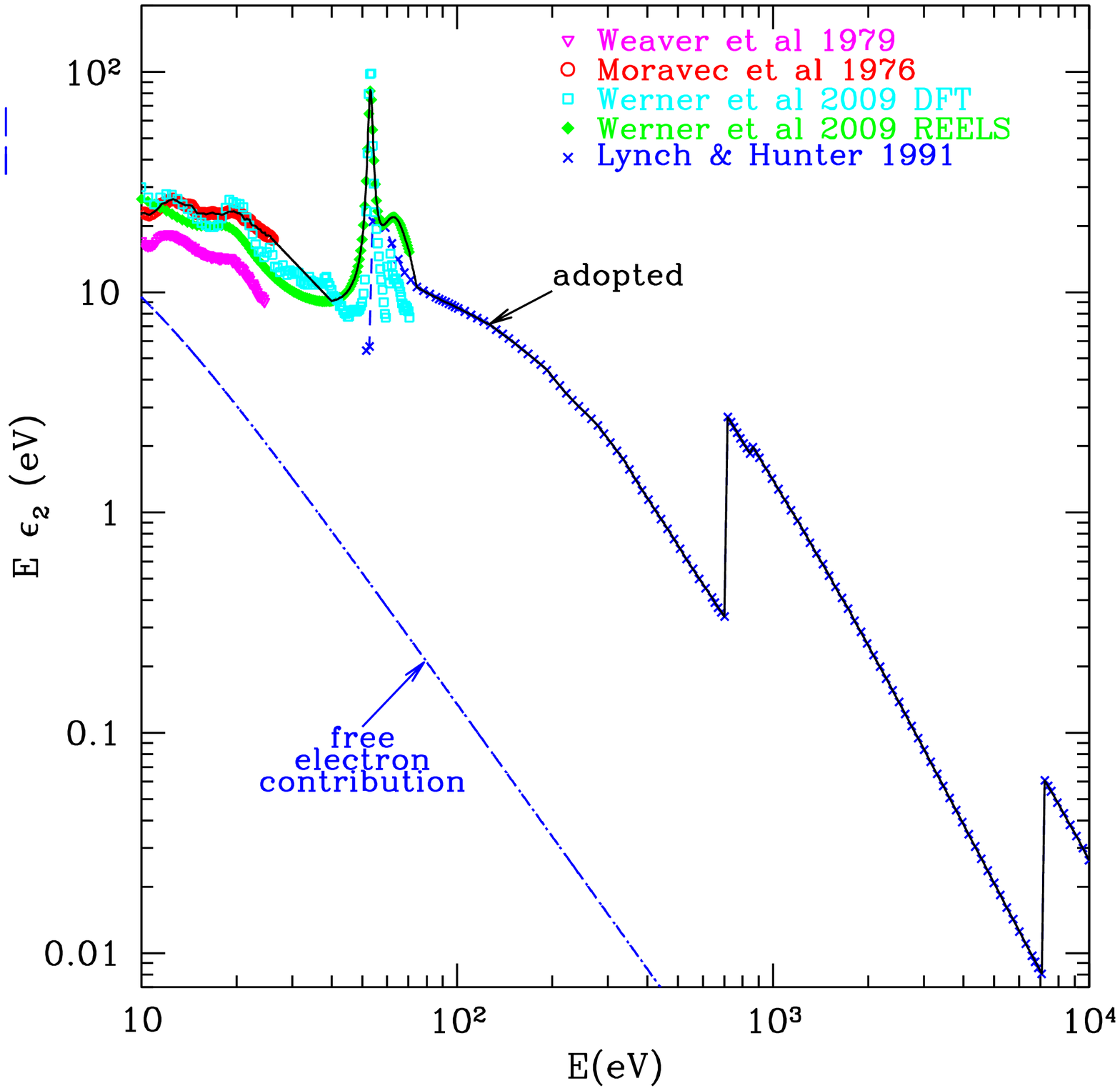}
\caption{\label{fig:experimental_eps2}
         \footnotesize
         $\hbar\omega\epsilon_2(\omega)$ for metallic Fe from
         \citet{Moravec+Rife+Dexter_1976},
         \citet{Weaver+Colavita+Lynch+Rosei_1979},
         \citet{Ordal+Bell+Alexander+etal_1985,Ordal+Bell+Alexander+etal_1988},
         \citet{Lynch+Hunter_1991},
         and
         \citet{Werner+Glantschnig+Ambrosch-Draxl_2009}.
         The solid curve shows $\hbar\omega\epsilon_2(\omega)$
         adopted for this study.  The dotted curves show the
         three free-electron contributions that together 
         (dashed curve) reproduce the adopted $\hbar\omega\epsilon_2(\omega)$
         at low energies.
         }
\end{center}
\end{figure}

Figure \ref{fig:experimental_eps2}
shows various determinations of $\epsilon_2$ for bulk Fe.
When different determinations of $\epsilon_2$ 
in the optical and vacuum ultraviolet region disagree, we have
tried to follow the data that appear to be most
reliable.  In the 1.24--4.75 eV range we take the mean of
the experimental results of \citet{Weaver+Colavita+Lynch+Rosei_1979} and the
density functional theory (DFT) estimates of
\citet{Werner+Glantschnig+Ambrosch-Draxl_2009}.
In the 5--26 eV range we adopt the values of
\citet{Moravec+Rife+Dexter_1976}.
From 45--70.5 eV we take the reflection energy loss spectroscopy
(REELS) results from \citet{Werner+Glantschnig+Ambrosch-Draxl_2009}.
Above 75 eV we use $\epsilon_2$ calculated by \citet{Lynch+Hunter_1991}
using parameters from \citet{Henke+Davis+Gullikson+Perera_1988}.

\begin{table}[bt]
\caption{\label{tab:drude_params}
         Free-electron model parameters for bulk Fe at $T=293\K$}
\begin{center}
{\footnotesize
\begin{tabular}{cccccc}
\hline
$j$ & $\hbar\tau_{b,j}^{-1}$ & $\hbar\omega_{p,j}^2\tau_{b,j}$ 
& $\hbar\omega_{p,j}$ & $\tau_{b,j}$ & $v_F\tau_{b,j}$\\
  & (eV) & (eV) & (eV) & (s) & (cm)\\
\hline
1 & 6.5    &  32 & 14.4   & $1.01\times10^{-16}$ & $2.00\times10^{-8}$\\
2 & 0.0165 & 600 & 3.146  & $3.99\times10^{-14}$ & $7.90\times10^{-6}$\\
3 & 0.010  &  50 & 0.707  & $6.59\times10^{-14}$ & $1.30\times10^{-5}$\\
\hline
\end{tabular}
}
\end{center}
\end{table}

The optical properties
of bulk Fe has been measured over
a wide range of frequencies \citep{Lynch+Hunter_1991}.
The dielectric function can be separated into contributions from free
electrons and bound electrons:
\beq
\epsilon(\omega) = 1 + \delta\epsilon^{\rm (f)}(\omega) + 
\delta\epsilon^{\rm (b)}(\omega)
~~~.
\eeq
The free-electron contribution 
can be approximated by a sum of three Drude free-electron components
\beq \label{eq:Drude model}
\delta\epsilon^{\rm (f)}(\omega) =  
i\sum_{j=1}^3 \frac{\omega_{p,j}^2\tau_j}{\omega (1-i\omega\tau_j)}
~~~.
\eeq
The free-electron contribution is taken to be large enough that
${\rm Im}(\delta\epsilon^{\rm (b)})\rightarrow 0$ as $\omega\rightarrow 0$.
The conductivity from the free-electron component is
\beqa
\sigma(\omega)&=&\frac{1}{4\pi}\omega \epsilon_2(\omega)
\\
&=& \frac{1}{4\pi}\sum_j 
\frac{\omega_{p,j}^2\tau_{b,j}}{1 + (\omega\tau_{j})^2}
~.
\eeqa
Taking $\tau_j$ 
to be the bulk values $\tau_{b,j}$ in Table \ref{tab:drude_params},
we obtain a zero-frequency conductivity
$\sigma(0)=8.24\times10^{16}\s^{-1}$, in good agreement with the
measured room temperature value $\sigma(0)=9.12\times10^{16}\s^{-1}$
\citep{Ho+Ackerman+Wu+etal_1983}.

The bound-electron contribution $\delta\epsilon_2^{\rm (b)}(\omega)$ is taken to
be the difference $\epsilon_2(\omega)-\delta\epsilon_2^{\rm (f)}(\omega)$.
Because $\epsilon_2$ is dominated by $\delta\epsilon_2^{\rm (f)}$ 
at low frequencies,
the low-frequency behavior of $\delta\epsilon_2^{\rm (b)}$
is unimportant.
For mathematical convenience, we
assume $\delta\epsilon_2^{\rm (b)}\propto \omega$ for $h\nu\ltsim 0.3\eV$.

\begin{figure}[ht]
\begin{center}
\includegraphics[width=7.0cm,angle=0]{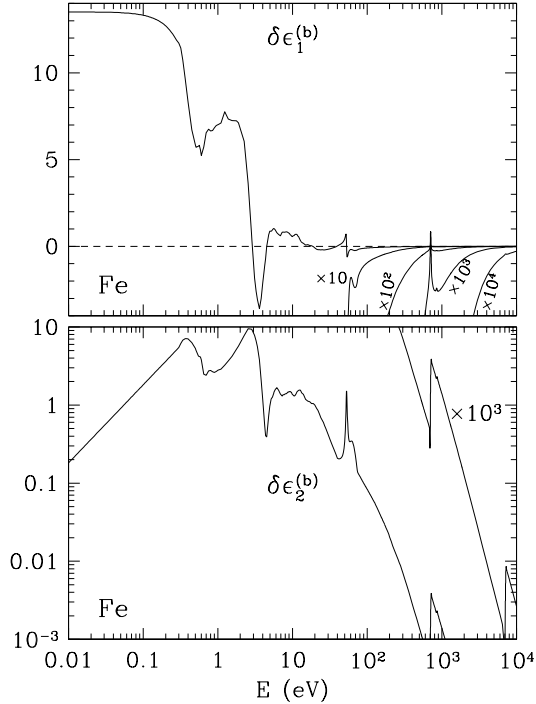}
\caption{\label{fig:delta_eps1_and_delta_eps2}
         \footnotesize
         Real and imaginary components of the bound-electron
         contribution $\delta\epsilon^{(b)}(\omega)$ to the
         dielectric function of Fe.
         }
\end{center}
\end{figure}

\subsection{Temperature Dependence and Effects of Impurities}

\citet{Fischera_2004} noted that the strong $T$ dependence of
the electrical conductivity $\sigma(\omega)$
of bulk Fe implies strong $T$ dependence of
the absorption cross section in the FIR.
The bulk electrical conductivity $\sigma(0)$ of pure Fe
increases by a factor 335 (from $9.2\times10^{16}\s^{-1}$ to
$3.1\times10^{19}\s^{-1}$) as $T$ is reduced
from $293\K$ to $20\K$ \citep{Ho+Ackerman+Wu+etal_1983}.
Thermal contraction causes the $\omega_{p,j}^2$ to increase by
only about 0.2\%, and therefore the large increase in $\sigma(0)$ is almost
entirely due to a decrease in the scattering rates $\tau_j^{-1}$.
Pure Ni shows a similar increase in conductivity with decreasing $T$.

The solar ratio of Fe:Ni is 95:5.
If Fe grains exist in the ISM, it seems likely that they will
contain appreciable amounts of Ni.
Even a few \% of Ni alloyed with Fe strongly limits the rise
in electrical
electrical conductivity at low temperatures: 1\% Ni by mass limits the
$T=20\K$ conductivity to $4.7\times10^{17}\s^{-1}$, and
5\% Ni limits it to $1.1\times10^{17}\s^{-1}$
\citep{Ho+Ackerman+Wu+etal_1983}.
The $T=20\K$ conductivity of a Fe:Ne::95:5 alloy is only $2.4$ times larger
than the $T=293\K$ value, and is similar to the $T=293\K$ conductivity
of pure Fe.

In view of the likely importance of impurity scattering, we will 
neglect $T$-dependence of the electrical
conductivity and $\epsilon(\omega)$, and we will use the
room-temperature properties of pure Fe as a reasonable approximation to the
properties of interstellar Fe:Ni particles at low ($10$ -- $50\K$)
temperatures.

\subsection{Small Particle Effects}

The Fermi velocity in Fe is $v_{\rm F}=1.98\times10^8\cm\s^{-1}$
\citep{Ashcroft+Mermin_1976}. For Drude components 2 and 3 
(see Table \ref{tab:drude_params}), the
electron mean free path $v_{\rm F}\tau_j$ may be comparable to or larger than
the grain size, and therefore the dielectric function must be modified
to allow for electron scattering at the grain surface.
To allow for this, we follow previous workers
\citep{Draine+Lee_1984,Fischera_2004} and set
\beq
\tau_j^{-1} = \tau_{b,j}^{-1} + \frac{v_{\rm F}}{a}
~~~.
\eeq
This modification of the free-electron contribution to the dielectric
function has no noticeable effect at optical frequencies or higher,
but, because of the relatively long mean free path $v_{\rm F}\tau_{b,j}$ for
free-electron components $j=2$ and $3$ (see Table \ref{tab:drude_params}), 
electron scattering from the
grain surface can affect the far-infrared (FIR) absorption when the
particle size drops below $\sim 1\micron$.

\section{\label{app:epsilon_Fe3O4}
         Dielectric Function and Conductivity of Magnetite Fe$_3$O$_4$}

Unfortunately, there does not appear to be a published dielectric function
for magnetite extending from the far-infrared to the far-UV.
In the infrared, there are often large disagreements between different
studies.
Here we attempt to synthesize a complete dielectric function for magnetite from
microwave to X-ray frequencies.

\begin{figure}[ht]
\begin{center}
\includegraphics[angle=0,width=9.5cm]{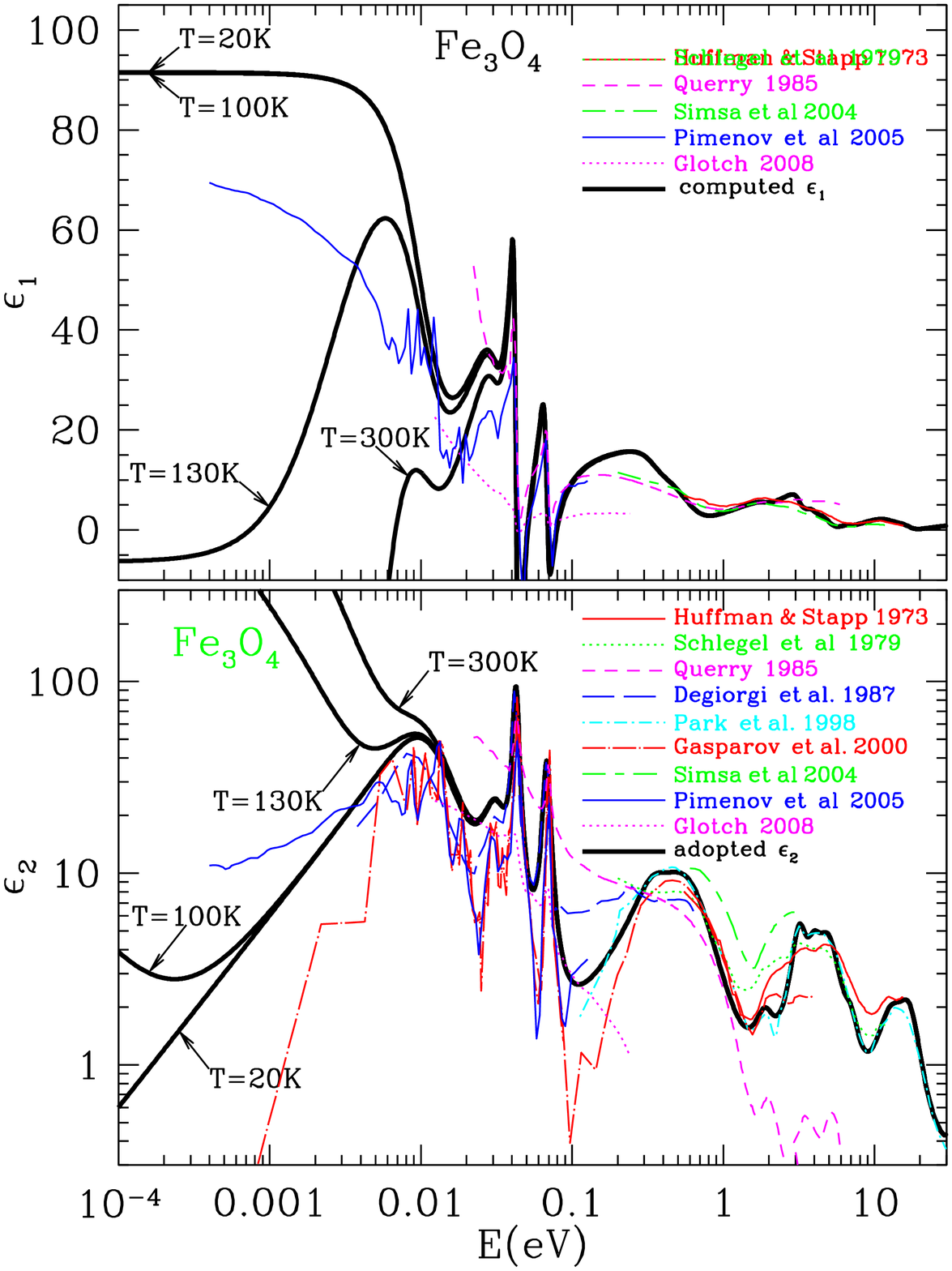}
\caption{\label{fig:epsilon_for_magnetite}\footnotesize
         Lower: solid line $=\epsilon_2(\omega)$ for bulk
         magnetite at $T=300\K$, $130\K$, $100\K$, and $20\K$ (see text)
         together with $\epsilon_2$ from
         \citet{Huffman+Stapp_1973},
         \citet{Schlegel+Alvarado+Wachter_1979}, 
         \citet{Querry_1985},
         \citet{Degiorgi+Wachter+Ihle_1987},
         \citet{Park+Ishikawa+Tokura_1998},
         \citet{Gasparov+Tanner+Romero+etal_2000},
         \citet{Simsa+Thailhades+Presmanes+Bonningue_2002},
         \citet{Pimenov+Tachos+Rudolf+etal_2005}, and
         \citet{Glotch_2008}.
         For $T<80\K$, the contribution of free electrons is negligible.
         Upper: solid line $=\epsilon_1(\omega)$ obtained from $\epsilon_2$
         using the Kramers-Kronig relation (\ref{eq:KK_integral}),
         together with values from the literature.
         }
\end{center}
\end{figure}

At room temperature, magnetite is a conductor, with d.c.\ conductivity
$\sigma_{\rm dc}\approx 250\ohm^{-1}\cm^{-1}$$= 2.3\times10^{14}\s^{-1}$, 
but $\sigma_{\rm dc}$
decreases abruptly, by a factor $\sim$$10^2$, upon cooling
through the Verwey transition temperature $T_V=119\K$.
At lower temperatures it behaves like
a semiconductor.
For high-quality samples, the d.c.\ conductivity is
\beqa
\sigma_{\rm dc}
&\approx&
\left(3\times10^{16}e^{-1330\K/T}+3\times10^{12}e^{-740\K/T}\right)\s^{-1}
\hspace*{1.0em}{\rm for~}45\K < T < T_V
\\
&\approx&
2.3\times10^{14}\s^{-1}\left(\frac{T-100\K}{160\K}\right) 
\hspace*{8em}{\rm for~} T_V < T \ltsim 260\K
\\
&\approx&2.3\times10^{14}\s^{-1} 
\hspace*{14.4em}{\rm for~}260\K < T < 450\K
\eeqa
\citep{Miles+Westphal+vonHippel_1957,Degiorgi+Wachter+Ihle_1987,
Tsuda+Nasu+Yanase+Siratori_1991}.
Thus $\sigma_{\rm dc} \ltsim 3\times10^4\s^{-1}$ at the $T\leq 40\K$
temperatures of interstellar grains.

The ``free-electron'' contribution to the dielectric function is
(see eq.\ \ref{eq:Drude model})
\beq \label{eq:free-electron}
\delta \epsilon^{\rm (f)} = 
\frac{4\pi i\sigma_{\rm dc}}{\omega} \times \frac{1}{(1-i\omega\tau)}
~~~,
\eeq
where $\tau$ is the mean-free-time between scatterings by phonons,
impurities, or boundaries.  
\citet{Degiorgi+Wachter+Ihle_1987} estimated 
$\tau_{\rm bulk}=3.1\times10^{-13}\s$ at
$T=130\K$; we use this value for $T<130\K$.
For a Fermi speed $v_{\rm F}\approx10^8\cm\s^{-1}$
this corresponds to a
mean-free-path $v_{\rm F}\tau_{\rm bulk}\approx3\times10^{-5}\cm$.
For nanoparticles we take 
\beqa
\tau^{-1}&=&\tau_{\rm bulk}^{-1}+\frac{v_{\rm F}}{a}
\\
\delta\epsilon^{\rm (f)} &=& 
\frac{4\pi i \sigma_{\rm dc}\tau}{\omega\tau_{\rm bulk}}\times
\frac{1}{1-i\omega\tau}
\eeqa

At room temperature $\sigma_{\rm dc}$ is 
large enough that $\delta\epsilon^{\rm (f)}$
makes a significant contribution to $\epsilon(\omega)$ for
$\omega < 10^{14}\s^{-1}$.
For $T<40\K$, $\sigma_{\rm dc}$ is small enough that $\delta\epsilon^{\rm (f)}$
can be neglected for
$\omega/2\pi \gtsim 1 \GHz$.

\begin{table}[bt]
\begin{center}
\caption{\label{tab:magnetite}
         Resonance parameters for Fe$_3$O$_4$}
{\footnotesize
\begin{tabular}{l c c c c c}
\hline
$j$   &$\hbar\omega_j (\eV)$
                & $\lambda_j (\micron)$ 
                        & $\gamma_j$ 
                                   & $S_j$ & ref \\
\hline
1     &  0.0115 & 108.  & 1.226    & 54.9  & \citet{Degiorgi+Wachter+Ihle_1987}
\\
2     &  0.0315 &  39.  & 0.448    &  7.51 & \citet{Degiorgi+Wachter+Ihle_1987}
\\
3     &  0.0425 &  29.  & 0.0965   &  8.47 & \citet{Degiorgi+Wachter+Ihle_1987}
\\
4     &  0.068  &  18.  & 0.1162   &  4.20 & \citet{Degiorgi+Wachter+Ihle_1987}
\\
5     &  0.35   &  3.5  & 0.60     & 2.0   & this work\\
6     &  0.60   &  2.1  & 1.00     & 8.5   & this work\\
7     &  1.9    &  0.65 & 0.35     & 0.30  & this work\\
8     &  3.2    &  0.39 & 0.25     & 0.90  & this work\\
9     &  4.0    &  0.31 & 0.25     & 0.45  & this work\\
10    &  5.0    &  0.25 & 0.42     & 1.60  & this work\\
11    &  7.0    &  0.177 & 0.28    & 0.21  & this work\\
12    &  13.    &  0.095 & 0.47    & 0.55  & this work\\
13    &  17.    &  0.073 & 0.40    & 0.60  & this work\\
\hline
\end{tabular}
}
\end{center}
\end{table}

Guided by available data (shown in Figure \ref{fig:epsilon_for_magnetite})
we adopt the function $\epsilon_2(\omega)$ shown in 
Figure \ref{fig:epsilon_for_magnetite}.
At optical and infrared frequencies, 
the dielectric function is approximated by 
\beq
\epsilon(\omega) = 1 + \Delta\epsilon + \delta\epsilon^{\rm (f)} + 
\sum_{j=1}^{13} \frac{S_j}{1-(\omega/\omega_j)^2 - i\gamma_j(\omega/\omega_j)}
\eeq
where the $\omega_j$, $\gamma_j$, and $S_j$ are 
listed in Table \ref{tab:magnetite}
and $\Delta\epsilon=0.375$ is the expected contribution to ${\rm Re}(\epsilon)$
from absorption at $h\nu\gtsim30\eV$.
Resonance parameters for $j\leq 4$ are taken from the $T=130\K$ results
of \citet{Degiorgi+Wachter+Ihle_1987}.
Adding resonances $5\leq j\leq 13$ generates a dielectric function that
is consistent with a subset of the
data in Figure \ref{fig:epsilon_for_magnetite}.

For $h\nu>30\eV$
we estimate $\epsilon_2$ from the sum of the absorption
cross sections for 3 neutral Fe atoms and 4 neutral O atoms.
Our adopted $\epsilon_2(\omega)$, together with
the various experimental determinations,
are shown in Figure \ref{fig:epsilon_for_magnetite}.
We calculate the real part $\epsilon_1$ from our adopted $\epsilon_2(\omega)$
using the Kramers-Kronig relation (\ref{eq:KK_integral}).
The resulting $\epsilon_1$ is shown in Fig. \ref{fig:epsilon_for_magnetite}.

\citet{Tikhonov+Boyarskii+Polyakova+etal_2010} measured the 
12--145$\GHz$ dielectric function of magnetite at $T\approx300\K$.
At 140$\GHz$ they found $m\approx20+0.1i$, or $\epsilon=m^2\approx
400+4i$.
With the room temperature conductivity $\sigma_{\rm dc}=2.3\times10^{14}\s^{-1}$
we would expect ${\rm Im}(\epsilon)=4\pi\sigma_{\rm dc}/\omega\approx3300$
at 140$\GHz$ -- almost 3 orders of magnitude larger than the value
found by \citet{Tikhonov+Boyarskii+Polyakova+etal_2010}.
The results of \citet{Tikhonov+Boyarskii+Polyakova+etal_2010}
are inconsistent with the results of
\citet{Degiorgi+Wachter+Ihle_1987} and
\citet{Pimenov+Tachos+Rudolf+etal_2005} but
the reason for the discrepancy is unclear.

\begin{figure}[t]
\begin{center}
\includegraphics[angle=0,width=7.0cm]{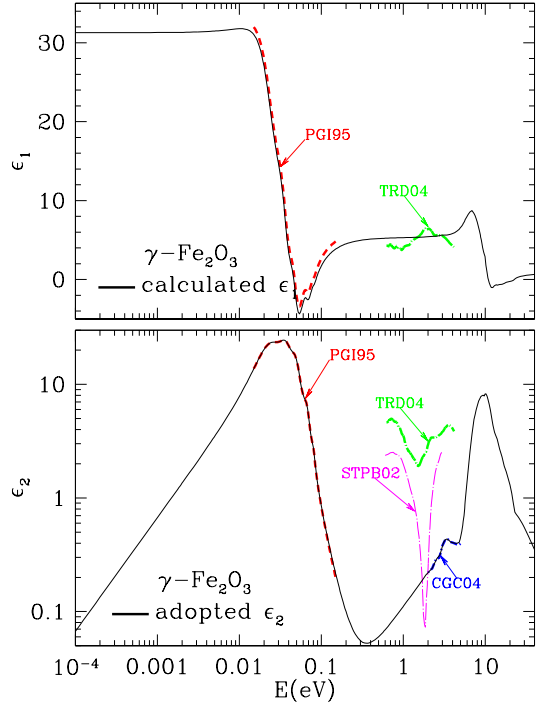}
\caption{\label{fig:epsilon_for_maghemite}\footnotesize
         Lower: solid line $=\epsilon_2(\omega)$ adopted for
         maghemite.
         Upper: solid line $=\epsilon_1(\omega)$ derived from the
         adopted $\epsilon_2$.
         Experimental results (see text):
         PGI95 \citep{Pecharroman+Gonzalez-Carreno+Iglesias_1995};
         STPB02 \citep{Simsa+Thailhades+Presmanes+Bonningue_2002};
         CGC04 \citep{Chakrabarti+Ganguli+Chaudhuri_2004};
         TRD04 \citep{Tepper+Ross+Dionne_2004}.
         }
\end{center}
\end{figure}
\section{Dielectric Function of Maghemite $\gamma$-Fe$_2$O$_3$
         \label{app:epsilon_maghemite}}

Maghemite ($\gamma$-Fe$_2$O$_3$)
is considered to be a semiconductor.
The band gap is variously estimated to be $E_g=2.03\eV$
\citep{Cornell+Schwertmann_2003}
and $2.43\eV$ \citep{Chakrabarti+Ganguli+Chaudhuri_2004}.
Maghemite has a spinel structure.
Each unit cell 
contains vacancies that may or may not have long-range order.  
If the vacancies are located randomly, the material is said to have
$Fd\bar{3}m$ symmetry.
The infrared dielectric function of $Fd\bar{3}m$ $\gamma$-Fe$_2$O$_3$ has
been determined by \citet{Pecharroman+Gonzalez-Carreno+Iglesias_1995}
from 0.015 -- $0.15\eV$ reflectivity measurements.
\citet{Pecharroman+Gonzalez-Carreno+Iglesias_1995} fit the IR behavior
with 5 damped oscillators.

\citet{Chakrabarti+Ganguli+Chaudhuri_2004} have measured the 
absorption coefficient $\alpha(h\nu)$ in $\gamma$-Fe$_2$O$_3$ films
over the range $1.95 < h\nu/{\rm eV} < 4.95$.
For $1.95<(h\nu/\eV)<2.5$, 
the measured absorption coefficient greatly exceeded that
expected for a semiconductor with a $2.43\eV$ bandgap.

We obtain $\epsilon_2$ from the measured $\alpha$
\beq \label{eq:eps2_from_alpha}
\epsilon_2 = 2{\rm Re}(m) {\rm Im}(m) 
= \frac{{\rm Re}(m)\alpha\lambda_{\rm vac}}{2\pi}
~~~,
\eeq
where $m$ is the complex refractive index, and
${\rm Re}(m)\approx 2.5$ in the optical
\citep{Pecharroman+Gonzalez-Carreno+Iglesias_1995}.
We assumed the electronic contribution to $\epsilon_2 \propto h\nu$
for $h\nu<2\eV$.
At $h\nu\approx1\eV$ this extrapolation approximately reproduces
the measured absorption in amorphous Fe$_2$O$_3$ films
\citep{Ozer+Tepehan_1999}.

\citet{Tepper+Ross+Dionne_2004} have measured the optical constants
of pulsed-laser-deposited iron oxide films between
$0.6$ and $4.1\eV$.  X-ray diffraction studies indicate the films
to consist of polycrystalline $\gamma$-Fe$_2$O$_3$, but the
measured absorption was 5--10 times
stronger than in the ``pure $\gamma$-Fe$_2$O$_3$'' nanoparticles
studied by \citet{Chakrabarti+Ganguli+Chaudhuri_2004}.
\citet{Tepper+Ross+Dionne_2004} found the absorption to depend
on the deposition conditions, weakening if oxygen is present,
and they attribute the strong absorption 
to an excess of Fe$^{2+}$ ions present in the films.
Given the availability of oxygen in the ISM, we will adopt the
\citet{Chakrabarti+Ganguli+Chaudhuri_2004} results, but the
\citet{Tepper+Ross+Dionne_2004} data indicate the large uncertainty
concerning the optical absorption properties of $\gamma$-Fe$_2$O$_3$ material.

Unfortunately, there do not appear to be published measurements
of $\gamma-$Fe$_2$O$_3$ above $5\eV$.
For $E>25\eV$ we estimate the absorption as the sum of the
photoelectric absorption by isolated Fe and O atoms, and
obtain $\epsilon_2$ from (\ref{eq:eps2_from_alpha})
with ${\rm Re}(m)\approx 1$.
For $h\nu>5\eV$ we adopt an entirely ad-hoc $\epsilon_2(E)$
that has sufficiently strong absorption in the 5--25$\eV$ range to be
consistent with the measured $\epsilon_1$ in the visible.

The resulting dielectric function should be reasonably accurate in the
infrared and optical, but at vacuum-UV energies ($h\nu>5\eV$) should
be regarded as merely illustrative of the expected strong absorption.
Figure \ref{fig:epsilon_for_maghemite} shows the adopted dielectric
function for $\gamma-$Fe$_2$O$_3$.

\section{\label{app:polarization_for_spinning_grain}
         Time-Averaged Orientations for a Spinning Grain}

Define a Cartesian coordinate system $\bxhat\byhat\bzhat$ where
$\bxhat$ is the direction to the observer.
Let $\Phi$ be the angle between the static magnetic field $\bH_0$
and the line of sight.
Without loss of generality, assume that 
$\bH_0$ is in the $\bxhat\byhat$ plane:
\beq
\bH_0=H_0 \left( \bxhat\cos\Phi + \byhat\sin\Phi \right)
~~~.
\eeq
Consider a prolate spheroid, spontaneously magnetized along the
long axis.  The grain is assumed to be spinning with the long
axis perpendicular to the angular momentum $\bJ$.
The angular momentum $\bJ$ will precess around $\bH_0$.
Let $\psi$ be the angle between $\bJ$ and $\bH_0$,
and let $\zeta$ be a precession angle:
\beq
\bJhat=(\bxhat\cos\Phi+\byhat\sin\Phi)\cos\psi
+ \sin\psi\left[
  \left(\bxhat\sin\Phi-\byhat\cos\Phi\right)\cos\zeta
  +
  \bzhat\sin\zeta
  \right]
~~~.
\eeq
Let $\bzmhat$ be the direction of magnetization of the grain; the
$\bxmhat$,$\bymhat$ directions are perpendicular to the magnetization.
Let $\bxmhat$ be parallel to $\bJ$; $\bymhat$ and $\bzmhat$ will
spin around $\bJ$ with the grain's angular velocity $\Omega$:
\beqa
\bxmhat &=& 
\left( \cos\Phi\cos\psi+\sin\Phi\sin\psi\cos\zeta \right)\bxhat +
\left( \sin\Phi\cos\psi-\cos\Phi\sin\psi\cos\zeta \right)\byhat +
\left(\sin\psi\sin\zeta\right)\bzhat ~~~~~
\\
\bymhat &=& \nonumber
\left[ -\sin\Phi\sin\zeta\cos\Omega t + 
     \left( \sin\Phi\cos\psi\cos\zeta -\cos\Phi\sin\psi \right)\sin\Omega t
\right]\bxhat +
\\
&& \nonumber
\left[ \cos\Phi\sin\zeta\cos\Omega t - 
     \left(\cos\Phi\cos\psi\cos\zeta+\sin\Phi\sin\psi \right)\sin\Omega t
\right]\byhat +
\\
&&
\left[ \cos\zeta\cos\Omega t + \cos\psi\sin\zeta\sin\Omega t
\right]\bzhat
\\
\bzmhat &=& \nonumber
\left[ \sin\Phi\sin\zeta\sin\Omega t + 
     \left( \sin\Phi\cos\psi\cos\zeta -\cos\Phi\sin\psi \right)\cos\Omega t
\right]\bxhat +
\\
&& \nonumber
\left[ -\cos\Phi\sin\zeta\sin\Omega t - 
     \left(\cos\Phi\cos\psi\cos\zeta+\sin\Phi\sin\psi \right)\cos\Omega t
\right]\byhat +
\\
&&
\left[- \cos\zeta\sin\Omega t + \cos\psi\sin\zeta\cos\Omega t
\right]\bzhat
~~~.
\eeqa
Thus, noting that $\langle\sin^2\zeta\rangle=\langle\cos^2\zeta\rangle
= \langle\sin^2\Omega t\rangle = \langle\cos^2\Omega t\rangle = 1/2$,
we have
\beqa
\langle (\byhat\cdot\bxmhat)^2\rangle &=&
\sin^2\Phi \cos^2\psi  + \frac{1}{2}\cos^2\Phi\sin^2\psi
\\
\langle (\byhat\cdot\bymhat)^2\rangle =
\langle (\byhat\cdot\bzmhat)^2\rangle &=&
\frac{1}{4}\cos^2\Phi (1 + \cos^2\psi) + \frac{1}{2}\sin^2\Phi\sin^2\psi
\\
\langle (\bzhat\cdot\bxmhat)^2\rangle &=&
\frac{1}{2}\sin^2\psi
\\
\langle (\bzhat\cdot\bymhat)^2\rangle =
\langle (\bzhat\cdot\bzmhat)^2\rangle &=&
\frac{1}{4}\left(1+\cos^2\psi\right)
~~~.
\eeqa
\section{\label{app:eff_med_theory}
         Effective Medium Theory}
Laboratory measurements of absorption in magnetic nanoparticles
generally study samples where the particles are dispersed in a
nonmagnetic dielectric matrix.  
Let the matrix have scalar dielectric function $\epsilon_\mat$
and magnetic permeability $\mu_\mat=1$.
Assume the magnetic nanoparticles to be spherical, with
volume filling factor $\ffill$, and to be characterized
by scalar dielectric function $\epsilon_{\rm m}$ and 
magnetic polarizability tensor $\balpha_{\rm m}$ given by
eq.\ (\ref{eq:alpha_mag}).

We wish to approximate the composite medium by a uniform
medium with an effective dielectric function $\epsilon_\eff(\omega)$
and effective permeability $\mu_\eff(\omega)$.
Unfortunately, there is no exact solution to this problem, with various
competing suggestions for how to estimate $\epsilon_\eff$ and $\mu_\eff$
\citep[see the discussion in][]{Bohren+Huffman_1983}.
The effective medium formulation due to Maxwell Garnett is often used,
with
\beq
\epsilon_\eff = 
\frac{\left(1-\ffill\right)\epsilon_\mat
      \left(\epsilon_{\rm m}+2\epsilon_\mat\right) + 
      3\ffill \epsilon_\mat \epsilon_{\rm m}
     }
     {\left(1-\ffill\right)
      \left(\epsilon_{\rm m}+2\epsilon_\mat\right) +
      3\ffill\epsilon_\mat
     }
~~~.
\eeq
To obtain $\mu_\eff$
we must take into account the magnetic anisotropy of the
magnetic nanoparticles.
Suppose that the directions of spontaneous magnetization are
randomly distributed.  In the dipole limit, this is the same
as if 1/6 of the particles have their static magnetization $\bM_0$
oriented in each of the
$\pm\bxhat$, $\pm\byhat$, and $\pm\bzhat$ directions.
Thus we model the system as consisting of a matrix with 6 types of
inclusions, $j=1-6$.  As in eq.\ (\ref{eq:magnetization_response}), 
each type of inclusion develops an internal
magnetization in response to the field in the matrix
$\bh_\mat$:
\beq
\bm_j = \left[\chi_+ \bhhat_{j+}\left(\bhhat_{j+}^*\cdot\bh_\mat\right)
             +\chi_- \bhhat_{j-}\left(\bhhat_{j-}^*\cdot\bh_\mat\right)
        \right]
~~~,
\eeq
where
\beqa \nonumber
\bhhat_{1\pm} &=& (\bxhat \pm i\byhat)/\sqrt{2} ~~~~{\rm for}~~ \bM_{0}=M_0\bzhat
\\ \nonumber
\bhhat_{2\pm} &=& (\byhat \pm i\bzhat)/\sqrt{2} ~~~~{\rm for}~~ \bM_{0}=M_0\bxhat
\\ \nonumber
\bhhat_{3\pm} &=& (\bzhat \pm i\bxhat)/\sqrt{2} ~~~~{\rm for}~~ \bM_{0}=M_0\byhat
\\ \nonumber
\bhhat_{4\pm} &=& (\bxhat \mp i\byhat)/\sqrt{2} ~~~~{\rm for}~~ \bM_{0}=-M_0\bzhat
\\ \nonumber
\bhhat_{5\pm} &=& (\byhat \mp i\bzhat)/\sqrt{2} ~~~~{\rm for}~~ \bM_{0}=-M_0\bxhat
\\ 
\bhhat_{6\pm} &=& (\bzhat \mp i\bxhat)/\sqrt{2} ~~~~{\rm for}~~ \bM_{0}=-M_0\byhat
~~~.
\eeqa
It is easily shown that
\beqa
\frac{1}{6}\sum_{j=1}^6 \bm_j &=& \left(\frac{\chi_++\chi_-}{3}\right)\bh_\mat
~~~.
\eeqa
The volume-averaged magnetization is
\beq
\langle\bm\rangle =
\frac{\ffill}{6}\sum_{j=1}^6  \bm_j
=
\ffill\frac{(\chi_++\chi_-)}{3}
\bh_\mat
~~~.
\eeq
The oscillating field within an inclusion of type $j$ is
\beq
\bh_j = \bh_\mat-D\bm_j
~~~,
\eeq
where $D$ is the demagnetization tensor [see eq.\ (\ref{eq:D_jj})],
with $D=4\pi/3$ for spherical inclusions.
The volume-averaged oscillating field is
\beqa
\langle \bh \rangle &=& (1-\ffill)\bh_\mat + 
\frac{\ffill}{6} 
\sum_{j=1}^6 (\bh_\mat - D\bm_j)
\\
&=& \left[1 - \ffill
D\frac{\left(\chi_++\chi_-\right)}{3}\right]\bh_\mat
~~~.
\eeqa
The effective permeability is
\beqa
\mu_\eff \equiv 1 + 4\pi\frac{\langle \bm\rangle}{\langle\bh\rangle}
&=& 1 + \frac{4\pi\ffill(\chi_++\chi_-)/3}
           {1-
            \ffill
            D(\chi_+\chi_-)/3}
\\ \label{eq:mu_eff_sphere}
&=& 1 + \frac{12\pi\ffill(\chi_++\chi_-)}
           {9-4\pi\ffill(\chi_++\chi_-)}
~~~.
\eeqa
where 
eq.\ (\ref{eq:mu_eff_sphere}) is for spherical inclusions.
The effective complex refractive index is
$m_\eff=\sqrt{\epsilon_\eff\mu_\eff}$.
A wave propagating through the material has an attenuation coefficient
\beq
\alpha = \frac{2\omega}{c} {\rm Im}\left(\sqrt{\epsilon_\eff\mu_\eff}\right)
~~~.
\eeq

\end{appendix}

\bibliography{btdrefs}

\end{document}